\documentclass[a4paper,10pt,fleqn,oneside]{article}

\usepackage[english]{babel}
\usepackage[utf8]{inputenc}
\usepackage[T1]{fontenc}

\usepackage{lmodern}

\usepackage[hidelinks]{hyperref}

\usepackage{amsmath,amsfonts,amssymb,mathtools,mathrsfs,dsfont,mathabx}
\usepackage{accents}

\usepackage[left=2cm,right=2cm,top=2cm,bottom=3cm]{geometry}

\usepackage{graphicx}
\graphicspath{{./}}

\usepackage{xcolor}

\usepackage{array,booktabs,multirow,makecell,tabu,colortbl}

\usepackage{enumerate}

\usepackage{subcaption}

\usepackage[numbers,sort&compress]{natbib}
\usepackage{doi}

\newcommand{\eb}{\boldsymbol{e}}

\newcommand{\xb}{\boldsymbol{x}}

\newcommand{\Ab}{\boldsymbol{A}}
\newcommand{\Bb}{\boldsymbol{B}}

\newcommand{\Eb}{\boldsymbol{E}}
\newcommand{\Fb}{\boldsymbol{F}}
\newcommand{\Gb}{\boldsymbol{G}}

\newcommand{\Xb}{\boldsymbol{X}}

\newcommand{\Bcal}{\mathcal{B}}
\newcommand{\Ccal}{\mathcal{C}}

\newcommand{\Gcal}{\mathcal{G}}

\newcommand{\Mcal}{\mathcal{M}}

\newcommand{\Wcal}{\mathcal{W}}

\newcommand{\etb}{\boldsymbol{\eta}}

\newcommand{\rhb}{\boldsymbol{\rho}}

\newcommand{\phb}{\boldsymbol{\phi}}

\newcommand{\dd}{\textrm{d}}
\newcommand{\de}{\partial}
\newcommand{\Nabla}{\boldsymbol{\nabla}}
\newcommand{\eps}{\varepsilon}
\newcommand{\imag}{\text{Im}\,}
\newcommand{\wb}{\widebar}
\newcommand{\wh}{\widehat}
\newcommand{\Rbb}{\mathbb{R}}
\newcommand{\ldot}[1]{\accentset{\mbox{\Large \hspace{0.25ex}.}}{#1}}

\title{Solving hyperbolic-elliptic problems on singular mapped disk-like domains
with the method of characteristics and spline finite elements}

\author{
Edoardo Zoni$^{\,a,b}$, Yaman Güçlü$^{\,a}$ \\[5mm]
{\small $^a$ {\it  Max-Planck-Institut für Plasmaphysik, Boltzmannstraße 2, 85748 Garching }} \\
{\small $^b$ {\it  Technische Universität München, Zentrum Mathematik, Boltzmannstraße 3, 85748 Garching }}}

\date{}

\begin{document}

\maketitle

\begin{abstract}
A common strategy in the numerical solution of partial differential equations is
to define a uniform discretization of a tensor-product multi-dimensional logical
domain, which is mapped to a physical domain through a given coordinate transformation.
By extending this concept to a multi-patch setting, simple and efficient numerical algorithms
can be employed on relatively complex geometries. The main drawback of such an approach
is the inherent difficulty in dealing with singularities of the coordinate transformation.

This work suggests a comprehensive numerical strategy for the common situation
of disk-like domains with a singularity at a unique pole, where one edge of the rectangular
logical domain collapses to one point of the physical domain (for example,
a circle). We present robust numerical methods for the solution of Vlasov-like
hyperbolic equations coupled to Poisson-like elliptic equations in such geometries.
We describe a semi-Lagrangian advection solver that employs a novel set of
coordinates, named pseudo-Cartesian coordinates, to integrate the characteristic
equations in the whole domain, including the pole, and a finite element elliptic
solver based on globally $\Ccal^1$ smooth splines (Toshniwal et al., 2017). The two
solvers are tested both independently and on a coupled model, namely the 2D guiding-center
model for magnetized plasmas, equivalent to a vorticity model for incompressible
inviscid Euler fluids. The numerical methods presented show high-order convergence
in the space discretization parameters, uniformly across the computational domain,
without effects of order reduction due to the singularity. Dedicated tests
show that the numerical techniques described can be applied straightforwardly
also in the presence of point charges (equivalently, point-like vortices),
within the context of particle-in-cell methods.
\end{abstract}

\section{Introduction}

This work is concerned with the solution of coupled hyperbolic and elliptic partial
differential equations (PDEs) on disk-like domains. These represent typically
an approximation of more complex non-circular physical domains, where the PDEs describing
the physical system under study need to be solved. It is sometimes useful to parametrize
such physical domains by curvilinear coordinates instead of Cartesian coordinates.
Such coordinates, which we refer to as logical coordinates, may allow, for example,
to describe more easily the boundary of the physical domain of interest. This can
be then obtained from the logical domain by applying a coordinate transformation,
which may introduce artificial singularities. In this work we denote by ${\wh\Omega
:=[0,1]\times[0,2\pi)}$ the logical domain and by $\Omega\subset\Rbb^2$ the physical
domain, which is the image of $\wh\Omega$ through a given coordinate mapping $\Fb:
\wh\Omega\to\Rbb^2$. In other words, $\Omega:=\Fb(\wh\Omega)$. Moreover, we denote
by ${\etb:=(s,\theta)\in\wh\Omega}$ and $\xb:=(x,y)\in\Omega$ the logical and Cartesian
coordinates, respectively: $s\in[0,1]$ is a radial-like coordinate and $\theta\in
[0,2\pi)$ is an angle-like $2\pi$-periodic coordinate. We consider, in particular,
logical domains with a singularity at a unique pole, where the edge $s=0$ of $\wh
\Omega$ collapses to the point $(x_0,y_0)$ of $\Omega$ (the pole) through the mapping
$\Fb$. The simplest example is a circular physical domain parametrized by polar coordinates
$(r,\theta)$ instead of Cartesian coordinates $(x,y)$: the polar transformation that
maps $(r,\theta)$~to~$(x,y)$ becomes singular (that is, not invertible) at the center
of the domain as $r\to0^+$. Our approach is alternative to other standard strategies,
such as, for example, employing Cartesian coordinates and treating boundary conditions
by means of the inverse Lax-Wendroff procedure~\citep{TanShu2010}. Our target
model is the 2D guiding-center model \citep{ONeil1985,DubinONeil1988}
\begin{equation}
\label{model_equations}
\begin{cases}
\dfrac{\de\rho}{\de t}-E^y\dfrac{\de\rho}{\de x}+E^x\dfrac{\de\rho}{\de y}=0 \,, \\[2mm]
-\Nabla\cdot\Nabla\phi=\rho \,,
\end{cases}
\quad \text{with} \quad
\begin{cases}
\rho(0,x,y)=\rho_\text{IN}(x,y) \,, \\[2mm]
\phi(t,x,y)=0 \textrm{ on } \de\Omega \,.
\end{cases}
\end{equation}
In the context of plasma physics, \eqref{model_equations} is typically used
to describe low-density non-neutral plasmas \citep{Davidson2001,Driscolletal2002,
SenguptaGanesh2014,SenguptaGanesh2015} in a uniform magnetic field $\Bb$ aligned
with the direction perpendicular to the $(x,y)$ plane. The unknowns in \eqref{model_equations}
are the density of the plasma particles $\rho$ and the electric scalar potential
$\phi$, related to the electric field via ${\Eb=(E^x,E^y)^T=-\Nabla\phi}$.
The advection field $(-E^y,E^x)^T$, responsible for the transport of $\rho$ in
\eqref{model_equations}, represents the $\Eb\times\Bb$ drift velocity. From a
mathematical point of view, this model is also equivalent to the 2D Euler equations for incompressible inviscid fluids,
with $-\rho$ representing the vorticity of the fluid and $\phi$ a stream function.
Indeed, \eqref{model_equations} has been investigated also in the fluid
dynamics community for a variety of problems related to vortex dynamics and
turbulence \citep{GaneshLee2002,SchecterDubin1999a,Schecteretal1999b,SchecterDubin2001}.

Regarding the numerical solution of \eqref{model_equations}, we are interested in
solving the hyperbolic part (transport advection equation for $\rho$) with the method
of characteristics and the elliptic part (Poisson's equation for $\phi$) with a finite
element method based on $B$-splines. More precisely, the advection equation is solved
by computing $\rho$ on a grid following the characteristics backward in time for a
single time step and interpolating at the origin of the characteristics using
neighboring grid values of $\rho$ at the previous time step. This procedure
is referred to as backward semi-Lagrangian method and was originally developed
within the context of numerical weather prediction \citep{Fjortoft1952,Fjortoft1955,
Wiin-Nielsen1959,Krishnamurti1962,Sawyer1963,Leith1964,Purnell1976} (see, for example,~\citep{StaniforthCote1991}
for a comprehensive review). The method was applied later
on to Vlasov-like transport equations and drift-kinetic and gyrokinetic models in
the context of plasma physics \citep{ChengKnorr1976,GagneShoucri1977,Sonnendrueckeretal1999,
Filbetetal2001,BesseSonnendruecker2003,Crouseillesetal2010}. One advantage of the
semi-Lagrangian method is to avoid any limitation related to the Courant-Friedrichs-Lewy
(CFL) condition \citep{Courantetal1928} in the region close to the pole, where the
grid cells become smaller and smaller. This scheme works on a structured logical
mesh, which is usually constructed to be conformal to the level curves of some given
function (in physical applications, they may correspond to magnetic field flux surfaces
for plasma models or level curves of the stream function for fluid models). Since
the method is based on the integration of the characteristics, the choice of coordinates
to be used while performing this integration turns out to be crucial: such coordinates
need indeed to be well-defined in the whole domain, including the pole. The
choice of coordinates that we propose, described in detail in section~\ref{sec_advection},
fulfills this aim without affecting the robustness, efficiency and accuracy
of the numerical scheme. The same coordinates can be used as well for the forward
time integration of the trajectories of point charges or point-like vortices.

On the other hand, the elliptic Poisson equation is solved with a finite element method
based on $B$-splines. We require the advection field $(-E^y,E^x)^T$ to be continuous
everywhere in the domain. This means that $\phi$, from which $\Eb$ is
obtained by means of derivatives, has to be $\Ccal^1$ smooth everywhere in the
domain. This is difficult to achieve at the pole. Therefore, we follow the approach
recently developed in \citep{Toshniwaletal2017} to define a set of globally $\Ccal^1$
smooth spline basis functions on singular mapped disk-like domains. A higher
degree of smoothness, consistent with the spline degree, may be imposed as well, if needed.

This paper is organized as follows. Sections \ref{sec_domains} and \ref{sec_discrete_mappings}
describe singular mapped disk-like domains in more detail, together
with their discrete representation. Section \ref{sec_advection} presents our numerical
strategy to solve advection problems on such domains, including numerical tests.
Section \ref{sec_Poisson} describes our finite element elliptic solver based on
globally $\Ccal^1$ smooth splines, including numerical tests. Section \ref{sec_guiding_center}
describes how to couple the two numerical schemes in order to solve a self-consistent
hyperbolic-elliptic problem and presents the results of different numerical tests
in various domains. \\

\textbf{Remark} (\textit{Notation}).
In this work, all quantities defined on the logical domain $\wh\Omega$ are denoted
by placing a hat over their symbols. On the other hand, the corresponding quantities
defined on the physical domain $\Omega$ are denoted by the same symbols without the
hat. For example, denoting by $\alpha$ a scalar quantity of interest, we have $\wh\alpha:
\wh\Omega\to\Rbb$ and $\alpha:\Omega\to\Rbb$, and the two functions are related via
$\wh\alpha=\alpha\circ\Fb$. For time-dependent quantities, the domain $\wh\Omega$
(respectively, $\Omega$) is replaced by $\Rbb^+\times\wh\Omega$ (respectively, $\Rbb^+
\times\Omega$). Moreover, for vector quantities, the codomain $\Rbb$ is replaced by $\Rbb^2$.

\section{Singular mapped disk-like domains}
\label{sec_domains}
As already mentioned, we denote by ${\wh\Omega:=[0,1]\times[0,2\pi)}$ the logical
domain and by $\Omega\subset\Rbb^2$ the physical domain, obtained from $\wh\Omega$ through a given
coordinate mapping $\Fb:\wh\Omega\to\Rbb^2$. Moreover, we denote by $\etb:=(s,\theta)
\in\wh\Omega$ and $\xb:=(x,y)\in\Omega$ the logical and Cartesian coordinates, respectively.
We consider, in particular, logical domains with a singularity at a unique pole,
where the edge $s=0$ of $\wh\Omega$ collapses to the point $(x_0,y_0)$ of $\Omega$
(the pole): $\Fb(0,\theta)=(x_0,y_0)$ for all $\theta$. In the following, two analytical examples
of such mappings are provided. The first mapping is defined in~\citep{Bouzatetal2018} as
\begin{equation}
\label{target}
\begin{aligned}
& x(s,\theta):=x_0+(1-\kappa)s\cos\theta-\Delta\,s^2 \,, \\
& y(s,\theta):=y_0+(1+\kappa)s\sin\theta \,,
\end{aligned}
\end{equation}
where $\kappa$ and $\Delta$ denote the elongation and the Shafranov shift, respectively.
For $s=0$ the mapping collapses to the pole $(x_0,y_0)$.
The Jacobian matrix of the mapping, denoted by $J_{\Fb}$, reads
\begin{equation*}
J_{\Fb}(s,\theta)=
\begin{bmatrix}
(1-\kappa)\cos\theta-2\,\Delta\,s & (\kappa-1)s\sin\theta \\
(1+\kappa)\sin\theta & (1+\kappa)s\cos\theta
\end{bmatrix} \,,
\end{equation*}
with determinant
\begin{equation*}
\det J_{\Fb}(s,\theta)=s(1+\kappa)[(1-\kappa)-2\,\Delta\,s\cos\theta] \,,
\end{equation*}
which vanishes at the pole. The Jacobian matrix of the inverse transformation reads
\begin{equation*}
J_{\Fb}^{-1}(s,\theta)=\frac{1}{\det J_{\Fb}(s,\theta)}
\begin{bmatrix}
(1+\kappa)s\cos\theta & (1-\kappa)s\sin\theta \\
-(1+\kappa)\sin\theta & (1-\kappa)\cos\theta-2\,\Delta\,s
\end{bmatrix} \,,
\end{equation*}
and it is singular at the pole. The second mapping is defined in \citep{CzarnyHuysmans2008} as
\begin{equation}
\label{czarny}
\begin{aligned}
& x(s,\theta):=\frac{1}{\eps}\bigg(1-\sqrt{1+\eps(\eps+2\,s\cos\theta)}\bigg) \,, \\
& y(s,\theta):=y_0+\frac{e\,\xi\,s\sin\theta}{2-\sqrt{1+\eps(\eps+2\,s\cos\theta)}}
=y_0+\frac{e\,\xi\,s\sin\theta}{1+\eps\,x(s,\theta)} \,,
\end{aligned}
\end{equation}
where $\eps$ and $e$ denote the inverse aspect ratio and the ellipticity,
respectively, and $\xi:=1/\sqrt{1-\eps^2/4}$. For $s=0$ the mapping collapses
to the pole $(x_0,y_0)=((1-\sqrt{1+\eps^2})/\eps,y_0)$.
The Jacobian matrix of the mapping reads
\begin{equation*}
J_{\Fb}(s,\theta)=
\dfrac{e\,\xi}{1+\eps\,x(s,\theta)}
\begingroup
\renewcommand*{\arraystretch}{2.5}
\begin{bmatrix}
-\dfrac{1+\eps\,x(s,\theta)}{1-\eps\,x(s,\theta)}\dfrac{\cos\theta}{e\,\xi} &
\dfrac{1+\eps\,x(s,\theta)}{1-\eps\,x(s,\theta)}\dfrac{s\sin\theta}{e\,\xi} \\
\sin\theta+\dfrac{\eps\,s\sin\theta\cos\theta}{1-\eps^2\,x^2(s,\theta)} &
s\cos\theta-\dfrac{\eps\,s^2\sin^2\theta}{1-\eps^2\,x^2(s,\theta)}
\end{bmatrix} \,,
\endgroup
\end{equation*}
with determinant
\begin{equation*}
\det J_{\Fb}(s,\theta)=\frac{s}{\eps\,x(s,\theta)-1} \,,
\end{equation*}
which vanishes at the pole. The Jacobian matrix of the inverse transformation reads
\begin{equation*}
J_{\Fb}^{-1}(s,\theta)=\frac{1}{\det J_{\Fb}(s,\theta)}
\begingroup
\renewcommand*{\arraystretch}{2.5}
\begin{bmatrix}
s\cos\theta-\dfrac{\eps\,s^2\sin^2\theta}{1-\eps^2\,x^2(s,\theta)} &
-\dfrac{1+\eps\,x(s,\theta)}{1-\eps\,x(s,\theta)}\dfrac{s\sin\theta}{e\,\xi} \\
-\sin\theta-\dfrac{\eps\,s\sin\theta\cos\theta}{1-\eps^2\,x^2(s,\theta)} &
-\dfrac{1+\eps\,x(s,\theta)}{1-\eps\,x(s,\theta)}\dfrac{\cos\theta}{e\,\xi}
\end{bmatrix} \,,
\endgroup
\end{equation*}
and it is again singular at the pole.
\begin{figure}
\centering
\includegraphics[width=0.4\linewidth,trim={3cm 0 4cm 0},clip]{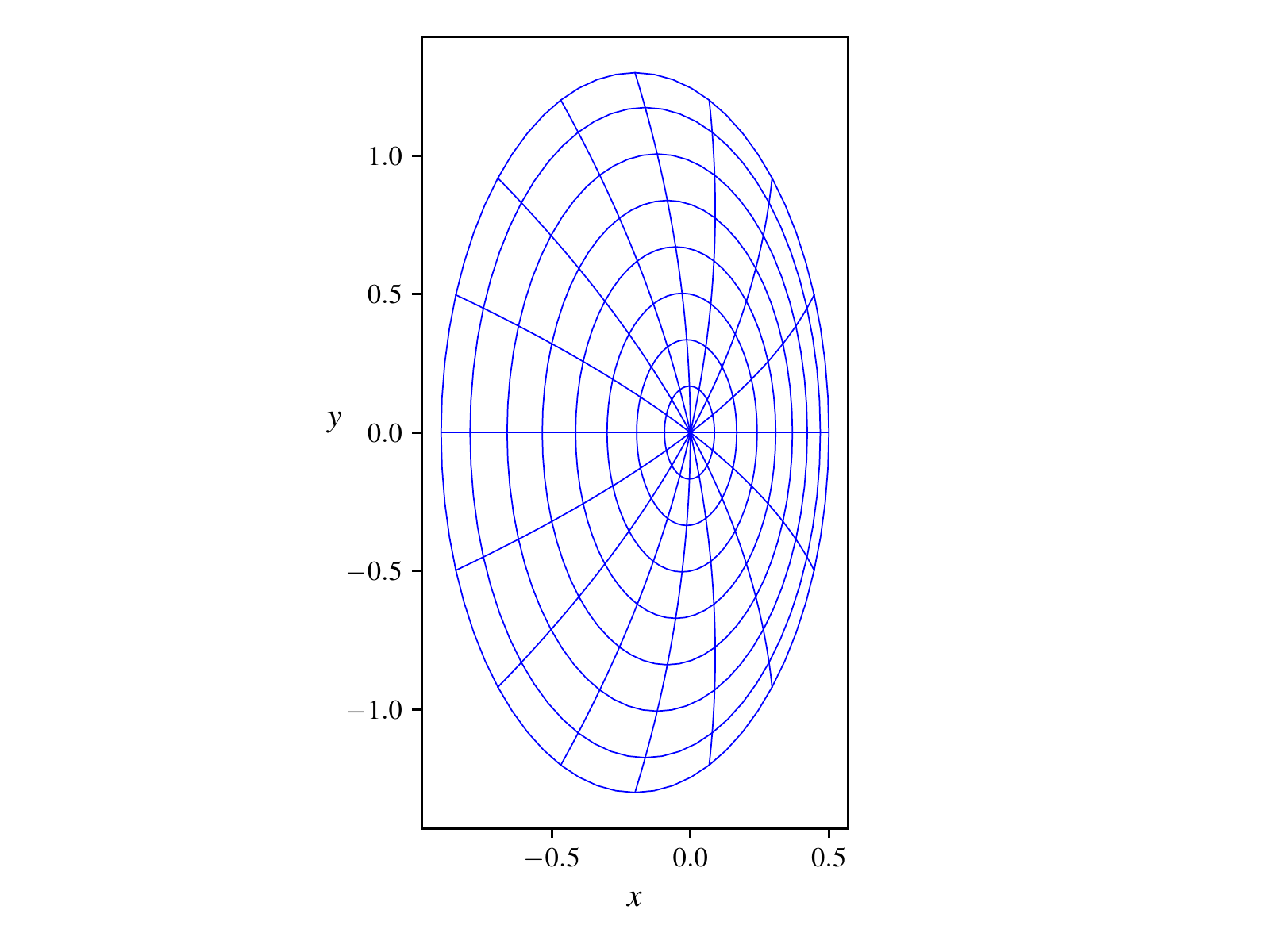}
\includegraphics[width=0.4\linewidth,trim={3cm 0 4cm 0},clip]{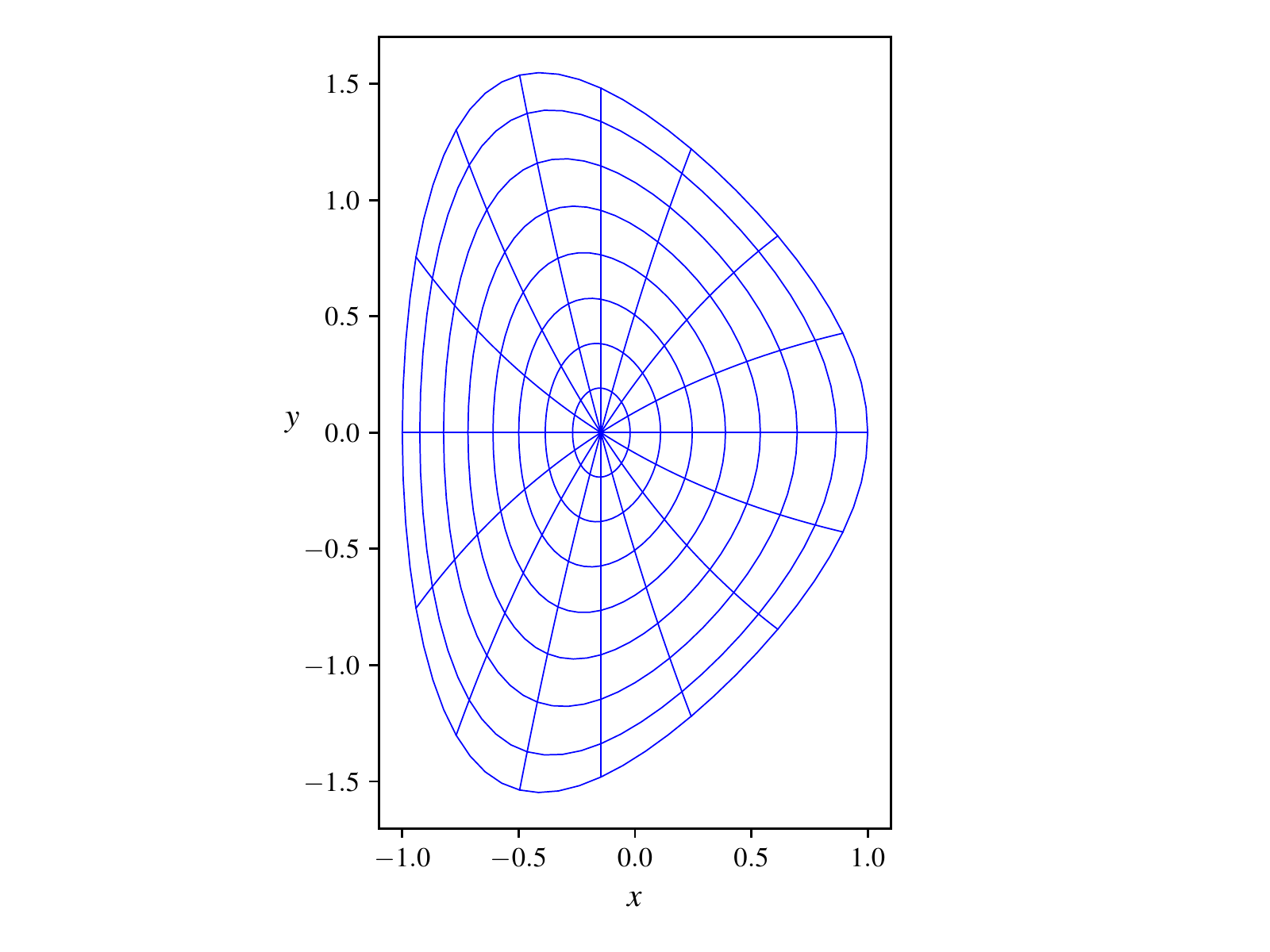}
\caption{Disk-like domains defined by the mappings \eqref{target} (left) and
\eqref{czarny} (right). Lines originating from the pole are isolines at constant $\theta$
and lines concentric around the pole are isolines at constant $s$.}
\label{fig_mappings}
\end{figure}
In all the numerical tests considered in this work, mapping \eqref{target}
is set up with the parameters
\begin{equation}
\label{target_params}
(x_0,y_0)=(0,0) \,, \quad \kappa=0.3 \,, \quad \Delta=0.2 \,,
\end{equation}
and mapping \eqref{czarny} is set up with the parameters
\begin{equation}
\label{czarny_params}
y_0=0 \,, \quad \eps=0.3 \,, \quad e=1.4 \,,
\end{equation}
which yield $x_0\approx -0.15$. 
Figure~\ref{fig_mappings} shows the physical domains obtained with these mappings.

\section{Discrete spline mappings}
\label{sec_discrete_mappings}
In practical applications it may not be possible to have an analytical description
of the mapping that represents the physical domain of interest, as in
the examples discussed above. Moreover, we are going to solve the elliptic equation
in \eqref{model_equations} with a finite element method based on $B$-splines.
Our numerical method is therefore based on a machinery inherently defined at the
discrete level. Hence, we need to have a discrete counterpart of the analytical
singular mapped disk-like domains discussed in the previous section.

We start by defining a 2D tensor-product spline basis of clamped $B$-splines
of degree $p_1$ in $s\in[0,1]$ and ${2\pi\text{-periodic}}$ $B$-splines of degree
$p_2$ in $\theta\in[0,2\pi)$: $\{\wh B_{i_1 i_2}(s,\theta):=\wh B_{i_1}^s(s)\wh B_{i_2}^\theta(\theta)\,
\}_{i_1,i_2=1}^{n_1,n_2}$. The domain along each direction is decomposed into 1D
intervals, also referred to as cells, whose limit points are named break points.
More precisely, the domain $[0,1]$ along $s$ is decomposed into ${n_1^c:=n_1-p_1}$
cells with ${n_1^b:=n_1^c+1}$ break points, denoted by $\{\wb s_{i_1}\}_{i_1=1}^{n_1^b}$,
and the domain $[0,2\pi)$ along $\theta$ is decomposed into ${n_2^c:=n_2}$ cells
with ${n_2^b:=n_2^c+1}$ break points, denoted by $\{\wb\theta_{i_2}\}_{i_2=1}^{n_2^b}$.
From the break points, we define a knot sequence $\{t^s_{i_1}\}_{i_1=1-p_1}^{n^b_1+p_1}$
of $n_1^b+2p_1$ open knots along $s$ and a knot sequence $\{t^\theta_{i_2}\}_{i_2=1-p_2}^{n^b_2+p_2}$
of $n_2^b+2p_2$ periodic knots along $\theta$:
\begin{equation*}
t^s_{i_1}:=
\begin{dcases}
\wb s_1       & i_1=1-p_1,\dots,0 \\
\wb s_{i_1}   & i_1=1,\dots,n^b_1 \\
\wb s_{n^b_1} & i_1=n^b_1+1,\dots,n^b_1+p_1
\end{dcases}
\quad
t^\theta_{i_2}:=
\begin{dcases}
\wb\theta_{n^b_2+i_2}-2\pi     & i_2=1-p_2,\dots,0 \\
\wb\theta_{i_2}        & i_2=1,\dots,n^b_2 \\
\wb\theta_{i_2-n^b_2}+2\pi & i_2=n^b_2+1,\dots,n^b_2+p_2
\end{dcases}
\end{equation*}
Due to the open knot sequence, the basis functions $\wh B_{i_1}^s(s)$ satisfy the following properties:
\begin{equation*}
\begin{alignedat}{3}
& \wh B_1^s && (0) = 1 \,, \quad && \wh B_{i_1}^s(0) = 0 \quad \text{for} \quad
i_1=2,\dots,n_1 \,, \\
& \wh B_{n_1}^s && (1) = 1 \,, \quad && \wh B_{i_1}^s(1) = 0 \quad \text{for} \quad
i_1=1,\dots,n_1-1 \,. \\
\end{alignedat}
\end{equation*}
Moreover, their derivatives satisfy the following property:
\begin{equation*}
(\wh B_1^s)'(0)=-(\wh B_2^s)'(0)\neq 0 \,, \quad
(\wh B_{i_1}^s)'(0)=0 \quad \text{for} \quad i_1=3,\dots,n_1 \,.
\end{equation*}
Based on this spline basis, we define a discrete representation of
our analytical singular mapped disk-like domains as
\begin{equation}
\label{discrete_mapping}
\begin{aligned}
& x(s,\theta):=\sum_{i_1=1}^{n_1}\sum_{i_2=1}^{n_2}c_{i_1 i_2}^x \wh B_{i_1}^s(s) \wh B_{i_2}^\theta(\theta)
=x_0\,\wh B_1^s(s)+\sum_{i_1=2}^{n_1}\sum_{i_2=1}^{n_2}c_{i_1 i_2}^x \wh B_{i_1}^s(s) \wh B_{i_2}^\theta(\theta) \,, \\
& y(s,\theta):=\sum_{i_1=1}^{n_1}\sum_{i_2=1}^{n_2}c_{i_1 i_2}^y \wh B_{i_1}^s(s) \wh B_{i_2}^\theta(\theta)
=y_0\,\wh B_1^s(s)+\sum_{i_1=2}^{n_1}\sum_{i_2=1}^{n_2}c_{i_1 i_2}^y \wh B_{i_1}^s(s) \wh B_{i_2}^\theta(\theta) \,.
\end{aligned}
\end{equation}
The control points $(c_{i_1 i_2}^x,c_{i_1 i_2}^y)$ are obtained by interpolating the
corresponding analytical mapping on the so-called Greville points \citep{GordonRiesenfeld1974,Farin1993},
defined as
\begin{equation}
\label{greville}
\begin{alignedat}{2}
& s_{i_1} && :=\frac{1}{p_1}\sum_{j_1=i_1+1-p_1}^{i_1}t^s_{j_1} \,, \quad
i_1=1,\dots,n_1 \,, \\
& \theta_{i_2} && :=\frac{1}{p_2}\sum_{j_2=i_2+1-p_2}^{i_2}t^\theta_{j_2} \,, \quad
i_2=1,\dots,n_2 \,.
\end{alignedat}
\end{equation}
Such points are averages of the knots generally lying near the values corresponding
to the maximum of the basis functions.
In the case of a periodic domain the Greville averages reduce to
either the break points themselves or the mid-points of each cell, depending on
whether the degree of the $B$-splines is odd or even, respectively. For more general
practical applications, for example in the context of magnetized fusion plasmas,
the control points $(c_{i_1 i_2}^x,c_{i_1 i_2}^y)$ could be given as an input from
any code capable of constructing a mesh conformal to the flux surfaces of a given
equilibrium magnetic field, such as, for example, the software Tokamesh \citep{Guillard2018}.
Finally, we note that all the control points at $i_1=1$ are equal to the pole,
$(c_{1 i_2}^x,c_{1 i_2}^y)=(x_0,y_0)$, which is another way of saying
that the edge $s=0$ of the logical domain collapses to the pole of the physical domain.

In order to compute integrals on the logical domain, $1+p_1$ Gauss-Legendre
quadrature points and weights are introduced in each cell of the domain $[0,1]$
along $s$ and $1+p_2$ Gauss-Legendre quadrature points and weights are introduced
in each cell of the domain $[0,2\pi)$ along $\theta$.

\section{Semi-Lagrangian advection solver}
\label{sec_advection}
We now consider the hyperbolic advection equation in the guiding-center model
\eqref{model_equations}:
\begin{equation}
\label{eq_transport_cart}
\dfrac{\de\rho}{\de t}-E^y\dfrac{\de\rho}{\de x}+E^x\dfrac{\de\rho}{\de y}=0 \,.
\end{equation}
We are interested in solving \eqref{eq_transport_cart} with the backward semi-Lagrangian
method, which we review briefly in the following. We first note that \eqref{eq_transport_cart}
can be also written as
\begin{equation}
\label{eq_transport_cart_consv}
\frac{\dd}{\dd t}\,\rho(t,\xb(t))=0 \,,
\end{equation}
with
\begin{equation}
\label{charac}
\frac{\dd\xb}{\dd t}=\Ab(t,\xb) \,,
\end{equation}
where we introduced the advection field $\Ab:=(-E^y,E^x)^T$. The characteristics of \eqref{eq_transport_cart_consv}
(and, equivalently, \eqref{eq_transport_cart}) are the solutions of the dynamical
system \eqref{charac} with given initial conditions.
The information contained in \eqref{eq_transport_cart_consv} is that its solution
$\rho$ is conserved along the characteristics \eqref{charac}. When we solve
\eqref{eq_transport_cart_consv} numerically, we are interested in knowing the value of
$\rho$ at a given time $t$ and a given mesh point $\xb_{ij}$ (assuming to have a
mesh in the Cartesian coordinates $\xb$), and the information at our disposal is
the set of values of $\rho$ at the previous time $t-\Delta t$ at each mesh point.
We then integrate the characteristics \eqref{charac} backward in time
to find the origin $\xb_{ij}^*$ at time $t-\Delta t$ of the characteristic passing
through $\xb_{ij}$ at time $t$, and set $\rho(t,\xb_{ij})=\rho(t-\Delta t,\xb_{ij}^*)$.
Typically, the point $\xb_{ij}^*$ does not coincide with a mesh point and the value
$\rho(t-\Delta t,\xb_{ij}^*)$, which is not immediately available, is obtained by
interpolating the values of $\rho$ at time $t-\Delta t$ and at mesh points in some
neighborhood of $\xb_{ij}^*$.

\begin{figure}
\centering
\includegraphics[width=0.4\linewidth,trim={3cm 0 4cm 0},clip]{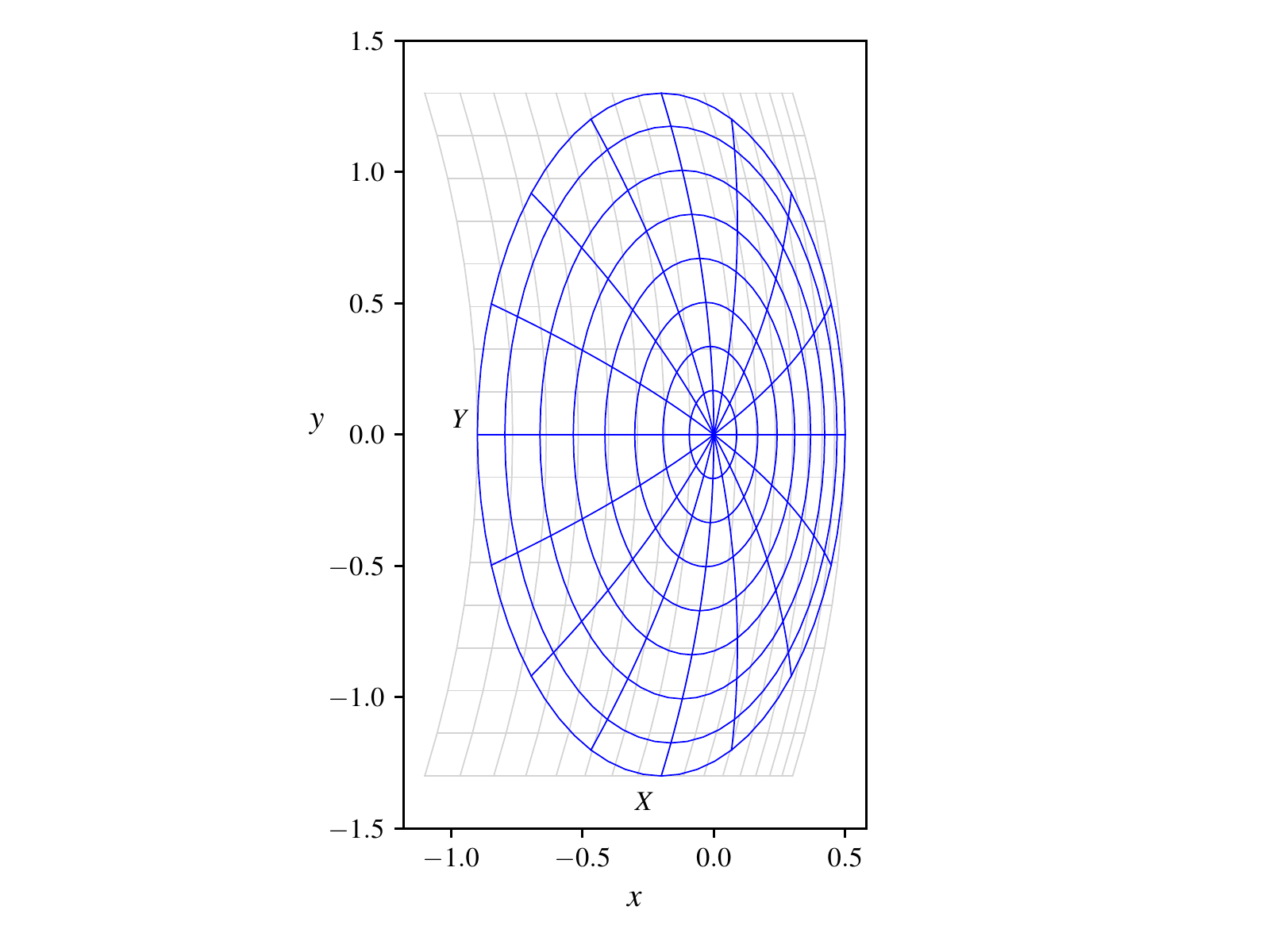}
\includegraphics[width=0.4\linewidth,trim={3cm 0 4cm 0},clip]{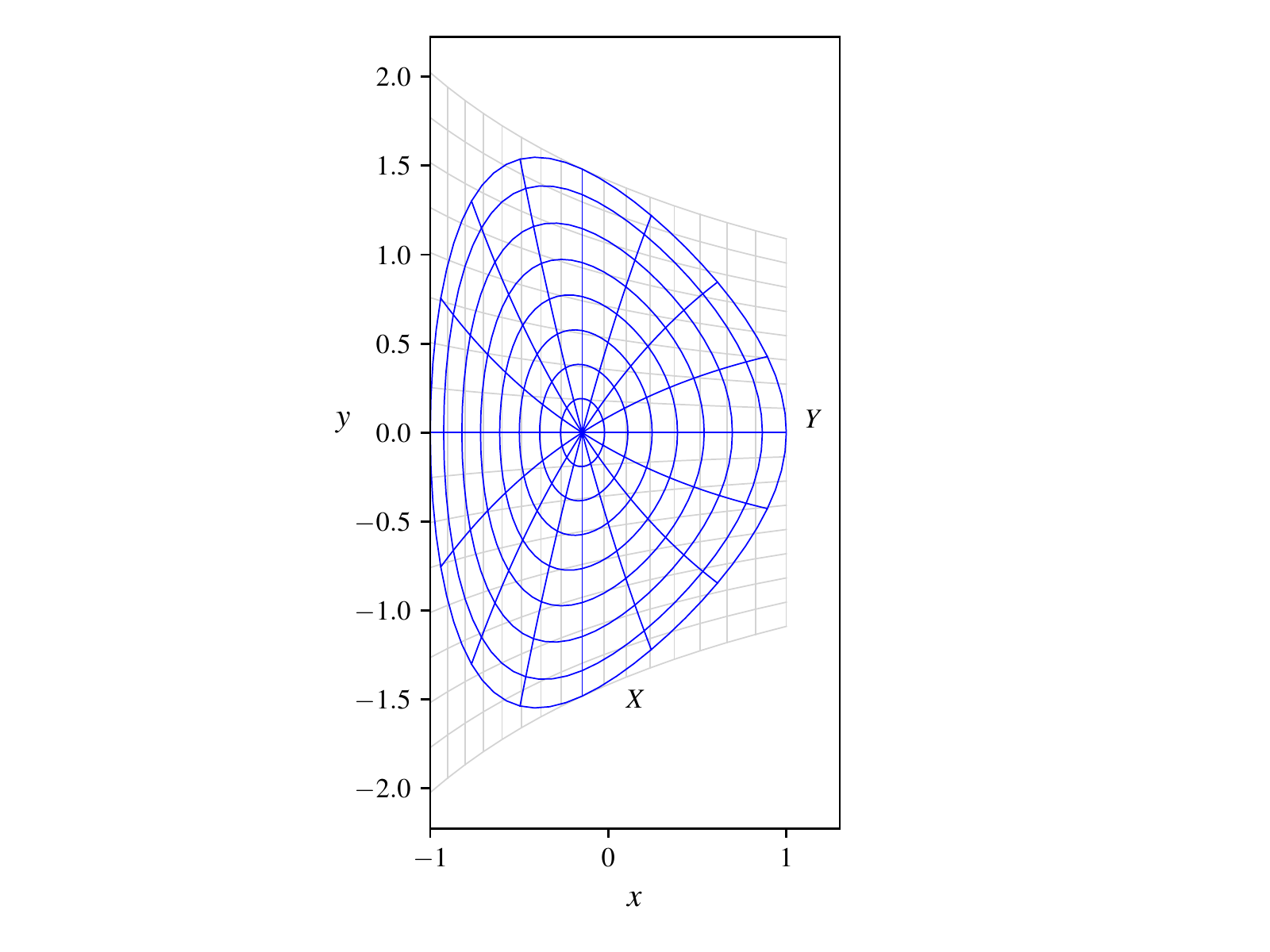}
\caption{Pseudo-Cartesian coordinates: the light-gray grids represent the grids
in the pseudo-Cartesian coordinates $(X,Y)$ for disk-like domains defined by
the mappings \eqref{target} (left) and \eqref{czarny} (right).}
\label{fig_mappings_pseudo}
\end{figure}

We now discuss the optimal choice of coordinates for the integration of the
characteristics \eqref{charac}. In the following we denote $\dd\xb/\dd t$ as $\ldot\xb$
and recall that the advection field available at the discrete level is the advection
field $\wh\Ab$ defined on the logical domain. It is natural to think of integrating
the characteristic equations in either Cartesian or logical coordinates. However,
both choices present some drawbacks. The characteristic equations in Cartesian
coordinates $\xb$ read
\begin{equation}
\label{charac_cart}
\ldot\xb=\wh\Ab(t,\Fb^{-1}(\xb)) \,.
\end{equation}
These equations are well-defined everywhere in the domain, but they become
computationally expensive if the mapping $\Fb$ is not easy to invert.
On the other hand, the characteristic equations in logical coordinates $\etb$ read
\begin{equation}
\label{charac_logical}
\ldot\etb=J_{\Fb}^{-1}\wh\Ab(t,\etb) \,.
\end{equation}
These equations are not defined at the pole $s=0$, because $J_{\Fb}^{-1}$
is singular there. We then suggest to introduce the new coordinates $\Xb:=(X,Y)$
defined by the polar transformation
\begin{equation}
\label{pseudo_cart}
\begin{aligned}
& X(s,\theta):=s\cos\theta \,, \\
& Y(s,\theta):=s\sin\theta \,,
\end{aligned}
\end{equation}
which we name pseudo-Cartesian coordinates. We denote by $\Gb:\wh\Omega\to\Rbb^2$
the new mapping defined by $\Gb(\etb):=\Xb$, and by $J_{\Gb}$ its Jacobian, given by
\begin{equation*}
J_{\Gb}(s,\theta)=
\begin{bmatrix}
\cos\theta & -s\sin\theta \\
\sin\theta & \phantom{-}s\cos\theta
\end{bmatrix} \,.
\end{equation*}
The pseudo-Cartesian coordinates for the domains defined by the mappings \eqref{target}
and \eqref{czarny} are shown in Figure~\ref{fig_mappings_pseudo}. In the simplest
case of a circular mapping, they reduce to standard Cartesian coordinates. The
characteristic equations in pseudo-Cartesian coordinates $\Xb$ read
\begin{equation}
\label{charac_pseudo_cart}
\ldot\Xb=(J_{\Fb}J_{\Gb}^{-1})^{-1}\wh\Ab(t,\Gb^{-1}(\Xb)) \,,
\end{equation}
where $J_{\Fb}J_{\Gb}^{-1}$ represents the Jacobian of the composite mapping
$\Fb\circ\Gb^{-1}$ defined by ${\Fb\circ\Gb^{-1}(\Xb)=\xb}$.
For a circular mapping, $\Fb\circ\Gb^{-1}$ reduces to the identity and \eqref{charac_pseudo_cart}
reduces to \eqref{charac_cart}, which works well because $\Fb^{-1}$ (inverse polar
transformation) is easy to compute. For more complex non-circular mappings, \eqref{charac_pseudo_cart}~is
more convenient than \eqref{charac_cart} because the mapping $\Gb$ is easier
to invert than the original mapping $\Fb$.
More precisely, the inverse mapping $\Gb^{-1}$ is analytical and reads
\begin{equation}
\label{quasi_cart_inverse}
\begin{alignedat}{2}
& s(X,Y) && =\sqrt{X^2+Y^2} \,, \\
& \theta(X,Y) && =\text{atan}2(Y,X) \,,
\end{alignedat}
\end{equation}
where $\text{atan}2(Y,X)$ returns the principal value of the argument function applied
to the complex number $X+iY$ in the range $(-\pi,\pi]$ (which then must be shifted
appropriately to the domain $[0,2\pi)$). Moreover, the inverse Jacobian matrix
$(J_{\Fb}J_{\Gb}^{-1})^{-1}$ in \eqref{charac_pseudo_cart} turns out to be well-behaved
everywhere in the physical domain, including the pole.
More precisely, the singularity of the inverse Jacobian matrix
\begin{equation*}
J_{\Gb}^{-1}(s,\theta)=
\begin{bmatrix}
\cos\theta & \sin\theta \\[2mm]
-\dfrac{1}{s}\sin\theta & \dfrac{1}{s}\cos\theta
\end{bmatrix} \,,
\end{equation*}
in the limit $s\to 0^+$, is canceled by the matrix elements of
$J_{\Fb}$. The product $J_{\Fb}J_{\Gb}^{-1}$ in general reads
\begin{equation}
\label{jac_char_sgqeps}
J_{\Fb}J_{\Gb}^{-1}(s,\theta)=
\begingroup
\renewcommand*{\arraystretch}{2.5}
\begin{bmatrix}
\dfrac{\de x}{\de s}\cos\theta-\dfrac{1}{s}\dfrac{\de x}{\de\theta}\sin\theta &
\dfrac{\de x}{\de s}\sin\theta+\dfrac{1}{s}\dfrac{\de x}{\de\theta}\cos\theta \\
\dfrac{\de y}{\de s}\cos\theta-\dfrac{1}{s}\dfrac{\de y}{\de\theta}\sin\theta &
\dfrac{\de y}{\de s}\sin\theta+\dfrac{1}{s}\dfrac{\de y}{\de\theta}\cos\theta
\end{bmatrix} \,.
\endgroup
\end{equation}
From an analytical point of view, \eqref{jac_char_sgqeps} holds for all values of $s$
except at the pole $s=0$. However, the products
$\dfrac{1}{s}\dfrac{\de x}{\de\theta}$~and~$\dfrac{1}{s}\dfrac{\de y}{\de\theta}$
are finite and well-defined in the limit $s\to 0^+$.
From a numerical point of view, \eqref{jac_char_sgqeps} holds for all values of $s$
sufficiently far from the pole, as far as the factor $1/s$ does not become too large.
Therefore, we assume that \eqref{jac_char_sgqeps} holds for $s\geq\epsilon$, for a
given small $\epsilon$. More precisely, the derivatives $\de x/\de\theta$ and
$\de y/\de\theta$ vanish for $s=0$. Hence, expanding in $s$ around $s=0$, we have
\begin{equation*}
\begin{aligned}
& \frac{\de x}{\de\theta}(s,\theta)=\frac{\de x}{\de\theta}(0,\theta)
+s\,\frac{\de^2x}{\de s\,\de\theta}(0,\theta)+O(s^2)
=s\,\frac{\de^2x}{\de s\,\de\theta}(0,\theta)+O(s^2) \,, \\
& \frac{\de y}{\de\theta}(s,\theta)=\frac{\de y}{\de\theta}(0,\theta)
+s\,\frac{\de^2y}{\de s\,\de\theta}(0,\theta)+O(s^2)
=s\,\frac{\de^2y}{\de s\,\de\theta}(0,\theta)+O(s^2) \,,
\end{aligned}
\end{equation*}
which yields
\begin{equation*}
\begin{aligned}
\lim_{s\to 0^+}\frac{1}{s}\frac{\de x}{\de\theta}(s,\theta)=\frac{\de^2x}{\de s\,\de\theta}(0,\theta) \,, \\
\lim_{s\to 0^+}\frac{1}{s}\frac{\de y}{\de\theta}(s,\theta)=\frac{\de^2y}{\de s\,\de\theta}(0,\theta) \,.
\end{aligned}
\end{equation*}
Therefore, the product $J_{\Fb}J_{\Gb}^{-1}$ at the pole $s=0$ reads
\begin{equation}
\label{jac_qcart_pole}
J_{\Fb}J_{\Gb}^{-1}(0,\theta)=
\begingroup
\renewcommand*{\arraystretch}{2.5}
\begin{bmatrix}
\dfrac{\de x}{\de s}(0,\theta)\cos\theta-\dfrac{\de^2x}{\de s\,\de\theta}(0,\theta)\sin\theta &
\dfrac{\de x}{\de s}(0,\theta)\sin\theta+\dfrac{\de^2x}{\de s\,\de\theta}(0,\theta)\cos\theta \\
\dfrac{\de y}{\de s}(0,\theta)\cos\theta-\dfrac{\de^2y}{\de s\,\de\theta}(0,\theta)\sin\theta &
\dfrac{\de y}{\de s}(0,\theta)\sin\theta+\dfrac{\de^2y}{\de s\,\de\theta}(0,\theta)\cos\theta
\end{bmatrix} \,.
\endgroup
\end{equation}
For example, in the case of mapping \eqref{target} we get
\begin{equation}
\label{jac_qcart_pole_target}
(J_{\Fb}J_{\Gb}^{-1})^{-1}(0,\theta)=
\begin{bmatrix}
\dfrac{1}{1-\kappa} & 0 \\[2mm]
0 & \dfrac{1}{1+\kappa}
\end{bmatrix} \,,
\end{equation}
and, similarly, in the case of mapping \eqref{czarny} we get
\begin{equation}
\label{jac_qcart_pole_czarny}
(J_{\Fb}J_{\Gb}^{-1})^{-1}(0,\theta)=
\begin{bmatrix}
-\sqrt{1+\eps^2} & 0 \\[2mm]
0 & \dfrac{2-\sqrt{1+\eps^2}}{e\,\xi}
\end{bmatrix} \,.
\end{equation}
In order to connect \eqref{jac_char_sgqeps} and \eqref{jac_qcart_pole}
in a smooth way, for $0<s<\epsilon$ we interpolate linearly the value at the pole
$s=0$ and the value at $s=\epsilon$, obtaining
\begin{equation*}
(J_{\Fb}J_{\Gb}^{-1})^{-1}(s,\theta)=\bigg(1-\frac{s}{\epsilon}\bigg)
(J_{\Fb}J_{\Gb}^{-1})^{-1}(0,\theta)
+\frac{s}{\epsilon}(J_{\Fb}J_{\Gb}^{-1})^{-1}(\epsilon,\theta) \,.
\end{equation*}
We remark that the result obtained in \eqref{jac_qcart_pole} needs to be single-valued,
and hence should not depend on the angle-like variable $\theta$. This is true if we consider
analytical mappings such as \eqref{target}~and~\eqref{czarny}, as demonstrated by
\eqref{jac_qcart_pole_target}~and~\eqref{jac_qcart_pole_czarny}, respectively. If
we consider, instead, a discrete representation of the aforementioned mappings, defined,
for example, in terms of splines, we observe a residual dependence of \eqref{jac_qcart_pole} on
$\theta$.
It is possible to measure the discrepancy between the matrix
elements of $(J_{\Fb}J_{\Gb}^{-1})^{-1}(0,\theta)$, computed by inverting \eqref{jac_qcart_pole}
(with the derivatives evaluated from the discrete spline mapping \eqref{discrete_mapping}), and
the corresponding analytical $\theta$-independent matrix elements.
As a measure of the error, we consider the maximum among all matrix elements and
all values of $\theta$ for a given interpolation grid. The results in Table~\ref{tab_jac}
show that such errors become asymptotically small as the computational mesh is refined
(that is, as the number of interpolation points is increased). Such errors do not
constitute a problem if they turn out to be smaller than the overall numerical accuracy
of our scheme. However, we suggest to guarantee that \eqref{jac_qcart_pole} is truly
single-valued by taking an average of \eqref{jac_qcart_pole}
over all values of $\theta$ in the interpolation grid. This may become particularly
useful if implicit integration schemes are used, when the magnitude of the aforementioned
errors may become comparable to the tolerances chosen for the implicit methods of choice.
{
\renewcommand{\arraystretch}{1.5}
\begin{table}
\begin{center}
\begin{tabular}{ccccccc}
\hline
& \multicolumn{2}{|c}{Circular mapping} & \multicolumn{2}{|c|}{Mapping \eqref{target}}
& \multicolumn{2}{c}{Mapping \eqref{czarny}} \\
\hline
$n_1\times n_2$ & Error & Order & Error & Order & Error & Order \\
\hline
$16\times32$    & $8.30\times10^{-6}$  &        & $1.19\times10^{-5}$  &        & $8.66\times10^{-6}$  & \\
$32\times64$    & $5.17\times10^{-7}$  & $4.01$ & $7.38\times10^{-7}$  & $4.01$ & $5.39\times10^{-7}$  & $4.01$ \\
$64\times128$   & $3.23\times10^{-8}$  & $4.00$ & $4.61\times10^{-8}$  & $4.00$ & $3.37\times10^{-8}$  & $4.00$ \\
$128\times256$  & $2.02\times10^{-9}$  & $4.00$ & $2.88\times10^{-9}$  & $4.00$ & $2.94\times10^{-9}$  & $3.52$ \\
$256\times512$  & $1.26\times10^{-10}$ & $4.00$ & $1.80\times10^{-10}$ & $4.00$ & $3.69\times10^{-10}$ & $3.00$ \\
\hline
\end{tabular}
\end{center}
\caption{Convergence of the product $(J_{\Fb}J_{\Gb}^{-1})^{-1}$ to the corresponding
$\theta$-independent analytical values for a circular mapping and for the mappings
\eqref{target} and \eqref{czarny}.}
\label{tab_jac}
\end{table}}
We also remark that the parameter $\epsilon$ can be chosen arbitrarily small,
as far as it avoids underflows and overflows in floating point arithmetic.
For all the numerical tests discussed in this work we set $\epsilon=10^{-12}$. We note
that the origin of a given characteristic may be located arbitrarily close to the
pole (with the pole itself being indeed the first point of our computational mesh
in the radial-like direction $s$). Therefore, a numerical strategy for the computation
of the product $(J_{\Fb}J_{\Gb}^{-1})^{-1}$ at the pole and in the region close to it,
where the factor $1/s$ appearing in \eqref{jac_char_sgqeps} is numerically too large,
cannot be avoided.

\subsection{Numerical tests}
We test the advection solver for the stationary rotating advection field
\begin{equation}
\label{rotating_adv}
\Ab(x,y):=\omega
\begin{pmatrix}
y_c-y \\
x-x_c
\end{pmatrix} \,,
\end{equation}
with $\omega=2\pi$ and $(x_c,y_c)=(0.25,0)$. The numerical test is performed
on mapping~\eqref{czarny} with the parameters in~\eqref{czarny_params}.
The flow field corresponding to the advection field \eqref{rotating_adv} can be
computed analytically and reads
\begin{equation}
\label{flow_field}
\begin{aligned}
& x(t+\Delta t)=x_c+\left(x(t)-x_c\right)\cos(\omega\Delta t)-\left(y(t)-y_c\right)\sin(\omega\Delta t) \,, \\
& y(t+\Delta t)=y_c+\left(x(t)-x_c\right)\sin(\omega\Delta t)+\left(y(t)-y_c\right)\cos(\omega\Delta t) \,,
\end{aligned}
\end{equation}
where $\Delta t$ denotes the time step. Therefore, the numerical solution
can be compared to the exact one obtained from the analytical flow field
by the method of characteristics, $\rho_{\text{ex}}(t,x(t),y(t))=\rho(0,x(0),y(0))$,
where the initial position $(x(0),y(0))$ is obtained from~\eqref{flow_field}
with $\Delta t=-t$. The initial condition is set to a superposition of cosine bells
with elliptical cross sections:
\begin{equation*}
\rho(0,x,y):=\frac{1}{2}\bigg[\Gcal\big(r_1(x,y)\big)+\Gcal(r_2(x,y)\big)\bigg] \,,
\end{equation*}
with $\Gcal(r)$ defined as
\begin{equation*}
\Gcal(r):=
\begin{dcases}
\cos\bigg(\frac{\pi r}{2a}\bigg)^4 & r < a \,, \\
0 & \text{elsewhere} \,,
\end{dcases}
\end{equation*}
with $a=0.3$, and $r_1(x,y)$ and $r_2(x,y)$ defined as
\begin{equation*}
r_1(x,y):=\sqrt{(x-x_0)^2+8(y-y_0)^2} \,, \quad
r_2(x,y):=\sqrt{8(x-x_0)^2+(y-y_0)^2} \,.
\end{equation*}
This test case is inspired by one presented in \citep[section 5.2]{Gucluetal2014}: the
non-Gaussian initial condition allows us to possibly detect any deformation of the
initial density perturbation while rotating under the action of the advection field
\eqref{rotating_adv}.
\begin{figure}
\centering
\includegraphics[width=0.5\linewidth,trim={4cm 0 0 0},clip]{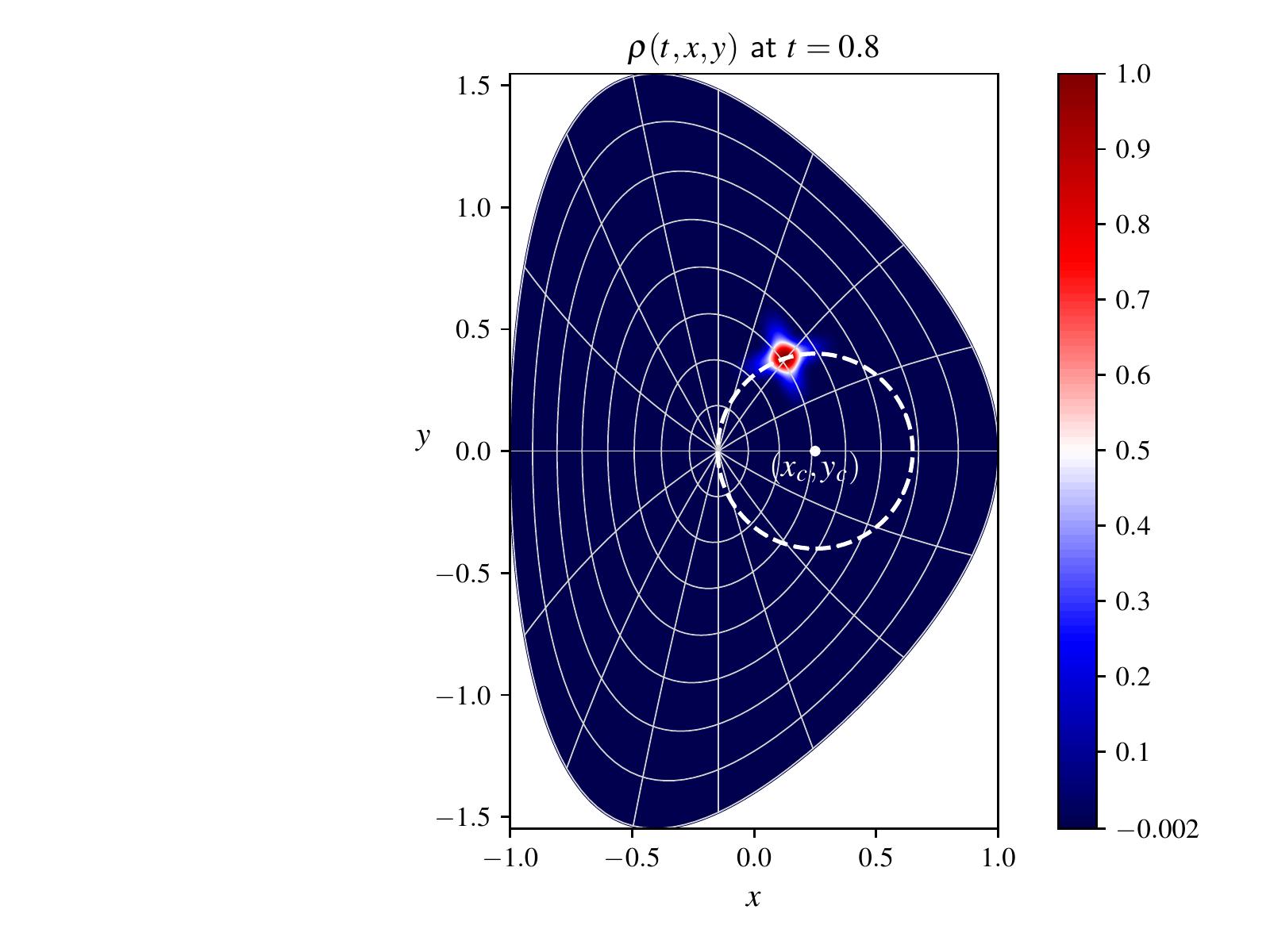}
\caption{Numerical test of the advection solver: contour plot of the density $\rho$
at a given time. The dashed circle represents the trajectory that the initial density
perturbation is expected to follow under the action of the rotating advection field
\eqref{rotating_adv}.
\label{fig_advection}}
\end{figure}
Denoting by ${\Delta\rho:=\rho-\rho_{\text{ex}}}$ the numerical error, that
is, the difference between the numerical solution and the exact one, measures of
the error of our numerical scheme are obtained by taking the $L^\infty$-norm in
time of the spatial $L^2$-norm of $\Delta\rho$,
\begin{equation*}
\max_t||\Delta\rho||_{L^2}:=\max_t\left(\sqrt{\int_\Omega\dd x\,\dd y\,
\left[\Delta\rho(t,x,y)\right]^2}\,\right)
=\max_t\left(\sqrt{\int_{\wh\Omega}\dd s\,\dd\theta\,|\det J_{\Fb}(s,\theta)|\,
\left[\Delta\wh\rho(t,s,\theta)\right]^2}\,\right) \,,
\end{equation*}
computed using the Gauss-Legendre quadrature points and weights mentioned in section
\ref{sec_discrete_mappings}, and the $L^\infty$-norm in time of the spatial $L^\infty$-norm
of $\Delta\rho$,
\begin{equation*}
\max_t||\Delta\rho||_{L^\infty}
:=\max_t\max_{(x,y)\in\Omega}\,\left|\Delta\rho(t,x,y)\right|
=\max_t\max_{(s,\theta)\in\wh\Omega}\,\left|\Delta\wh\rho(t,s,\theta)\right| \,,
\end{equation*}
computed on the Greville points \eqref{greville}.
We remark that the pole is included when we estimate
the spatial ${L^\infty\text{-norm}}$.
Table \ref{tab_conv_time_and_space_adv} shows the convergence of our numerical
scheme while decreasing the time step $\Delta t$ and correspondingly refining
the spatial mesh by increasing the number of points $n_1$ in the direction $s$ and
the number of points $n_2$ in the direction $\theta$ (in order to keep the CFL number
constant), using cubic splines and an explicit third-order Runge-Kutta method for the
integration of the characteristics.
We note that there are no effects of order reduction due to the singularity
at the pole. Standard tensor-product spline interpolation turns out to work
well in the presence of analytical advection fields, provided our choice of coordinates
for the time integration of the characteristics.
The time integration algorithm is as follows. Starting from a mesh point
$\etb_{ij}:=(s_i,\theta_j)$ with pseudo-Cartesian coordinates $\Xb_{ij}:=\Gb(\etb_{ij})$,
we compute the first-stage, second-stage and third-stage derivatives and solutions
\begin{equation*}
\begin{alignedat}{5}
& 1. \quad \ldot\Xb_{ij}^{(1)}:=(J_{\Fb}J_{\Gb}^{-1})^{-1}(\etb_{ij})\,\wh\Ab(\etb_{ij}) && \,, \quad
&& \Xb_{ij}^{(1)}:=\Xb_{ij}-\frac{\Delta t}{2}\ldot\Xb_{ij}^{(1)} && \,, \quad
&& \etb_{ij}^{(1)}:=\Gb^{-1}(\Xb_{ij}^{(1)}) \,; \\
& 2. \quad \ldot\Xb_{ij}^{(2)}:=(J_{\Fb}J_{\Gb}^{-1})^{-1}(\etb_{ij}^{(1)})\,\wh\Ab(\etb_{ij}^{(1)}) && \,, \quad
&& \Xb_{ij}^{(2)}:=\Xb_{ij}-\Delta t\left[2\ldot\Xb_{ij}^{(2)}-\ldot\Xb_{ij}^{(1)}\right] && \,, \quad
&& \etb_{ij}^{(2)}:=\Gb^{-1}(\Xb_{ij}^{(2)}) \,; \\
& 3. \quad \ldot\Xb_{ij}^{(3)}:=(J_{\Fb}J_{\Gb}^{-1})^{-1}(\etb_{ij}^{(2)})\,\wh\Ab(\etb_{ij}^{(2)}) && \,, \quad
&& \Xb_{ij}^{(3)}:=\Xb_{ij}-\frac{\Delta t}{6}\left[\ldot\Xb_{ij}^{(1)}+4\ldot\Xb_{ij}^{(2)}+\ldot\Xb_{ij}^{(3)}\right] && \,, \quad
&& \etb_{ij}^{(3)}:=\Gb^{-1}(\Xb_{ij}^{(3)}) \,.
\end{alignedat}
\end{equation*}
The logical coordinates $\etb_{ij}^{(3)}$ obtained represent the origin of the
characteristic at time $t-\Delta t$ passing through the point $\etb_{ij}$ at time $t$.
{
\renewcommand{\arraystretch}{1.5}
\begin{table}[!h]
\begin{center}
\begin{tabular}{cccccc}
\hline
$\Delta t$ & $n_1\times n_2$ & $\max_t||\Delta\rho||_{L^2}$
& Order & $\max_t||\Delta\rho||_{L^\infty}$ & Order \\
\hline
$0.1$    & $64\times128$    & $3.25\times10^{-2}$ & &      $3.53\times10^{-1}$ &        \\
$0.1/2$  & $128\times256$   & $4.10\times10^{-3}$ & 2.99 & $4.34\times10^{-2}$ & $3.02$ \\
$0.1/4$  & $256\times512$   & $5.11\times10^{-4}$ & 3.00 & $5.09\times10^{-3}$ & $3.09$ \\
$0.1/8$  & $512\times1024$  & $6.39\times10^{-5}$ & 3.00 & $6.13\times10^{-4}$ & $3.05$ \\
$0.1/16$ & $1024\times2048$ & $7.98\times10^{-6}$ & 3.00 & $7.52\times10^{-5}$ & $3.03$ \\
\hline
\end{tabular}
\end{center}
\caption{Third-order convergence of the advection solver using cubic splines
and an explicit third-order Runge-Kutta method for the integration of the characteristics.}
\label{tab_conv_time_and_space_adv}
\end{table}}

\section{Finite element elliptic solver}
\label{sec_Poisson}
We now focus on the elliptic Poisson equation in the guiding-center model
\eqref{model_equations}:
\begin{equation}
\label{eq_poisson}
-\Nabla\cdot\Nabla\phi=\rho \,.
\end{equation}
We want to solve this equation with a finite element method based on $B$-splines.
Following an isogeometric analysis approach, we use the same spline basis used
to construct the discrete spline mappings as a basis for our finite element method.
The aim is to obtain a potential $\phi$ which is $\Ccal^1$ smooth everywhere
in the physical domain, including the pole, so that the corresponding advection fields
for the transport of $\rho$ are continuous. This is achieved by imposing appropriate
$\Ccal^1$ smoothness constraints on the spline basis while solving the linear system
obtained from the weak form of \eqref{eq_poisson}. A systematic approach to define a
set of globally $\Ccal^1$ smooth spline basis functions on singular mapped
disk-like domains was developed in \citep{Toshniwaletal2017} and we now recall its
basic ideas (\citep{Toshniwaletal2017} actually suggests a more general procedure
valid also for higher-order smoothness, consistent with the spline degree).

\subsection{$\Ccal^1$ smooth polar splines}
The idea is to satisfy the $\Ccal^1$ smoothness requirements by imposing
appropriate constraints on the $2n_2$ degrees of freedom corresponding to
$i_1=1,2$ for all $i_2$. More precisely, the $2n_2$ basis functions corresponding
to these degrees of freedom are replaced by only three new basis functions,
defined as linear combinations of the existing ones. In order to guarantee the properties
of partition of unity and positivity, \citep{Toshniwaletal2017} suggests to use barycentric
coordinates to construct these linear combinations. Taking an equilateral triangle
enclosing the pole and the first row of control points $(c_{2\,i_2}^x,c_{2\,i_2}^y)$,
with vertices
\begin{equation*}
V_1:=(x_0+\tau,y_0) \,, \quad
V_2:=\left(x_0-\frac{\tau}{2},y_0+\frac{\sqrt{3}}{2}\tau\right) \,, \quad
V_3:=\left(x_0-\frac{\tau}{2},y_0-\frac{\sqrt{3}}{2}\tau\right) \,,
\end{equation*}
where $(x_0,y_0)$ denotes the Cartesian coordinates of the pole
and $\tau$ is defined as
\begin{equation*}
\tau:=\max\left[
\max_{i_2}\left(-2(c_{2\,i_2}^x-x_0)\right),
\max_{i_2}\left((c_{2\,i_2}^x-x_0)-\sqrt{3}(c_{2\,i_2}^y-y_0)\right),
\max_{i_2}\left((c_{2\,i_2}^x-x_0)+\sqrt{3}(c_{2\,i_2}^y-y_0)\right)\right] \,,
\end{equation*}
we denote by $(\lambda_1,\lambda_2,\lambda_3)$ the barycentric coordinates of any
point with respect to the vertices of this triangle:
\begin{equation*}
\begin{aligned}
& \lambda_1(x,y):=\frac{1}{3}+\frac{2}{3}\frac{1}{\tau}(x-x_0) \,, \\
& \lambda_2(x,y):=\frac{1}{3}-\frac{1}{3}\frac{1}{\tau}(x-x_0)
+\frac{\sqrt{3}}{3}\frac{1}{\tau}(y-y_0) \,, \\
& \lambda_3(x,y):=\frac{1}{3}-\frac{1}{3}\frac{1}{\tau}(x-x_0)
-\frac{\sqrt{3}}{3}\frac{1}{\tau}(y-y_0) \,.
\end{aligned}
\end{equation*}
Then, the three new basis functions are denoted by $\wh\Bcal_l$,
for $l=1,2,3$, and defined as
\begin{equation*}
\wh\Bcal_l(s,\theta):=\sum_{i_1=1}^2\sum_{i_2=1}^{n_2}
\lambda_l(c_{i_1 i_2}^x,c_{i_1 i_2}^y)\wh B_{i_1}^s(s)\wh B_{i_2}^\theta(\theta) \,.
\end{equation*}
It is easy to show that these basis functions are positive, $\wh\Bcal_l(s,\theta)\geq0$
$\forall(s,\theta)$ and $\forall l$, and that they satisfy the partition of unity
property, namely
\begin{equation}
\sum_{l=1}^3\wh\Bcal_l(s,\theta)+\sum_{i_1=3}^{n_1}\sum_{i_2=1}^{n_2}
\wh B_{i_1}^s(s)\wh B_{i_2}^\theta(\theta)=1
\quad \forall (s,\theta) \,.
\end{equation}
Moreover, the new basis functions $\Bcal_l$, related to $\wh\Bcal_l$ via $\wh\Bcal_l
=\Bcal_l\circ\Fb$, are $\Ccal^1$ smooth everywhere in the physical domain.

\subsection{Finite element solver}
\label{sec_fem_solver}
We now consider a more general version of Poisson's equation \eqref{eq_poisson}
which includes a finite set of $n_c$ point charges, denoted with the
label $c$, of charges $q_c$ and positions $(x_c,y_c)$. Denoting
by $\rho_\text{SL}$ and $\rho_{\text{PIC}}$ the semi-Lagrangian density and the
particle density, respectively, we rewrite \eqref{eq_poisson} as
\begin{equation}
\label{eq_strong_poisson}
-\Nabla\cdot\Nabla\phi=\rho_\text{SL}+\rho_\text{PIC} \,,
\quad \text{with} \quad
\rho_\text{PIC}(x,y):=\sum_{c=1}^{n_c}q_c\,\delta(x-x_c)\delta(y-y_c) \,,
\end{equation}
with homogeneous Dirichlet boundary conditions $\phi(x,y)=0$ on $\de\Omega$
(omitting the time dependence of~$\phi$). We impose these boundary conditions by
removing the corresponding basis functions from both the test space and the trial
space. More precisely, we choose as test and trial spaces the space defined by the
tensor-product spline basis
${\{\wh B_{i_1 i_2}(s,\theta):=\wh B_{i_1}^s(s)\wh B_{i_2}^\theta(\theta)\,\}_{i_1,i_2=1}^{n_1-1,n_2}}$,
where we remove the last $n_2$ basis functions corresponding to $i_1=n_1$.
Hence, the weak form of \eqref{eq_strong_poisson} reads
\begin{equation*}
\int_\Omega\dd x\,\dd y\,\Nabla\phi\cdot\Nabla B_{i_1 i_2}
=\int_\Omega\dd x\,\dd y\,\rho_\text{SL}\,B_{i_1 i_2}
+\sum_{c=1}^{n_c}q_c B_{i_1 i_2}(x_c,y_c) \,,
\quad \forall i_1,i_2 \,.
\end{equation*}
We now expand $\phi$ on the trial space,
\begin{equation*}
\phi=\sum_{j_1=1}^{n_1-1}\sum_{j_2=1}^{n_2}\phi_{j_1 j_2}\,B_{j_1 j_2} \,,
\end{equation*}
and $\rho_\text{SL}$ on the full tensor-product space (without removing the last
$n_2$ basis functions, as the space where $\rho_\text{SL}$ is defined is completely
independent from the test and trial spaces),
\begin{equation*}
\rho_\text{SL}=\sum_{k_1=1}^{n_1}\sum_{k_2=1}^{n_2}\rho_{k_1 k_2}\,B_{k_1 k_2} \,.
\end{equation*}
To sum up, the following integer indices are being used:
\begin{equation*}
\begin{alignedat}{3}
& i_1=1,\dots,n_1-1 \qquad && i_2=1,\dots,n_2 \qquad \text{(test space)} \\
& j_1=1,\dots,n_1-1 \qquad && j_2=1,\dots,n_2 \qquad \text{(trial space)} \\
& k_1=1,\dots,n_1   \qquad && k_2=1,\dots,n_2 \qquad \text{(space of $\rho_\text{SL}$)}
\end{alignedat}
\end{equation*}
Hence, we obtain
\begin{equation*}
\sum_{j_1=1}^{n_1-1}\sum_{j_2=1}^{n_2}\phi_{j_1 j_2}
\int_\Omega\dd x\,\dd y\,\Nabla B_{j_1 j_2}\cdot\Nabla B_{i_1 i_2}=
\sum_{k_1=1}^{n_1}\sum_{k_2=1}^{n_2}\rho_{k_1 k_2}
\int_\Omega\dd x\,\dd y\,B_{k_1 k_2}\,B_{i_1 i_2}
+\sum_{c=1}^{n_c}q_c B_{i_1 i_2}(x_c,y_c) \,,
\quad \forall i_1,i_2 \,.
\end{equation*}
We now introduce the tensors
\begin{equation}
\label{stiffness_mass_tensors}
\begin{alignedat}{3}
& S_{i_1 i_2 j_1 j_2} && :=\int_\Omega\dd x\,\dd y\,\Nabla B_{j_1 j_2}\cdot\Nabla B_{i_1 i_2}
&& =\int_{\wh\Omega}\dd s\,\dd\theta\,|\det J_{\Fb}|\,
\wh\Nabla\wh B_{j_1 j_2}\cdot g^{-1}\cdot\wh\Nabla\wh B_{i_1 i_2} \,, \\
& M_{i_1 i_2 k_1 k_2} && :=\int_\Omega\dd x\,\dd y\,B_{k_1 k_2}\,B_{i_1 i_2}
&& =\int_{\wh\Omega}\dd s\,\dd\theta\,|\det J_{\Fb}|\,\wh B_{k_1 k_2}\,\wh B_{i_1 i_2} \,,
\end{alignedat}
\end{equation}
where $\wh\Nabla$ denotes the gradient in the logical domain, defined as
$\wh\Nabla\wh f:=\left(\dfrac{\de\wh f}{\de s},\dfrac{\de\wh f}{\de\theta}\right)^T$ for
any function $\wh f\in\Ccal^1(\wh\Omega)$, and $g^{-1}$ denotes
the inverse metric tensor of the logical coordinate system. Such integrals are computed
using the Gauss-Legendre quadrature points and weights mentioned in section \ref{sec_discrete_mappings}.
We then obtain
\begin{equation}
\label{eq_weak_poisson_tensor}
\sum_{j_1=1}^{n_1-1}\sum_{j_2=1}^{n_2}S_{i_1 i_2 j_1 j_2}\,\phi_{j_1 j_2}
=\sum_{k_1=1}^{n_1}\sum_{k_2=1}^{n_2}M_{i_1 i_2 k_1 k_2}\,\rho_{k_1 k_2}
+\sum_{c=1}^{n_c}q_c\wh B_{i_1 i_2}(s_c,\theta_c) \,,
\quad \forall  i_1,i_2 \,.
\end{equation}
Here, the basis functions $\wh B_{i_1 i_2}$ in the last term are evaluated at the
positions $(s_c,\theta_c)=\Fb^{-1}(x_c,y_c)$ of the point charges in the logical
domain. We remark that, when Poisson's equation is coupled to the advection equation
for $\rho$ in the guiding-center model, $\Fb^{-1}(x_c,y_c)$
needs only to be computed at the beginning of a simulation: later on, the particle
equations of motion are integrated using the pseudo-Cartesian coordinates $(X_c,Y_c)$
and therefore ${(s_c,\theta_c)=\Gb^{-1}(X_c,Y_c)}$.
Equation \eqref{eq_weak_poisson_tensor} can be written in matrix form as follows.
Defining the new integer indices
\begin{equation*}
\begin{alignedat}{3}
& i:=(i_1-1)n_2+i_2 && =1,\dots,(n_1-1)n_2 \qquad && \text{(test space)} \\
& j:=(j_1-1)n_2+j_2 && =1,\dots,(n_1-1)n_2 \qquad && \text{(trial space)} \\
& k:=(k_1-1)n_2+k_2 && =1,\dots,n_1 n_2    \qquad && \text{(space of $\rho_\text{SL}$)}
\end{alignedat}
\end{equation*}
we can write \eqref{eq_weak_poisson_tensor} as
\begin{equation}
\label{eq_poisson_matrix}
S\phb=M\rhb_\text{SL}+\rhb_\text{PIC} \,,
\end{equation}
where we introduced the matrices $S$ and $M$ with elements
$(S)_{ij}:=S_{i_1 i_2 j_1 j_2}$ and $(M)_{ik}:=M_{i_1 i_2 k_1 k_2}$, and the
vectors $\phb$, $\rhb_\text{SL}$ and $\rhb_\text{PIC}$ with elements $(\phb)_j:=\phi_{j_1 j_2}$,
$(\rhb_\text{SL})_k:=\rho_{k_1 k_2}$ and ${(\rhb_\text{PIC})_i:=\sum_{c=1}^{n_c}
q_c\wh B_{i_1 i_2}(s_c,\theta_c)}$. The $\Ccal^1$ smoothness constraint is imposed
by applying to the tensor-product spline basis of the test and trial spaces
the restriction operator (using a notation similar to \citep[section 3.3]{Toshniwaletal2017})
\begin{equation*}
E:=
\begin{pmatrix}
\wb E & 0 \\
0 & I
\end{pmatrix} \,,
\end{equation*}
where $\wb E$ contains the barycentric coordinates of the pole and of
the first row of control points. More precisely, $\wb E$ is a $2n_2\times 3$
matrix with elements $\wb E_{il}:=\lambda_l(c^x_{i_1 i_2},c^y_{i_1 i_2})$ and $I$
is the identity matrix of size $[(n_1-3)n_2]\times[(n_1-3)n_2]$. Hence, the restriction
operator $E$ is a matrix of size $[(n_1-1)n_2]\times[3+(n_1-3)n_2]$.
Therefore, \eqref{eq_poisson_matrix} becomes
\begin{equation}
\label{eq_poisson_matrix_smooth}
\wb S\wb\phb=E^T(M\rhb_\text{SL}+\rhb_\text{PIC}) \,,
\end{equation}
where $\wb S:=E^T S E$ and the solution vector $\wb\phb$ is of size $[3+(n_1-3)n_2]$.
The matrix $\wb S$ is symmetric and positive-definite, hence we can solve the linear
system \eqref{eq_poisson_matrix_smooth} with the conjugate gradient method \citep{HestenesStiefel1952,
Quarteronietal2010}. The resulting solution is then prolonged back to the
trial space via $\phb=E\wb\phb$.

\subsection{Numerical tests}
\begin{figure}
\centering
\includegraphics[width=0.45\linewidth,trim={5cm 0 0 0},clip]{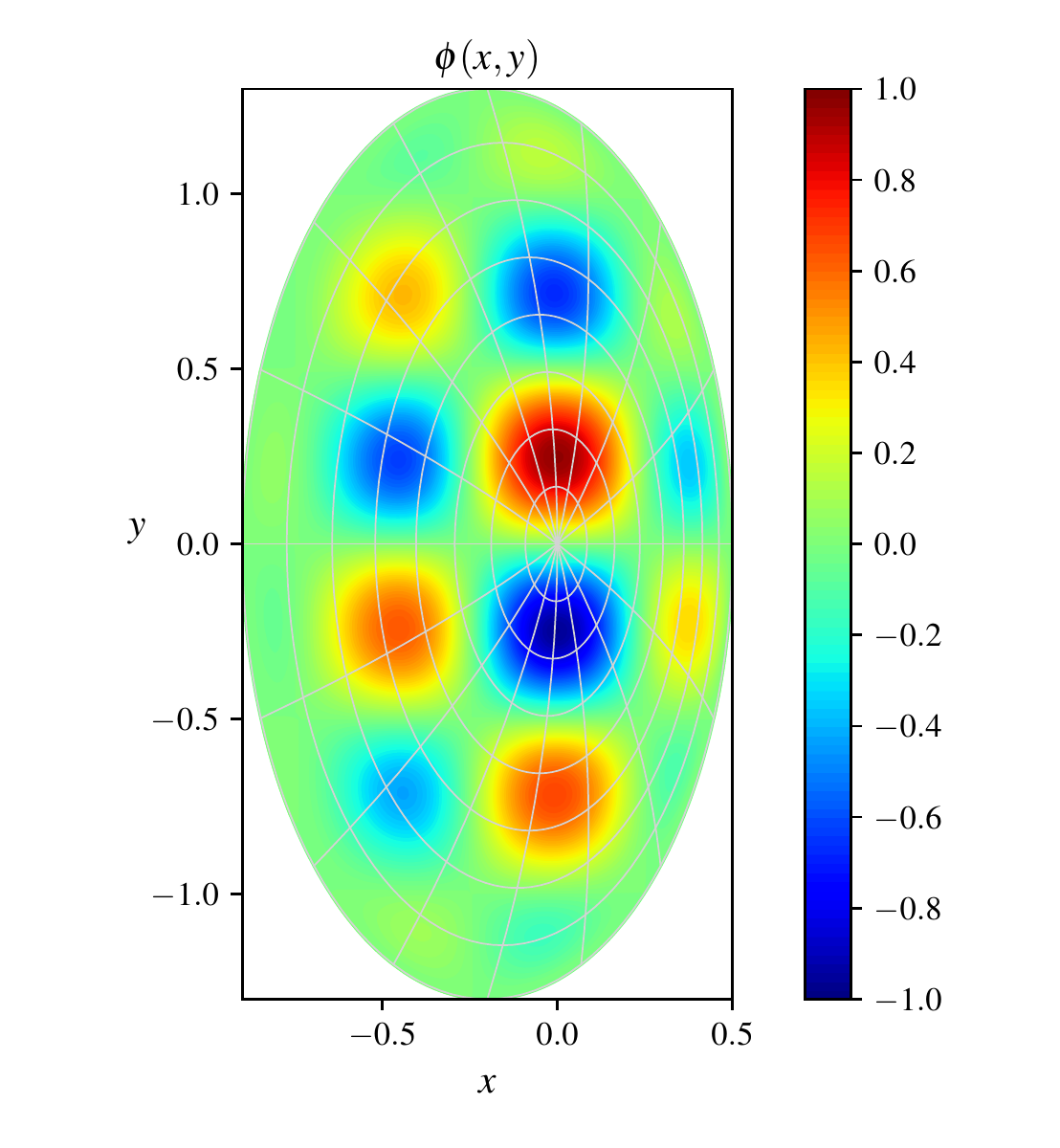}
\includegraphics[width=0.45\linewidth,trim={5cm 0 0 0},clip]{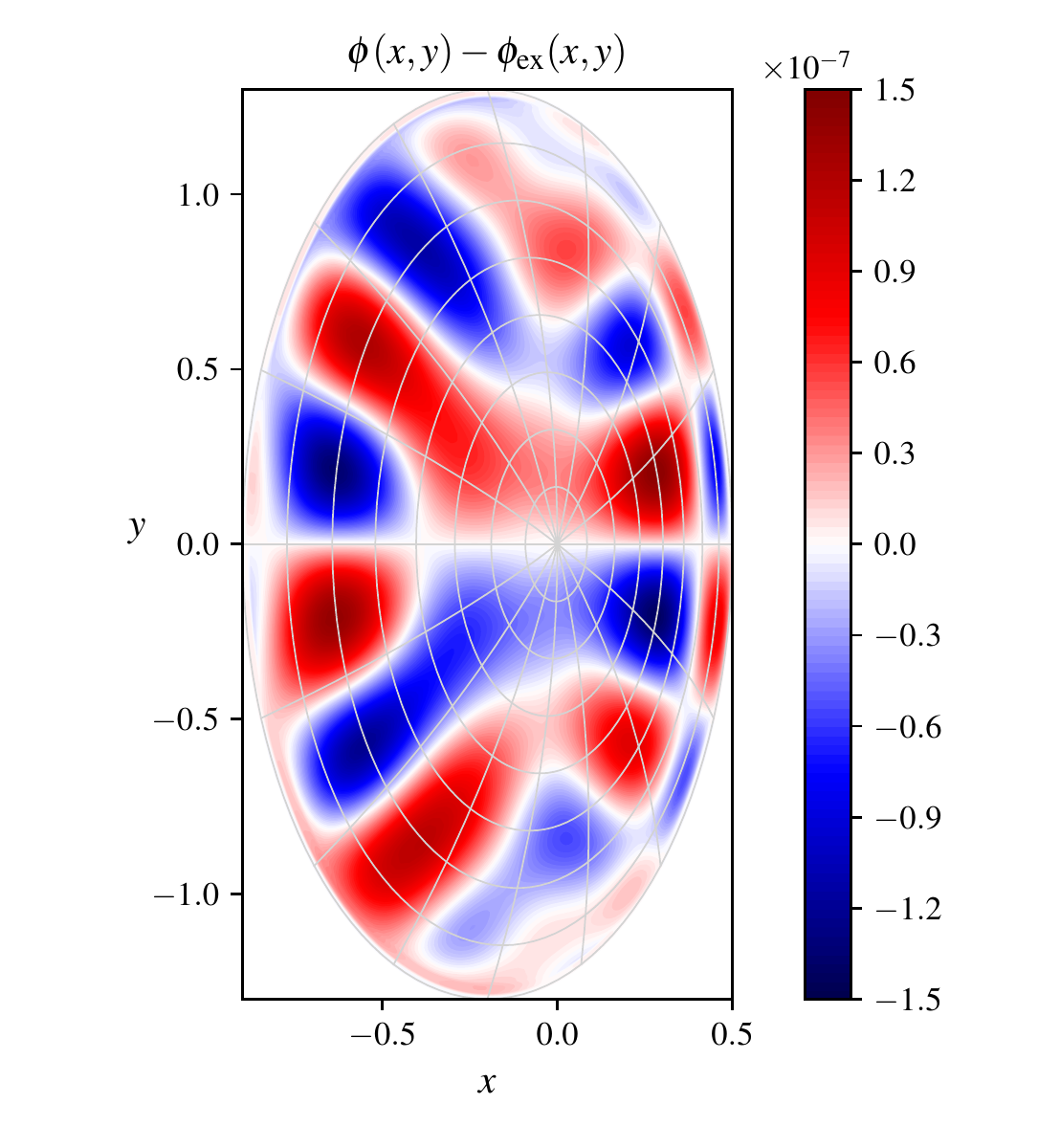}
\caption{Numerical solution of Poisson's equation on a disk-like domain defined
by mapping~\eqref{target} with the parameters in~\eqref{target_params}: contour
plots of the numerical solution (left) and error (right), obtained with $n_1\times n_2
=128\times 256$ and cubic splines.}
\label{fig_poisson}
\end{figure}

We test the Poisson solver with the method of manufactured solutions,
looking for an exact solution of the form
\begin{equation*}
\wh\phi_\text{ex}(s,\theta):=
(1-s^2)\cos(2\pi\,x(s,\theta))\sin(2\pi\,y(s,\theta)) \,,
\end{equation*}
on the physical domain defined by mapping~\eqref{target}. Denoting by ${\Delta
\phi:=\phi-\phi_{\text{ex}}}$ the numerical error, that is the difference between
the numerical solution and the exact one, measures of the error are obtained by
computing the spatial $L^2$-norm of $\Delta\phi$,
\begin{equation*}
||\Delta\phi||_{L^2}:=\sqrt{\int_\Omega\dd x\,\dd y\,\left[\Delta\phi(x,y)\right]^2}
=\sqrt{\int_{\wh\Omega}\dd s\,\dd\theta\,|\det J_{\Fb}(s,\theta)|\,\left[\Delta\wh\phi(s,\theta)\right]^2} \,,
\end{equation*}
computed using the Gauss-Legendre quadrature points and weights mentioned in section
\ref{sec_discrete_mappings}, and the spatial $L^\infty$-norm of $\Delta\phi$,
\begin{equation*}
||\Delta\phi||_{L^\infty}:=\max_{(x,y)\in\Omega}\,\left|\Delta\phi(x,y)\right|
=\max_{(s,\theta)\in\wh\Omega}\,\left|\Delta\wh\phi(s,\theta)\right| \,,
\end{equation*}
computed on the Greville points \eqref{greville}.
We remark again that the pole is included when we estimate
the spatial $L^\infty$-norm. Numerical results are shown in Figure~\ref{fig_poisson}.
Table~\ref{tab_conv_space_poisson_target} shows the
convergence of the solver while increasing the mesh size using cubic splines.
{
\renewcommand{\arraystretch}{1.5}
\begin{table}[!h]
\begin{center}
\begin{tabular}{ccccc}
\hline
$n_1\times n_2$ & $||\Delta\phi||_{L^2}$ & Order & $||\Delta\phi||_{L^\infty}$ & Order \\
\hline
$32\times64$    & $7.10\times10^{-5}$  &        & $4.17\times10^{-5}$  &        \\
$64\times128$   & $3.87\times10^{-6}$  & $4.20$ & $2.31\times10^{-6}$  & $4.17$ \\
$128\times256$  & $2.33\times10^{-7}$  & $4.05$ & $1.41\times10^{-7}$  & $4.03$ \\
$256\times512$  & $1.44\times10^{-8}$  & $4.02$ & $8.78\times10^{-9}$  & $4.01$ \\
$512\times1024$ & $8.99\times10^{-10}$ & $4.00$ & $5.48\times10^{-10}$ & $4.00$ \\
\hline
\end{tabular}
\end{center}
\caption{Numerical solution of Poisson's equation on a disk-like domain
defined by mapping~\eqref{target} with the parameters in~\eqref{target_params}:
fourth-order convergence of the solver using cubic splines.}
\label{tab_conv_space_poisson_target}
\end{table}}

\subsection{Evaluation of the electric field}
The advection fields for the transport of $\rho$ in the guiding-center model
\eqref{model_equations} are obtained from the potential $\phi$ by means of derivatives.
This section suggests a strategy to evaluate the Cartesian components of the
gradient of $\wh\phi$ while taking into account the singularity at the pole.
We denote again by $\wh\Nabla\wh\phi$ the gradient of $\wh\phi$ in the logical
domain. The Cartesian components of the gradient are obtained from the logical
ones by applying the inverse of the transposed Jacobian matrix:
\begin{equation}
\label{grad_phys}
\Nabla\wh\phi(s,\theta)=(J_{\Fb}^{-1})^T(s,\theta)\,\wh\Nabla{\wh\phi}(s,\theta) \,.
\end{equation}
From an analytical point of view, \eqref{grad_phys} holds for all values of $s>0$
and its limit as $s\to0^+$ is finite and unique. From a numerical point of
view, \eqref{grad_phys} holds for all values of $s$ sufficiently far from
the pole, as far as the inverse Jacobian does not become too large. Therefore, we
assume that \eqref{grad_phys} holds for $s\geq\epsilon$, for a given small $\epsilon$.
For $s=0$ the partial derivative with respect to $\theta$ vanishes
and all the information is contained in the partial derivative with respect to $s$,
which takes a different value for each value of $\theta$. Recalling
that a partial derivative has the geometrical meaning of a directional derivative along
a vector of the tangent basis, the idea is to combine two given values
corresponding to two different values of $\theta$ and extract from them the
Cartesian components of the gradient at the pole. The two chosen values of $\theta$
must correspond to linearly independent directions, so that from
\begin{equation*}
\begin{aligned}
& \frac{\de\wh\phi}{\de s}(0,\theta_1)=\Nabla\wh\phi\cdot\eb_s
=(\Nabla\wh\phi)_x\frac{\de x}{\de s}(0,\theta_1)+(\Nabla\wh\phi)_y\frac{\de y}{\de s}(0,\theta_1) \,, \\
& \frac{\de\wh\phi}{\de s}(0,\theta_2)=\Nabla\wh\phi\cdot\eb_s
=(\Nabla\wh\phi)_x\frac{\de x}{\de s}(0,\theta_2)+(\Nabla\wh\phi)_y\frac{\de y}{\de s}(0,\theta_2) \,,
\end{aligned}
\end{equation*}
the two components $(\Nabla\wh\phi)_x$ and $(\Nabla\wh\phi)_y$ can be obtained.
Each possible couple of linearly independent directions produces the same
result. In order to connect the two approaches
in a smooth way, for $0<s<\epsilon$ we interpolate linearly the value at the pole
$s=0$ and the value at $s=\epsilon$:
\begin{equation*}
\Nabla\wh\phi(s,\theta)=\bigg(1-\frac{s}{\epsilon}\bigg)\Nabla\wh\phi(0,\theta)
+\frac{s}{\epsilon}\Nabla\wh\phi(\epsilon,\theta) \,.
\end{equation*}
We remark that the parameter $\epsilon$ can be chosen arbitrarily small, as far as
it avoids underflows and overflows in floating point arithmetic. For all the numerical
tests discussed in this work we set $\epsilon=10^{-12}$.

\section{Self-consistent test cases: the guiding-center model}
\label{sec_guiding_center}
We now address the solution of the guiding-center model
\begin{equation}
\label{model_equations_bis}
\begin{cases}
\dfrac{\de\rho}{\de t}-E^y\dfrac{\de\rho}{\de x}+E^x\dfrac{\de\rho}{\de y}=0 \,, \\[2mm]
-\Nabla\cdot\Nabla\phi=\rho \,,
\end{cases}
\quad \textrm{with} \quad
\begin{cases}
\rho(0,x,y)=\rho_\text{IN}(x,y) \,, \\[2mm]
\phi(t,x,y)=0 \textrm{ on } \de\Omega \,.
\end{cases}
\end{equation}
Physical quantities conserved by the model are the total mass and energy
\begin{equation}
\label{invariants}
\begin{alignedat}{3}
& \Mcal(t):= && \int_\Omega\dd x\,\dd y\,\rho(t,x,y) &&
=\int_{\wh\Omega}\dd s\,\dd\theta\,|\det J_{\Fb}(s,\theta)|\,\wh\rho(t,s,\theta) \,, \\
& \Wcal(t):= && \int_\Omega\dd x\,\dd y\,|\Eb(t,x,y)|^2 &&
=\int_{\wh\Omega}\dd s\,\dd\theta\,|\det J_{\Fb}(s,\theta)|\,|\wh\Eb(t,s,\theta)|^2 \,.
\end{alignedat}
\end{equation}
These integrals are computed using the Gauss-Legendre quadrature points and weights mentioned
in section \ref{sec_discrete_mappings}. We define the relative errors for
the conservation of the invariants \eqref{invariants} as
\begin{equation}
\label{invariants_error}
\delta\Mcal(t):=\frac{|\Mcal(0)-\Mcal(t)|}{|\Mcal(0)|} \,, \quad
\delta\Wcal(t):=\frac{|\Wcal(0)-\Wcal(t)|}{|\Wcal(0)|} \,.
\end{equation}
Before describing the numerical tests considered for this model, we present
our time-advancing strategy and how we deal with the problem of defining an equilibrium
density on complex mappings while initializing our simulations.

\subsection{Time integration}
We present here two different time integration schemes, one explicit and one
implicit, that may be chosen according to the particular physical dynamics described
by model \eqref{model_equations_bis}. Both integration schemes
are based on a predictor-corrector procedure. In the numerical tests discussed in this
section, the explicit scheme is our default choice, because of its low computational
cost. However, there are situations (as, for example, the test case simulating the
merger of two macroscopic vortices presented in section \ref{sec_vortex_merger})
where the dynamics described by model \eqref{model_equations_bis} is such that the
explicit scheme would require very small time steps in order to produce correct
results. Instead, the implicit trapezoidal scheme that we describe here has proven
capable of capturing the correct dynamics with much larger time steps, thanks to
its symmetry and adjoint-symplecticity \citep{Haireretal2006}.

\subsubsection{Second-order explicit scheme}
\label{sec_explicit}
The explicit time integration scheme is the second-order integrator described
in \citep[section 2.2]{Xiongetal2018}. Since it will be used also for test cases
involving point charges, we denote here again by $\rho_\text{SL}$ and $\rho_\text{PIC}$
the semi-Lagrangian density and the particle density, respectively. Moreover, following
the notation of section~\ref{sec_advection}, we denote by $\Xb_{ij}:=\Gb(\etb_{ij})$
the pseudo-Cartesian coordinates of a given mesh point and by $\Xb_c:=\Gb(\etb_c)$
the pseudo-Cartesian coordinates of a given point charge, respectively. The first-order
prediction (superscript ``$(P)$'') is given by
\begin{equation*}
\begin{alignedat}{5}
& \ldot\Xb_{ij}^{(P)}:=(J_{\Fb}J_{\Gb}^{-1})^{-1}(\etb_{ij})\,\wh\Ab(\etb_{ij}) && \,, \quad
&& \Xb_{ij}^{(P)}:=\Xb_{ij}-\Delta t\,\ldot\Xb_{ij}^{(P)} && \,, \quad
&& \etb_{ij}^{(P)}:=\Gb^{-1}(\Xb_{ij}^{(P)}) \,; \\
& \ldot\Xb_c^{(P)}:=(J_{\Fb}J_{\Gb}^{-1})^{-1}(\etb_c)\,\wh\Ab(\etb_c) && \,, \quad
&& \Xb_c^{(P)}:=\Xb_c+\Delta t\,\ldot\Xb_c^{(P)} && \,, \quad
&& \etb_c^{(P)}:=\Gb^{-1}(\Xb_c^{(P)}) \,.
\end{alignedat}
\end{equation*}
We then compute the intermediate semi-Lagrangian and particle densities $\rho^{(P)}_\text{SL}$
and $\rho^{(P)}_\text{PIC}$ and obtain the intermediate electric potential $\phi^{(P)}$
by solving Poisson's equation. Denoting by $\wh\Ab^{(P)}$ the corresponding intermediate
advection field, the second-order correction (superscript ``$(C)$'') is given by
\begin{equation*}
\begin{alignedat}{5}
& \ldot\Xb_{ij}^{(C)}:=(J_{\Fb}J_{\Gb}^{-1})^{-1}(\etb_{ij}^{(P)})\,\wh\Ab(\etb_{ij}^{(P)})
+(J_{\Fb}J_{\Gb}^{-1})^{-1}(\etb_{ij})\,\wh\Ab^{(P)}(\etb_{ij}) && \,, \quad
&& \Xb_{ij}^{(C)}:=\Xb_{ij}-\frac{\Delta t}{2}\ldot\Xb_{ij}^{(C)} && \,, \quad
&& \etb_{ij}^{(C)}:=\Gb^{-1}(\Xb_{ij}^{(C)}) \,; \\
& \ldot\Xb_c^{(C)}:=(J_{\Fb}J_{\Gb}^{-1})^{-1}(\etb_c)\,\wh\Ab(\etb_c)
+(J_{\Fb}J_{\Gb}^{-1})^{-1}(\etb_c^{(P)})\,\wh\Ab^{(P)}(\etb_c^{(P)}) && \,, \quad
&& \Xb_c^{(C)}:=\Xb_c+\frac{\Delta t}{2}\ldot\Xb_c^{(C)} && \,, \quad
&& \etb_c^{(C)}:=\Gb^{-1}(\Xb_c^{(C)}) \,. \\
\end{alignedat}
\end{equation*}
For point charges, this second-order scheme is equivalent to Heun's method (improved
Euler's method \citep{SueliMayers2003}).

\subsubsection{Second-order implicit scheme}
\label{sec_implicit}
The implicit time integration scheme is based on the implicit trapezoidal rule
and it will not be used for test cases involving point charges. We denote again by
$\rho_\text{SL}$ the semi-Lagrangian density and by $\Xb_{ij}:=\Gb(\etb_{ij})$ the
pseudo-Cartesian coordinates of a given mesh point. The second-order prediction
(superscript ``$(P)$'') is given by $\Xb_{ij}^{(P)}:=\Xb_{ij}^{(k)}$ and ${\etb_{ij}^{(P)}
:=\Gb^{-1}(\Xb_{ij}^{(P)})}$, where the $k$-th iteration is computed as
\begin{equation*}
\ldot\Xb_{ij}^{(k)}:=(J_{\Fb}J_{\Gb}^{-1})^{-1}(\etb_{ij})\,
\wh\Ab(\etb_{ij})+(J_{\Fb}J_{\Gb}^{-1})^{-1}(\etb_{ij}^{(k-1)})\,\wh\Ab(\etb_{ij}^{(k-1)}) \,, \quad
\Xb_{ij}^{(k)}:=\Xb_{ij}-\frac{\Delta t}{4}\ldot\Xb_{ij}^{(k)} \,, \quad
\etb_{ij}^{(k)}:=\Gb^{-1}(\Xb_{ij}^{(k)}) \,,
\end{equation*}
with $\Xb_{ij}^{(0)}:=\Xb_{ij}$ and $\etb_{ij}^{(0)}:=\etb_{ij}$, provided that $|\Xb_{ij}^{(k)}-\Xb_{ij}^{(k-1)}|^2\leq\tau^2$,
where the tolerance $\tau$ is defined as ${\tau:=\tau_A+\tau_R\,|\Xb_{ij}|}$, for given
absolute and relative tolerances $\tau_A$ and $\tau_R$. We then compute the intermediate
semi-Lagrangian density $\rho^{(P)}_\text{SL}$ and obtain the intermediate electric
potential $\phi^{(P)}$ by solving Poisson's equation. Denoting by $\wh\Ab^{(P)}$ the
corresponding intermediate advection field, the second-order correction (superscript
``$(C)$'') is given by $\Xb_{ij}^{(C)}:=\Xb_{ij}^{(k)}$ and ${\etb_{ij}^{(C)}:=
\Gb^{-1}(\Xb_{ij}^{(C)})}$, where the $k$-th iteration is computed as
\begin{equation*}
\ldot\Xb_{ij}^{(k)}:=(J_{\Fb}J_{\Gb}^{-1})^{-1}(\etb_{ij})\,
\wh\Ab^{(P)}(\etb_{ij})+(J_{\Fb}J_{\Gb}^{-1})^{-1}(\etb_{ij}^{(k-1)})\,\wh\Ab^{(P)}(\etb_{ij}^{(k-1)}) \,, \quad
\Xb_{ij}^{(k)}:=\Xb_{ij}-\frac{\Delta t}{2}\ldot\Xb_{ij}^{(k)} \,, \quad
\etb_{ij}^{(k)}:=\Gb^{-1}(\Xb_{ij}^{(k)}) \,,
\end{equation*}
with $\Xb_{ij}^{(0)}:=\Xb_{ij}$ and $\etb_{ij}^{(0)}:=\etb_{ij}$, provided that $|\Xb_{ij}^{(k)}-\Xb_{ij}^{(k-1)}|^2\leq\tau^2$.

\subsection{Numerical equilibria}
\label{num_equil}
\begin{figure}
\centering
\includegraphics[width=0.48\linewidth,trim={1cm 0 1cm 0},clip]{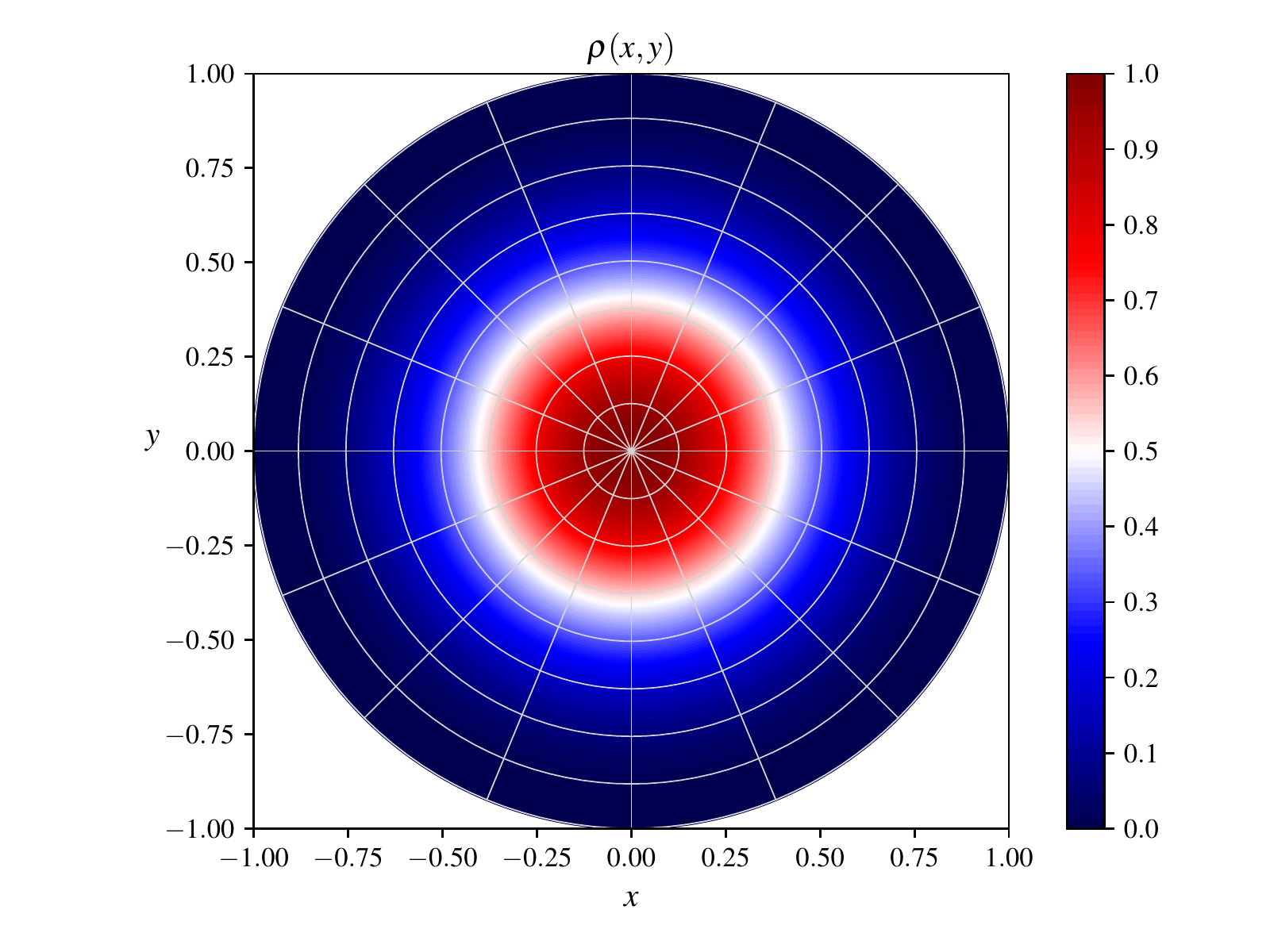}
\includegraphics[width=0.48\linewidth,trim={2cm 0 0 0},clip]{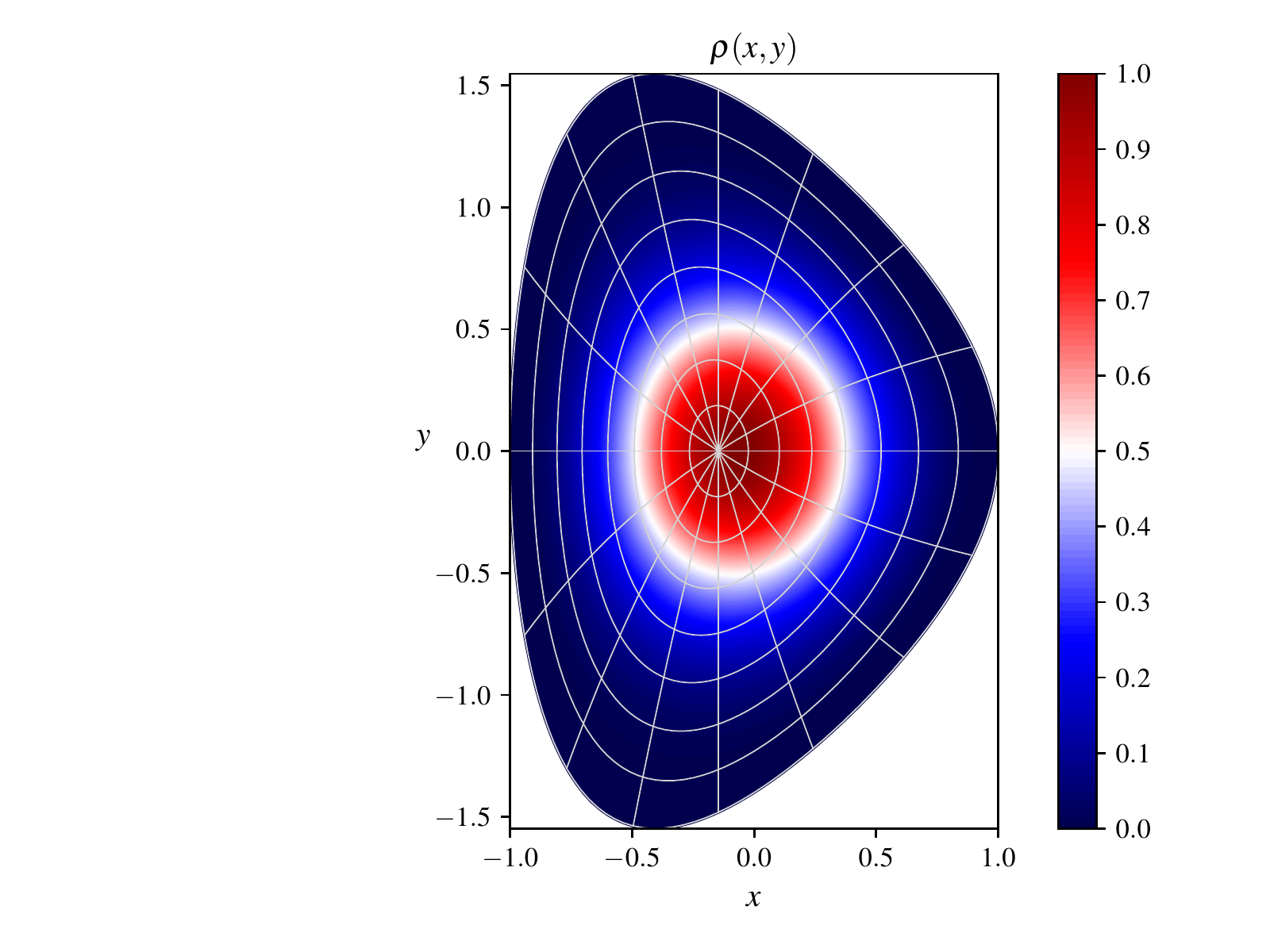}
\caption{Numerical equilibrium density $\rho$ obtained with $f(\phi)=\phi^2$ and
$\rho_\text{max}=1$ on disk-like domains defined by a circular mapping (left) and
by mapping~\eqref{czarny} with the parameters in~\eqref{czarny_params} (right).}
\label{fig_eq}
\end{figure}

Defining an equilibrium density $\rho$ and a corresponding
equilibrium potential $\phi$ for the system \eqref{model_equations_bis} becomes
non-trivial on domains defined by complex non-circular mappings, such as \eqref{target} and \eqref{czarny}.
In the case of circular mappings, any axisymmetric density independent of the angle
variable $\theta$ turns out to be an equilibrium for the transport equation in
\eqref{model_equations_bis}. For more complex mappings we follow the numerical procedure
suggested in \citep{TakedaTokuda1991}, and references therein, to compute an equilibrium
couple $(\rho,\phi)$. The equilibrium is determined by the eigenvalue problem of
finding $(\sigma,\phi)$ such that $-\Nabla\cdot\Nabla\phi=\sigma\,f(\phi)$,
with given $f$ such that $f'(\phi)\neq0$ in some limited domain. Given initial data
$(\sigma^{(0)},\phi^{(0)})$, the $i$-th iteration, with $i\geq1$, is computed with
the following steps:
\begin{enumerate}
\item compute $\rho^{(i)}:=\sigma^{(i-1)}f(\phi^{(i-1)})$;
\item compute $\phi^{(i)}_*$ by solving $-\Nabla\cdot\Nabla\phi^{(i)}_*=\rho^{(i)}$;
\item if a maximum value $\phi_\text{max}$ is given, compute $c^{(i)}$ by setting
$c^{(i)}:=\phi_\text{max}/||\phi^{(i)}_*||_{L^\infty}$; \\
if a maximum value $\rho_\text{max}$ is given, compute $c^{(i)}$ by solving
${c^{(i)}f(c^{(i)}\,||\phi^{(i)}_*||_{L^\infty})=\rho_\text{max}/\sigma^{(i-1)}}$;
\item compute $(\sigma^{(i)},\phi^{(i)}):=c^{(i)}\,(\sigma^{(i-1)},\phi^{(i)}_*)$.
\end{enumerate}
The iterative procedure stops when $|\sigma^{(i)}-\sigma^{(i-1)}|\leq\tau$, for a
given tolerance $\tau$.
The eigenvalue problem does not have a unique solution, but the algorithm is expected
to converge to the ground state, that is, the eigenstate with minimum eigenvalue. Figure
\ref{fig_eq} illustrates, for example, the equilibrium obtained in this way for $f(\phi)=\phi^2$
and $\rho_\text{max}=1$ on domains defined by a circular mapping and by
mapping~\eqref{czarny} with the parameters in \eqref{czarny_params}.

\subsection{Numerical test: diocotron instability}
\begin{figure}
\centering
\includegraphics[width=0.6\textwidth,trim={0 0 0 0},clip]{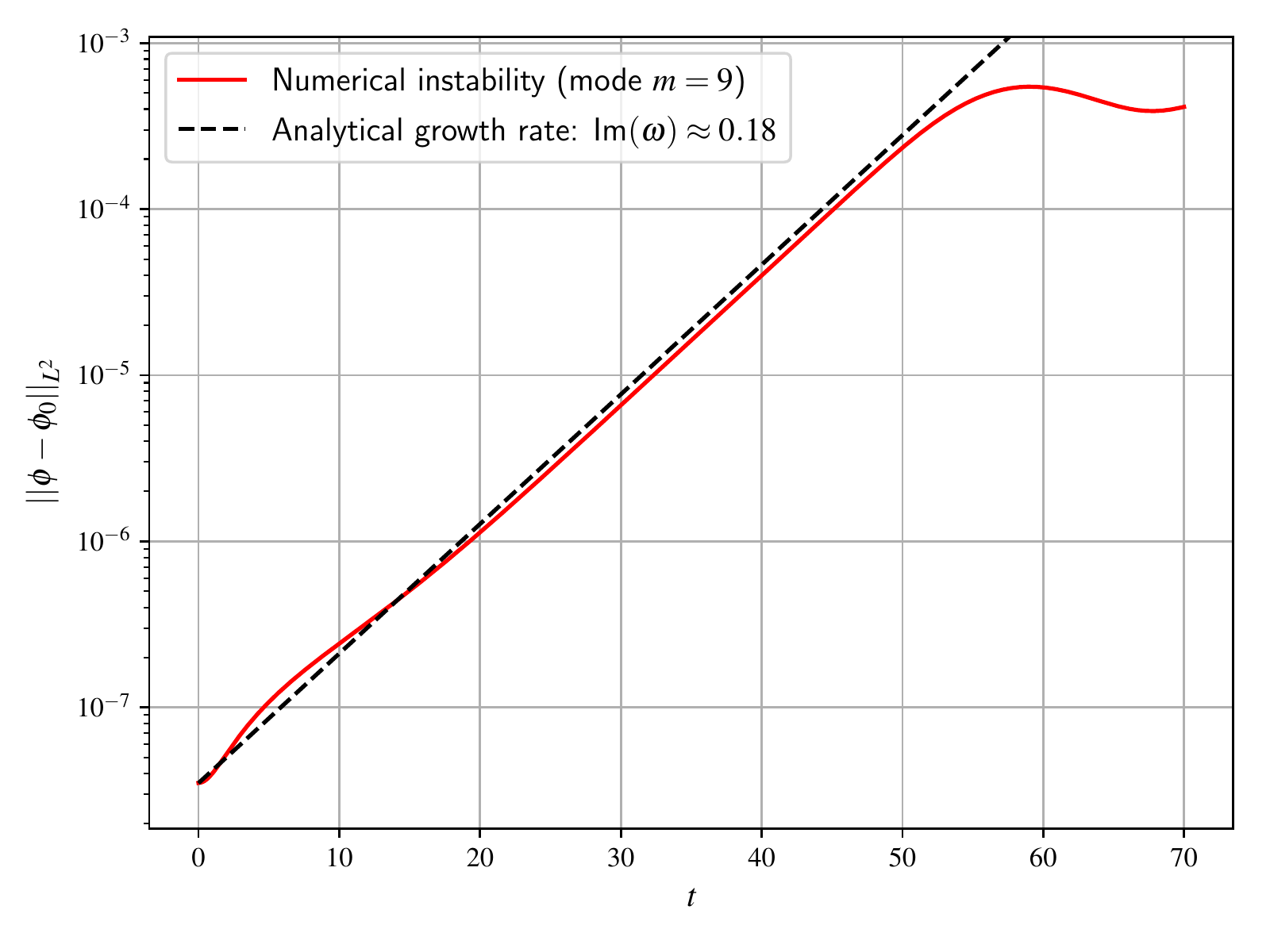}
\caption{Numerical simulation of the diocotron instability: $L^2$-norm of the
perturbed electric potential.}
\label{fig_diocotron1}
\end{figure}

\begin{figure}
\includegraphics[width=0.48\linewidth,trim={1cm 0 1cm 0},clip]{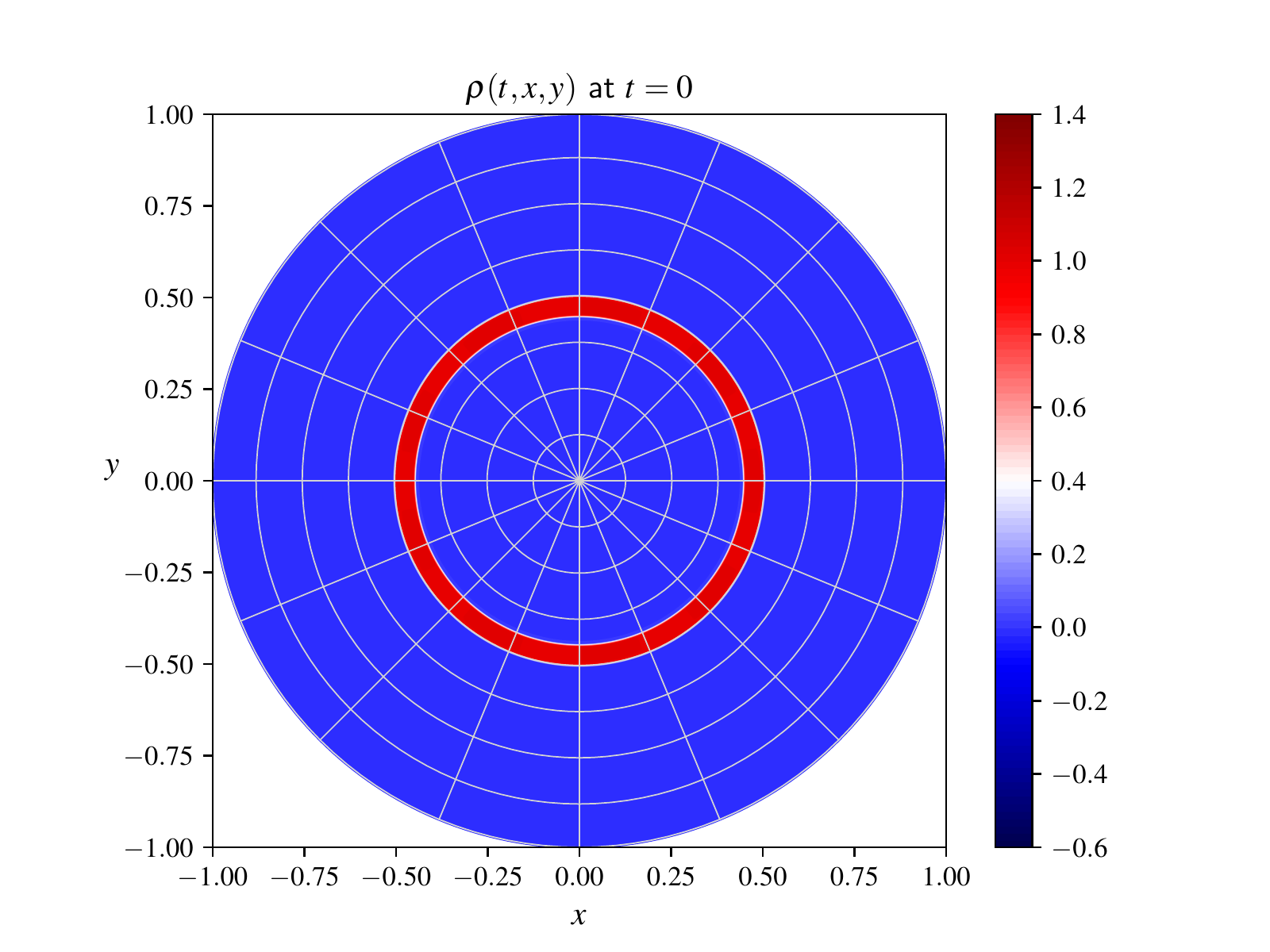}
\includegraphics[width=0.48\linewidth,trim={1cm 0 1cm 0},clip]{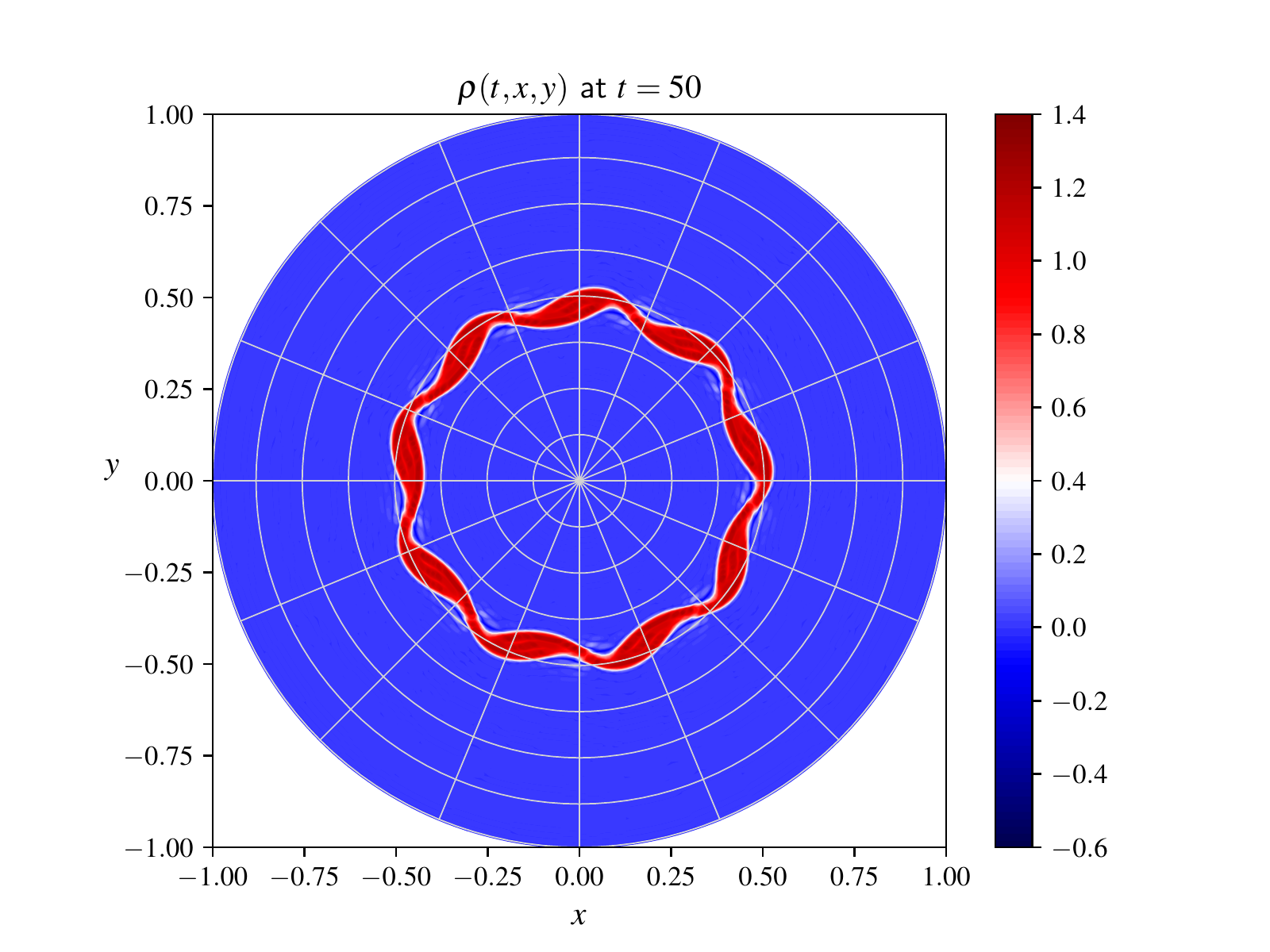}
\includegraphics[width=0.48\linewidth,trim={1cm 0 1cm 0},clip]{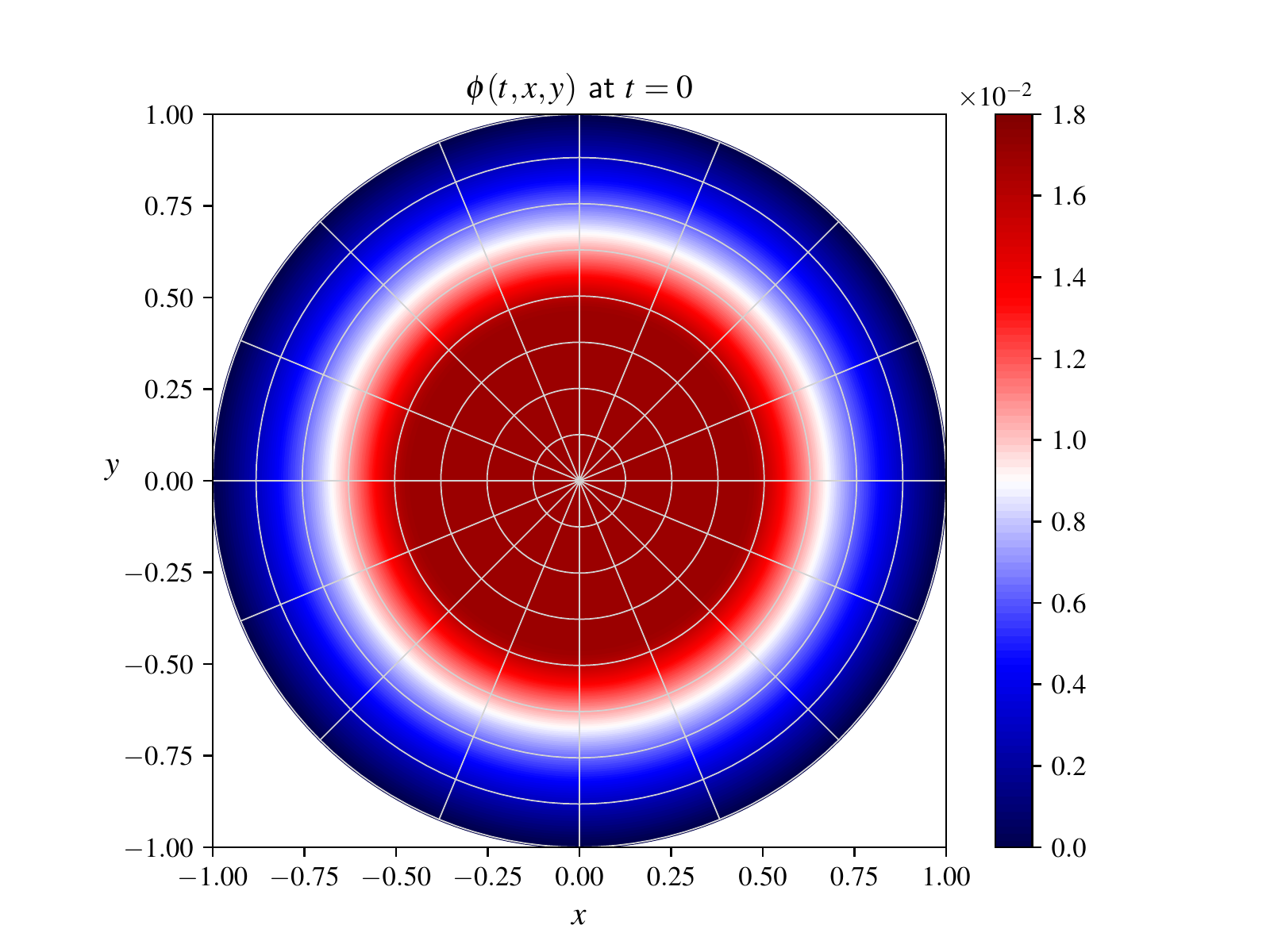}
\includegraphics[width=0.48\linewidth,trim={1cm 0 1cm 0},clip]{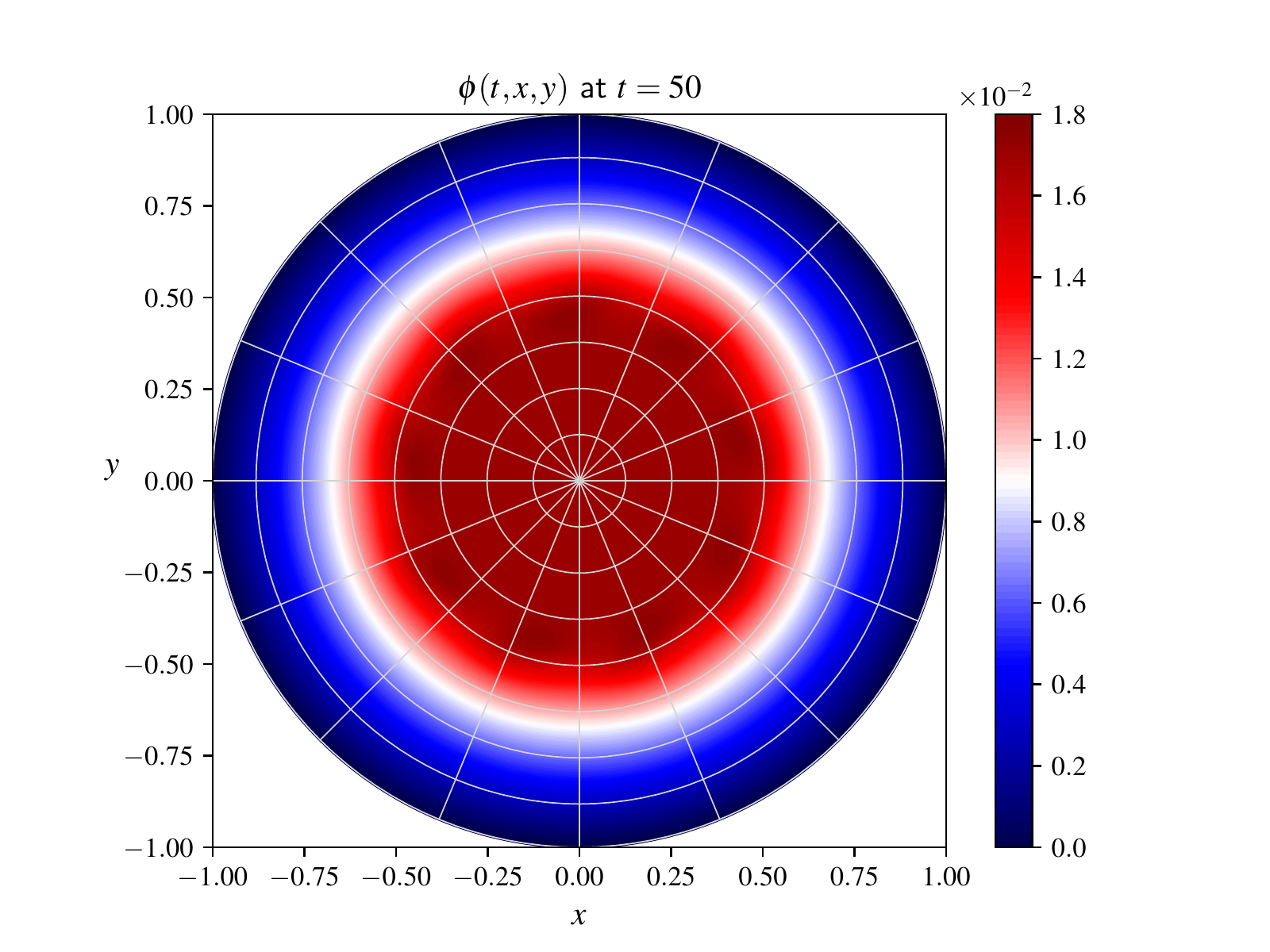}
\includegraphics[width=0.48\linewidth,trim={1cm 0 1cm 0},clip]{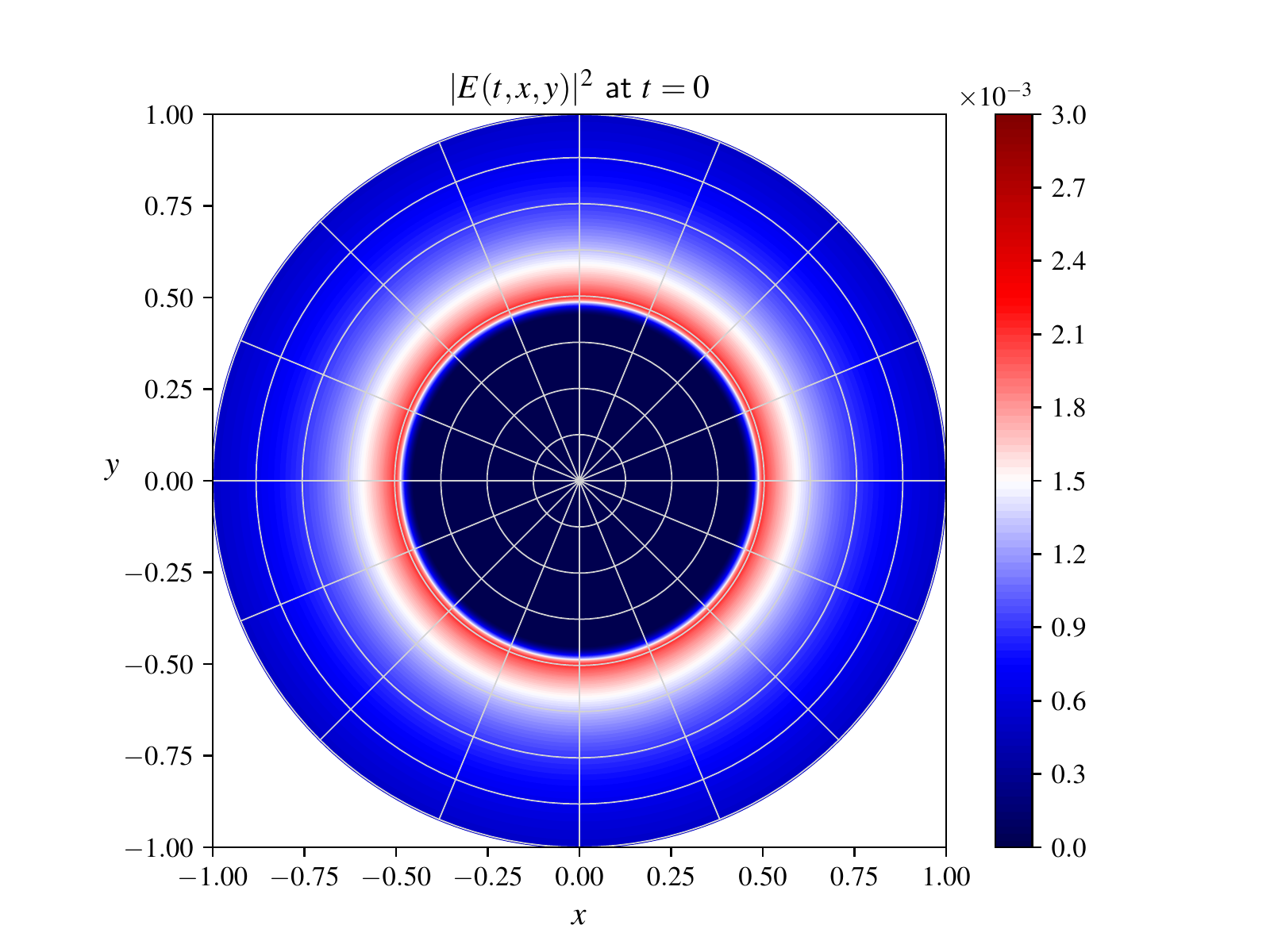}
\hspace{5mm}
\includegraphics[width=0.48\linewidth,trim={1cm 0 1cm 0},clip]{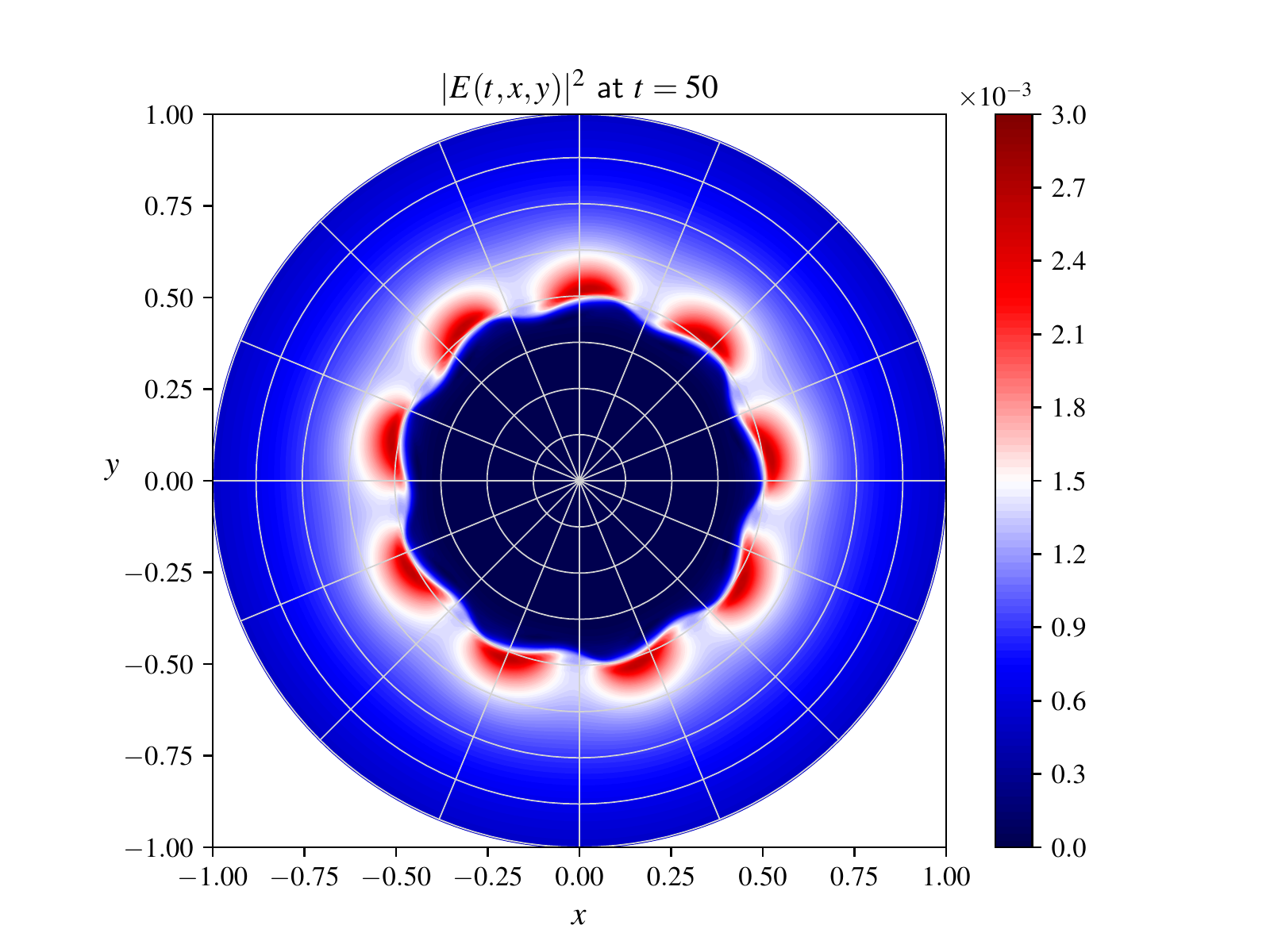}
\caption{Numerical simulation of the diocotron instability. From left to right,
from top to bottom: contour plots of the density $\rho(t,x,y)$, the electric potential
$\phi(t,x,y)$ and the electric energy density $|\Eb(t,x,y)|^2$ at times $t=0$
(beginning of the simulation) and $t=50$ (end of the linear phase).}
\label{fig_diocotron2}
\end{figure}

\begin{figure}
\centering
\includegraphics[width=0.48\linewidth,trim={0 0 0 0},clip]{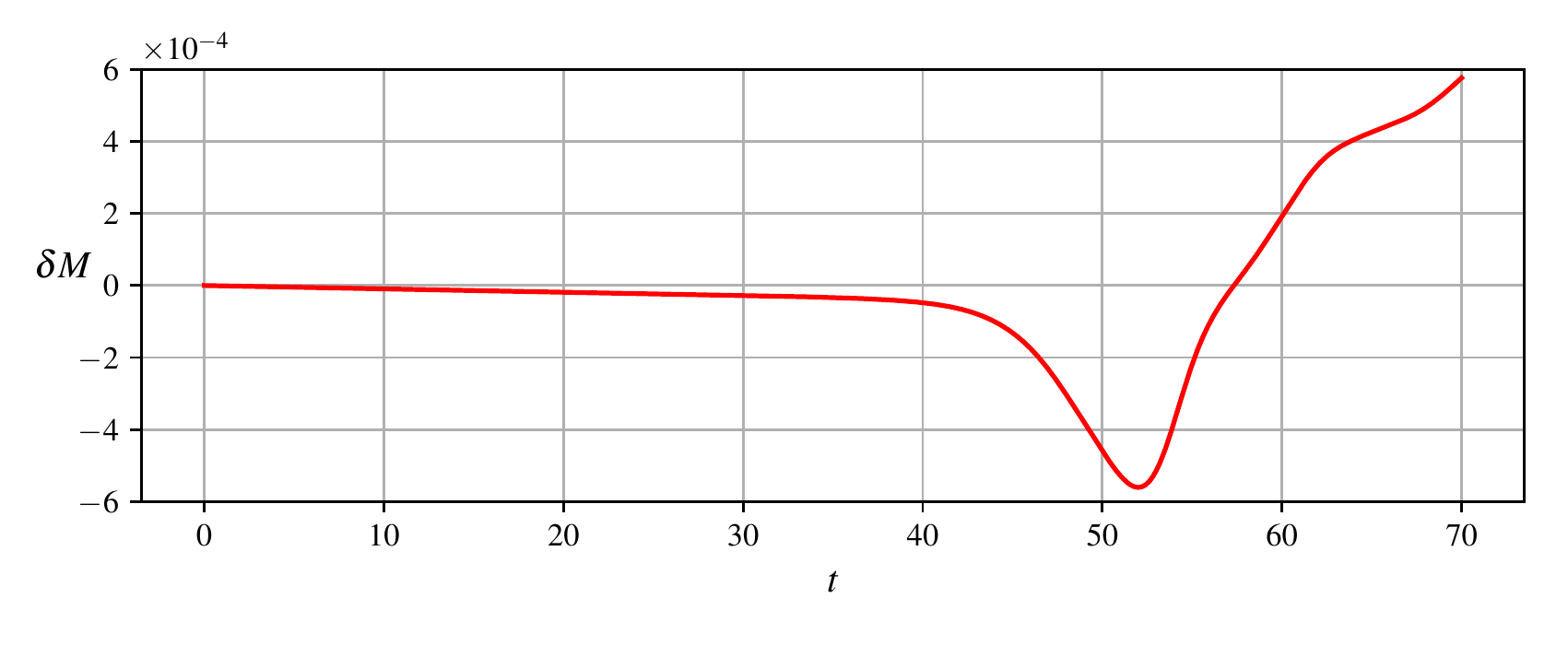}
\includegraphics[width=0.48\linewidth,trim={0 0 0 0},clip]{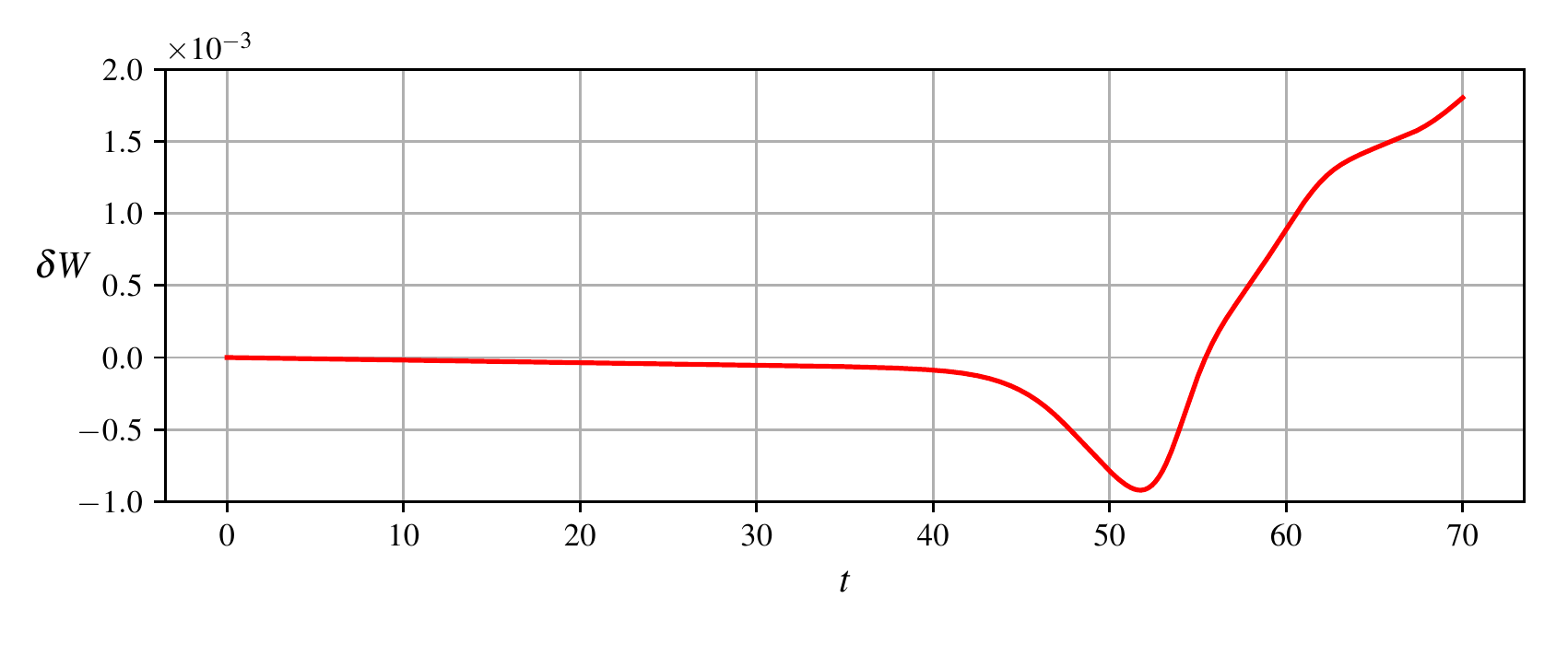}
\caption{Numerical simulation of the diocotron instability: time evolution of the
relative errors on the total mass (left) and energy (right).}
\label{fig_dioc_invariants}
\end{figure}

As a first test we investigate the evolution of the diocotron instability on
a domain defined by a circular mapping. From a physical point of view, this
corresponds to studying a non-neutral plasma in cylindrical geometry, where the
plasma particles are confined radially by a uniform axial magnetic field with a
cylindrical conducting wall located at the outer boundary \citep{Levy1965}.
Following \citep{Davidson2001}, we consider the initial density profile
\begin{equation}
\label{rho_initial}
\wh\rho(0,s,\theta):=\wh\rho_0(s)+\wh\rho_1(0,s,\theta):=
\begin{dcases}
1+\epsilon\cos(m\,\theta) & s^- \leq s \leq s^+ \,, \\
0 & \text{elsewhere} \,.
\end{dcases}
\end{equation}
This corresponds to a $\theta$-independent equilibrium $\wh\rho_0$ (an annular
charged layer) with a density perturbation $\wh\rho_1$ of azimuthal mode number $m$
and small amplitude $\epsilon$. The linear dispersion relation for a complex eigenfrequency
$\omega$ reads \citep{Davidson2001}
\begin{equation}
\label{dispersion_relation}
\left(\frac{\omega}{\omega_D}\right)^2-b_m\frac{\omega}{\omega_D}+c_m=0 \,,
\end{equation}
where $\omega_D$ is the diocotron frequency ($\omega_D=1/2$ in our units), and
$b_m$ and $c_m$ are defined as
\begin{equation*}
\begin{aligned}
& b_m:=m\left[1-\left(\frac{s^-}{s^+}\right)^2\right]+(s^+)^{2m}-(s^-)^{2m} \,, \\
& c_m:=m\left[1-\left(\frac{s^-}{s^+}\right)^2\right]\left[1-(s^-)^{2m}\right]
-\left[1-\left(\frac{s^-}{s^+}\right)^{2m}\right]\left[1-(s^+)^{2m}\right] \,.
\end{aligned}
\end{equation*}
If $4c_m>b_m^2$, then the oscillation frequencies resulting from \eqref{dispersion_relation}
form complex conjugate pairs. The solution with $\imag\omega>0$ corresponds to the
diocotron instability and describes how rapidly the electric potential grows.
The quantity of interest, in this regard, is the $L^2$-norm of the perturbed
electric potential
\begin{equation*}
||\phi-\phi_0||_{L^2}:=\sqrt{\int_\Omega\dd x\,\dd y\,
\left[\phi(t,x,y)-\phi_0(x,y)\right]^2}
=\sqrt{\int_{\wh\Omega}\dd s\,\dd\theta\,|\det J_{\Fb}(s,\theta)|\,
\left[\wh\phi(t,s,\theta)-\wh\phi_0(s,\theta)\right]^2} \,,
\end{equation*}
where $\phi_0$ denotes the equilibrium electric potential and the integration
is performed again on the Gauss-Legendre quadrature points and weights. In order to represent
the initial density in the finite-dimensional space of tensor-product splines,
we modify~\eqref{rho_initial} by a radial smoothing to avoid discontinuities:
\begin{equation}
\label{rho_initial_smooth}
\wh\rho(0,s,\theta):=\wh\rho_0(s)+\wh\rho_1(0,s,\theta):=
\begin{dcases}
\left[1+\epsilon\cos(m\,\theta)\right]
\exp\left[-\left(\frac{s-\wb s}{d}\right)^p\right]
& s^- \leq s \leq s^+ \,, \\
0 & \text{elsewhere} \,,
\end{dcases}
\end{equation}
with $\wb s:=(s^++s^-)/2$ and $d:=(s^+-s^-)/2$. If the smoothing layer is small enough,
we can still rely on the analytical result obtained for the dispersion relation in the
case of the sharp annular layer \eqref{rho_initial}. The numerical results have been
verified against the analytical dispersion relation for a perturbation with azimuthal
mode number $m=9$ and amplitude $\epsilon=10^{-4}$. The numerical growth rate is in
good agreement with the analytical one, $\imag\omega\approx 0.18$, for the time interval
$20 \lesssim t \lesssim 50$, which corresponds to the linear phase.
At time $t \approx 50$, the system enters its non-linear phase. The simulation is run
with $n_1\times n_2=128\times256$ and $\Delta t=0.1$, with the explicit
time integrator described in section \ref{sec_explicit}. Additional parameters
defining the initial condition \eqref{rho_initial_smooth} have been set to $s^-=0.45$,
$s^+=0.50$ and $p=50$. Numerical results are illustrated in Figures \ref{fig_diocotron1}
and~\ref{fig_diocotron2}.
For the conservation of mass and energy we get
\begin{equation*}
\max_{t\in[0,70]}\delta\Mcal(t)\approx 5.8\times10^{-4} \,, \quad
\max_{t\in[0,70]}\delta\Wcal(t)\approx 1.8\times10^{-3} \,.
\end{equation*}
The time evolution of the relative errors on these conserved quantities is shown in Figure \ref{fig_dioc_invariants}.
Figure \ref{fig_diocotron2} shows that in
this test case nothing significant happens in the region close to the pole. The effect
of using $\Ccal^1$ smooth polar splines in such situations is not particularly evident,
but they do ensure continuity of the advection field responsible for the transport
of $\rho$ (the electric field) everywhere in the domain.
Moreover, pseudo-Cartesian coordinates reduce to standard Cartesian coordinates,
as the physical domain is defined by a simple circular mapping. The interest of this
test case lies primarily in the fact that it provides the valuable possibility of
easily verifying the implementation of our numerical scheme by comparing the numerical
results with an analytical dispersion relation.

\subsection{Numerical test: vortex merger}
\label{sec_vortex_merger}
\begin{figure}[h]
\centering
\includegraphics[width=0.48\linewidth,trim={1cm 0 5mm 0},clip]{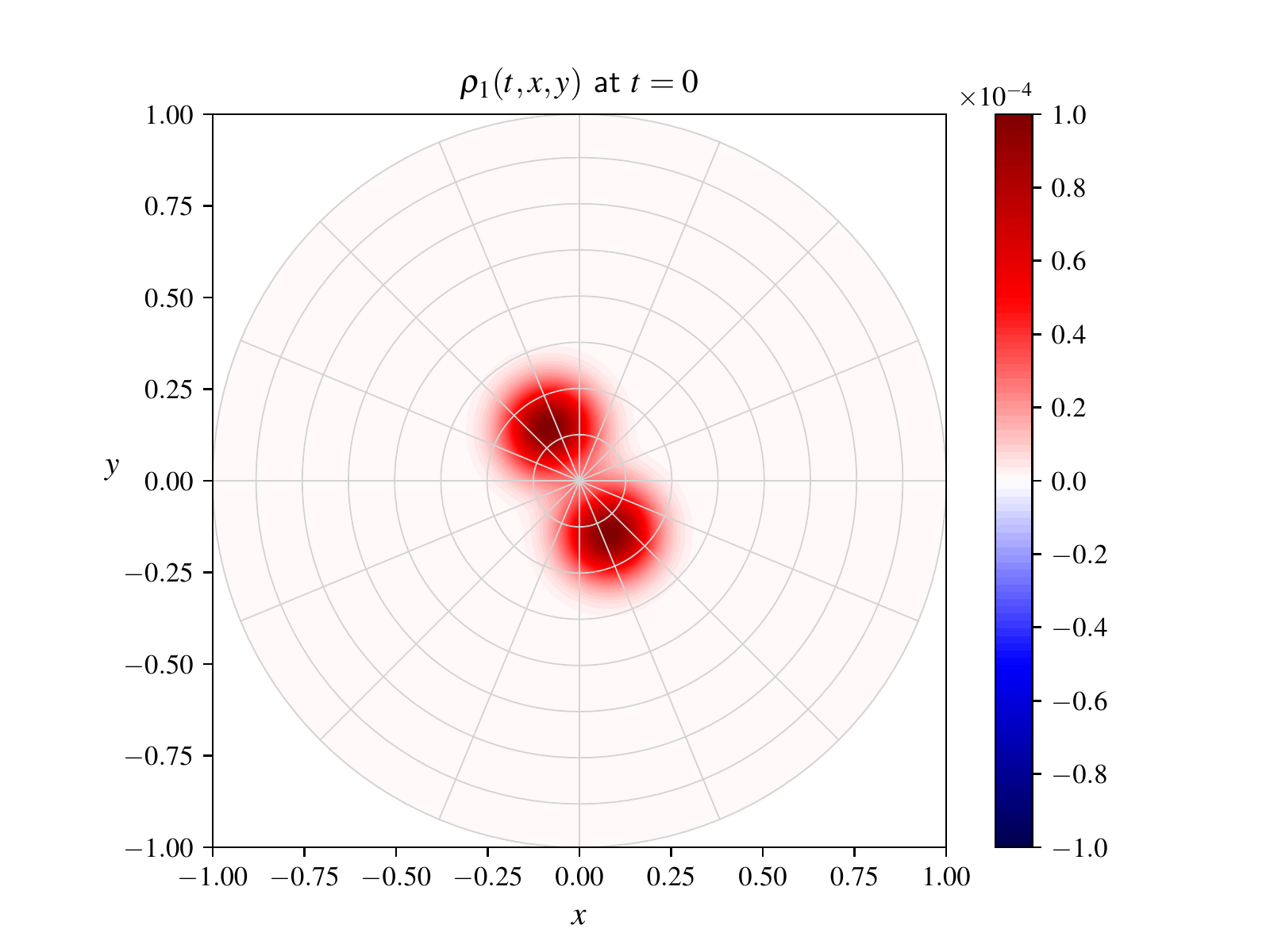}
\includegraphics[width=0.48\linewidth,trim={1cm 0 5mm 0},clip]{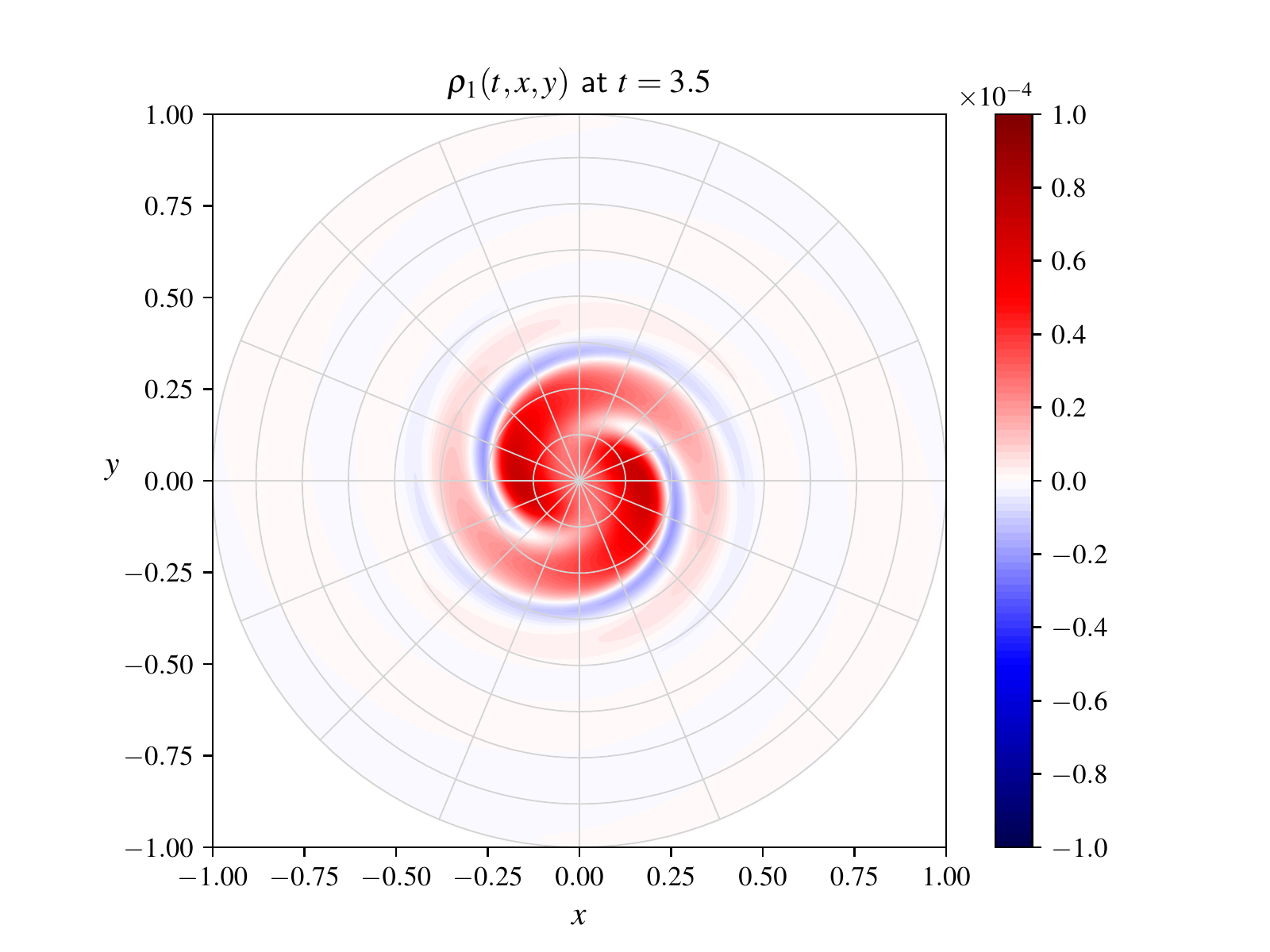}
\includegraphics[width=0.48\linewidth,trim={1cm 0 5mm 0},clip]{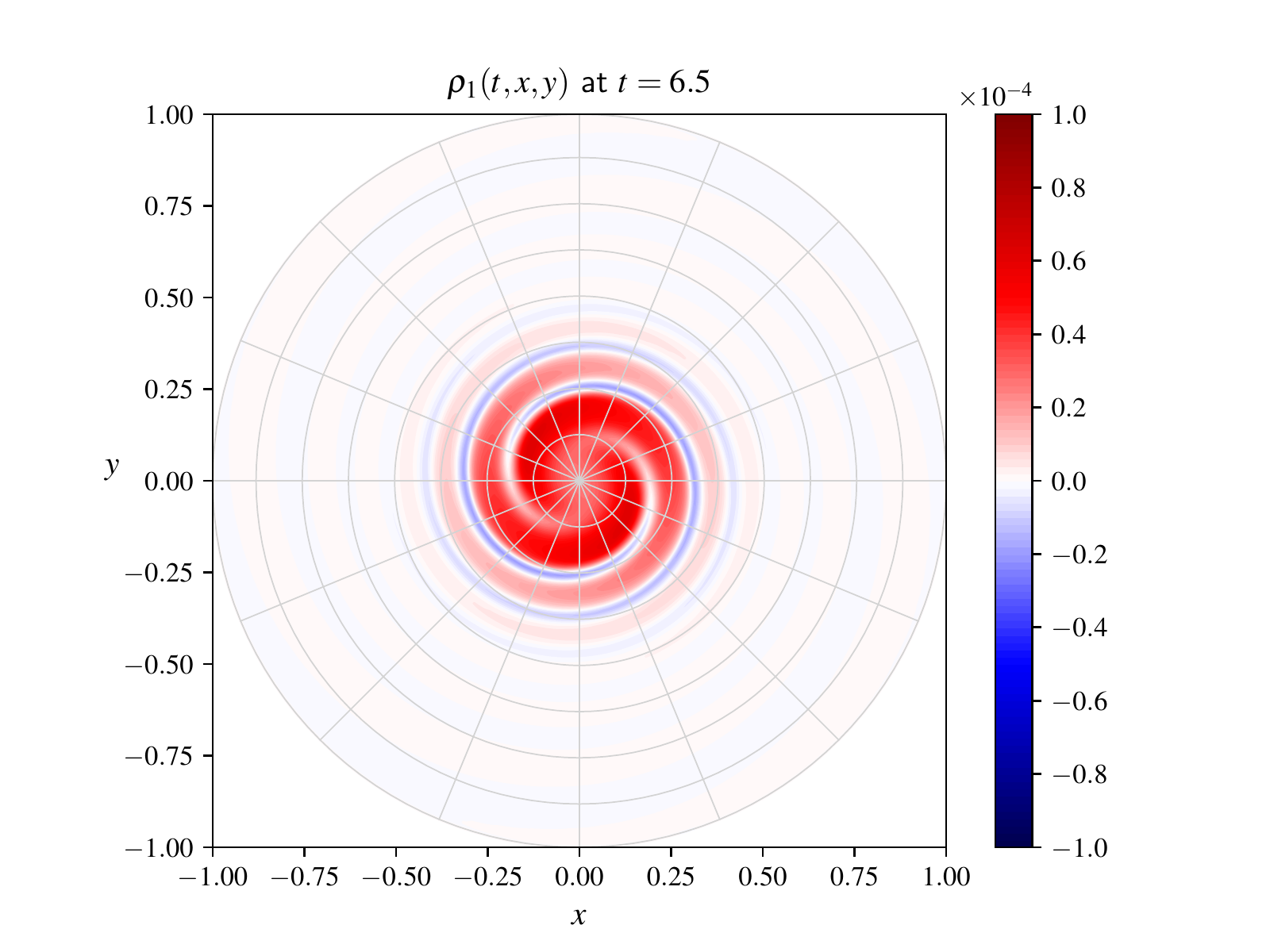}
\includegraphics[width=0.48\linewidth,trim={1cm 0 5mm 0},clip]{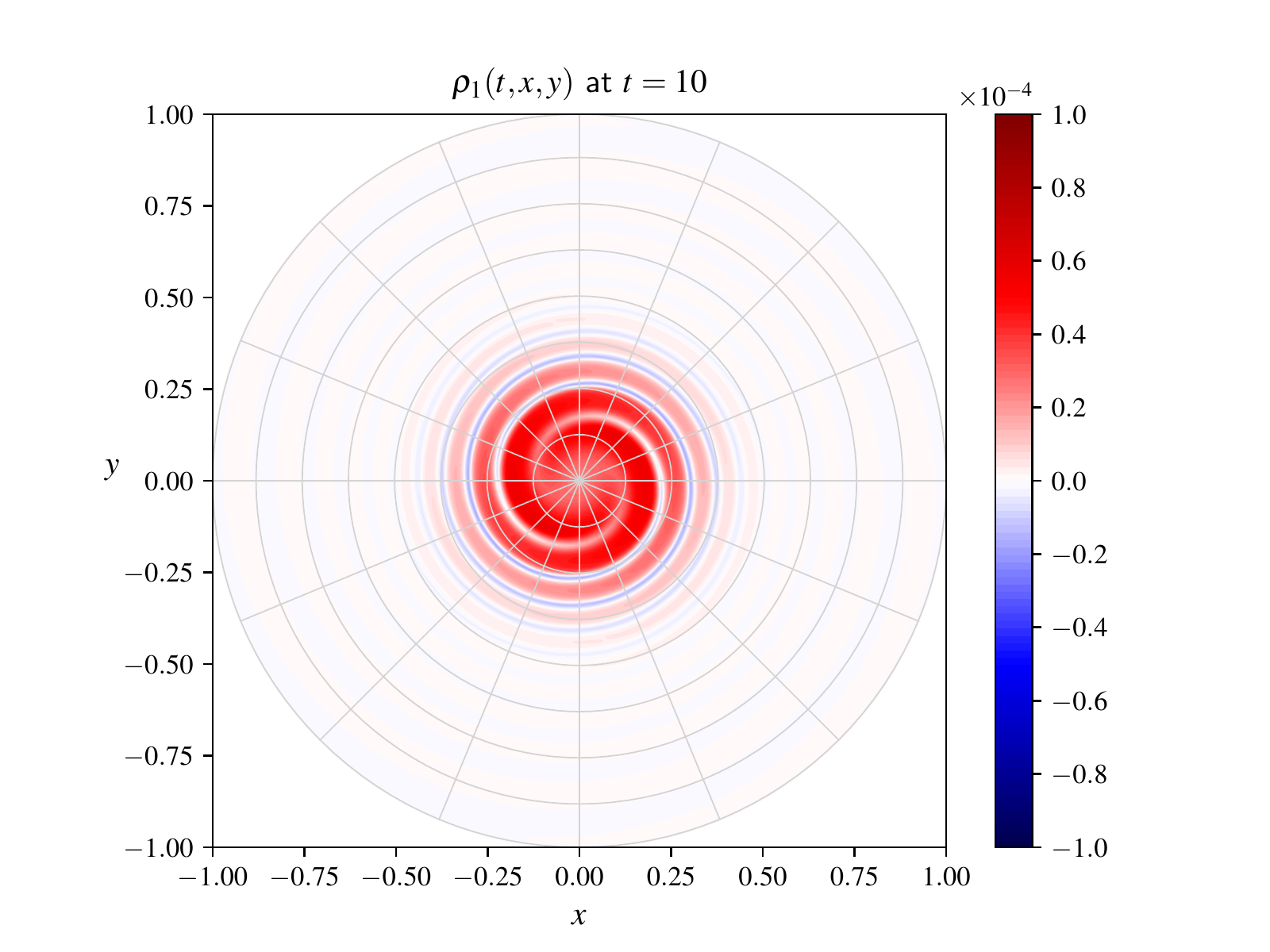}
\caption{Numerical simulation of the merger of two vortices: contour plots of
the vorticity perturbation $\rho_1(t,x,y)$ at different times.}
\label{fig_vortex_merger}
\end{figure}

\begin{figure}[h]
\centering
\includegraphics[width=0.48\linewidth,trim={0 0 0 0},clip]{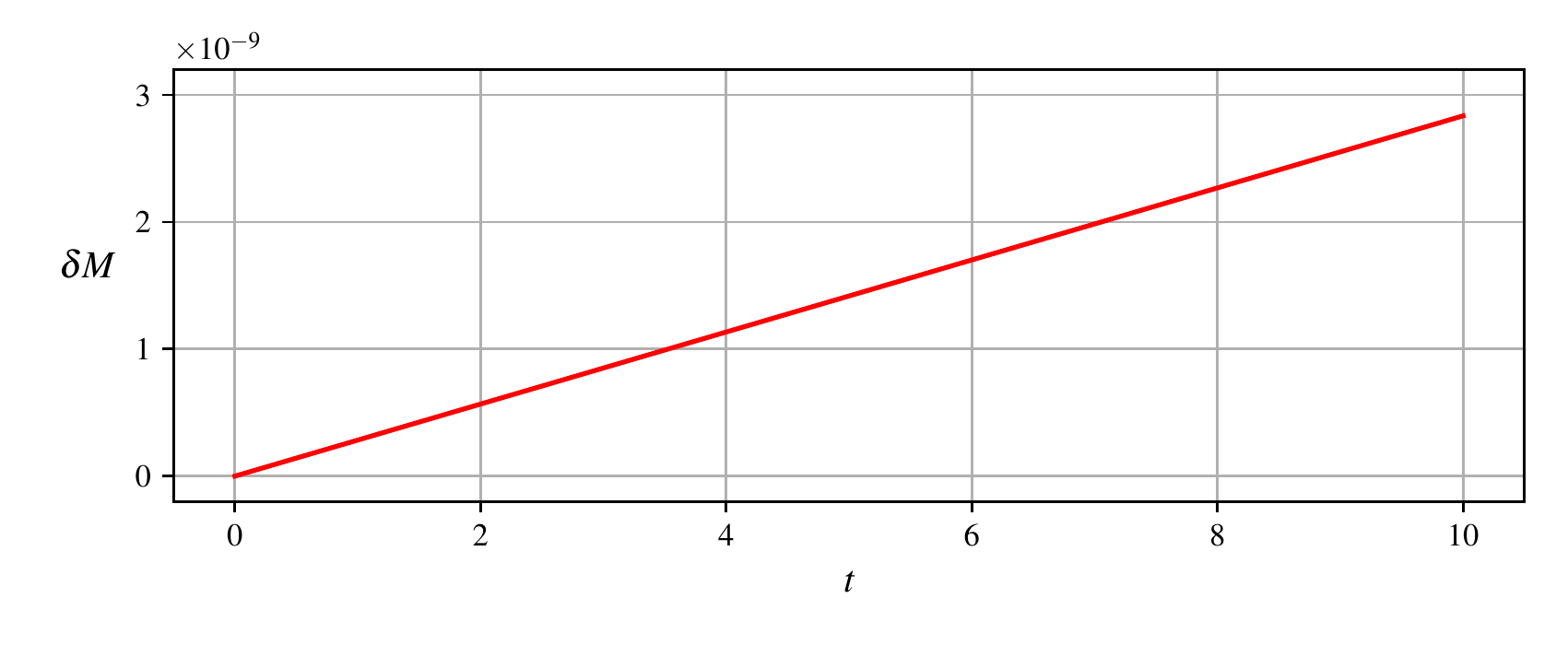}
\includegraphics[width=0.48\linewidth,trim={0 0 0 0},clip]{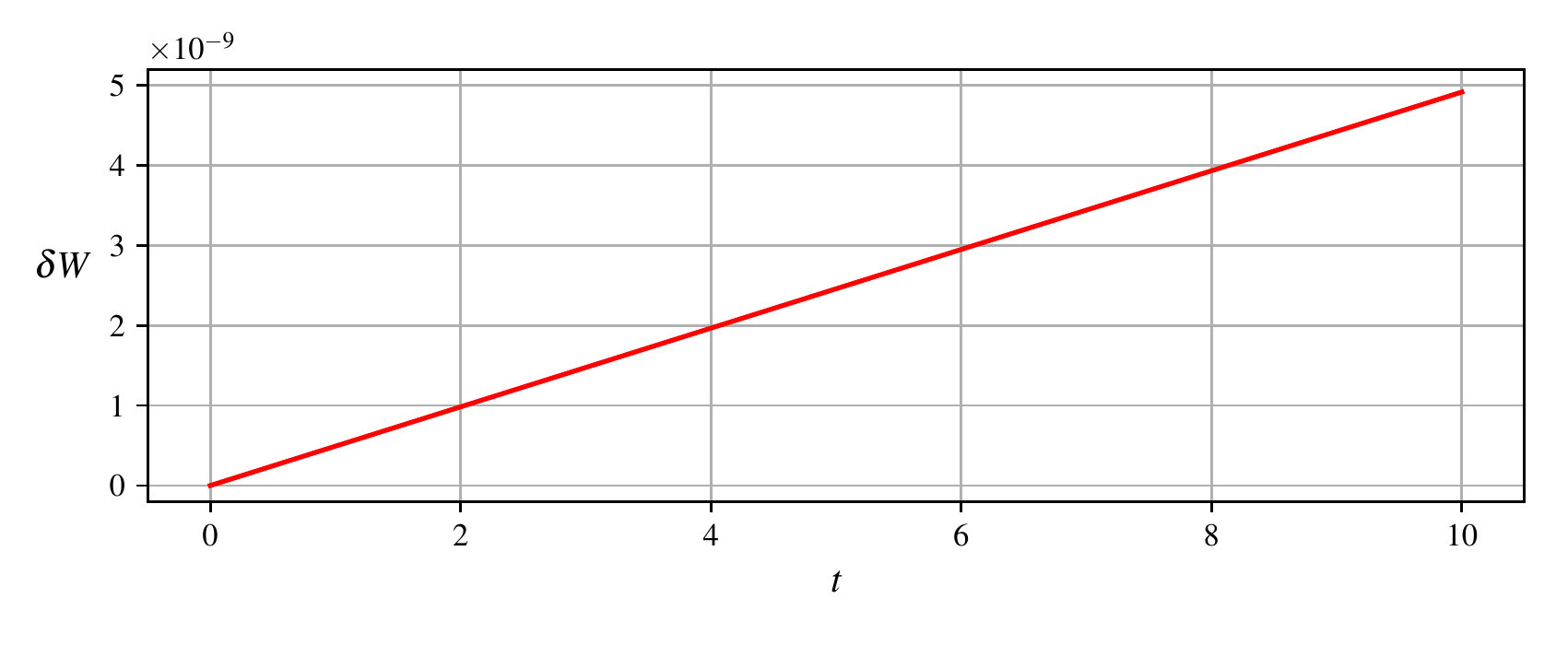}
\caption{Numerical simulation of the merger of two vortices: time evolution
of the relative errors on the total mass (left) and energy (right).}
\label{fig_vm_invariants}
\end{figure}

In the context of incompressible inviscid 2D Euler fluids, we simulate the merger
of two macroscopic vortices by setting up initial conditions qualitatively similar to
those described in \citep[section 3]{Driscolletal2002}. Unlike the diocotron
instability, the interest of this test case lies primarily in the fact that the
relevant dynamics occurs in a region close to the pole of the physical domain. We consider
an equilibrium $\rho_0$ obtained with the numerical procedure described in
section~\ref{num_equil} with $f(\phi)=\phi^2$ and $\phi_\text{max}=1$, and perturb
it with two Gaussian perturbations,
\begin{equation*}
\begin{split}
\rho(0,x,y) & :=\rho_0(x,y)+\rho_1(0,x,y) \\
& :=\rho_0(x,y)+\epsilon\left(\exp\left[-\frac{(x-x^*_1)^2+(y-y^*_1)^2}{2\sigma^2}\right]
+\exp\left[-\frac{(x-x^*_2)^2+(y-y^*_2)^2}{2\sigma^2}\right]\,\right) \,,
\end{split}
\end{equation*}
with amplitude $\epsilon=10^{-4}$, width $\sigma=0.08$ and centered in
$(x^*_1,y^*_1)=(+0.08,-0.14)$ and $(x^*_2,y^*_2)=(-0.08,+0.14)$.
The time evolution of the initial perturbation $\rho_1$ is shown in Figure~\ref{fig_vortex_merger}.
The simulation is run with ${n_1\times n_2=128\times256}$ and time step
$\Delta t=0.1$, with the second-order implicit time integrator described in section~\ref{sec_implicit}.
The explicit time integrator would require in this case very small time steps in
order to capture the correct dynamics. Two different aspects play a role in the choice
of the time integrator for this particular test case. On the one hand, the error
in the integration of the characteristics, which scales with $\Delta t^2$ for the
second-order explicit scheme described in section \ref{sec_explicit}, must not be
larger than the amplitude of the perturbation on the advection field caused by the
density perturbation. In other words, for the explicit scheme, the choice of the
time step would be dependent on the amplitude~$\epsilon$ of the density perturbation.
On the other hand, committing an error in the integration of closed trajectories
(as it would be when using the explicit scheme even for stationary advection fields)
seems to disrupt the dynamics, preventing the simulation from correctly predicting
the merger of the two macroscopic vortices. For the conservation of mass and energy we get
\begin{equation*}
\max_{t\in[0,10]}\delta\Mcal(t)\approx 2.8\times10^{-9} \,, \quad
\max_{t\in[0,10]}\delta\Wcal(t)\approx 4.9\times10^{-9} \,.
\end{equation*}
The time evolution of the relative errors on these conserved quantities is shown in Figure \ref{fig_vm_invariants}.
The results of a convergence analysis of the numerical results while decreasing
the time step are shown in Table~\ref{tab_vortex_merger_conv}, where $\Delta\rho$
denotes the difference between the vorticity $\rho$ and a reference vorticity obtained
by running a simulation with time step $\Delta t=0.1/16$.
{
\renewcommand{\arraystretch}{1.5}
\begin{table}[h]
\begin{center}
\begin{tabular}{ccccccc}
\hline
$\Delta t$ & $||\Delta\rho||_{L^\infty}$ & Order & $\max_t\delta\Mcal(t)$ & Order & $\max_t\delta\Wcal(t)$ & Order \\
\hline
$0.1$   & $3.04\times10^{-5}$ &        & $2.84\times10^{-9}$  &        & $4.92\times10^{-9}$  &    \\
$0.1/2$ & $8.44\times10^{-6}$ & $1.85$ & $1.42\times10^{-9}$  & $1.00$ & $2.47\times10^{-9}$  & $0.99$ \\
$0.1/4$ & $2.05\times10^{-6}$ & $2.04$ & $7.14\times10^{-10}$ & $0.99$ & $1.22\times10^{-9}$  & $1.02$ \\
$0.1/8$ & $4.12\times10^{-7}$ & $2.32$ & $3.46\times10^{-10}$ & $1.05$ & $6.08\times10^{-10}$ & $1.01$ \\
\hline
\end{tabular}
\end{center}
\caption{Convergence in time of the numerical results for the vortex merger with
respect to reference results obtained with time step $\Delta t=0.1/16$. The mesh
size $n_1\times n_2=128\times 256$ is kept fixed in this convergence analysis.}
\label{tab_vortex_merger_conv}
\end{table}}

\subsection{Numerical test: point-like vortex dynamics}
\label{sec_pointlike}
\begin{figure}
\centering
\includegraphics[width=0.48\linewidth,trim={1cm 0 2cm 0},clip]{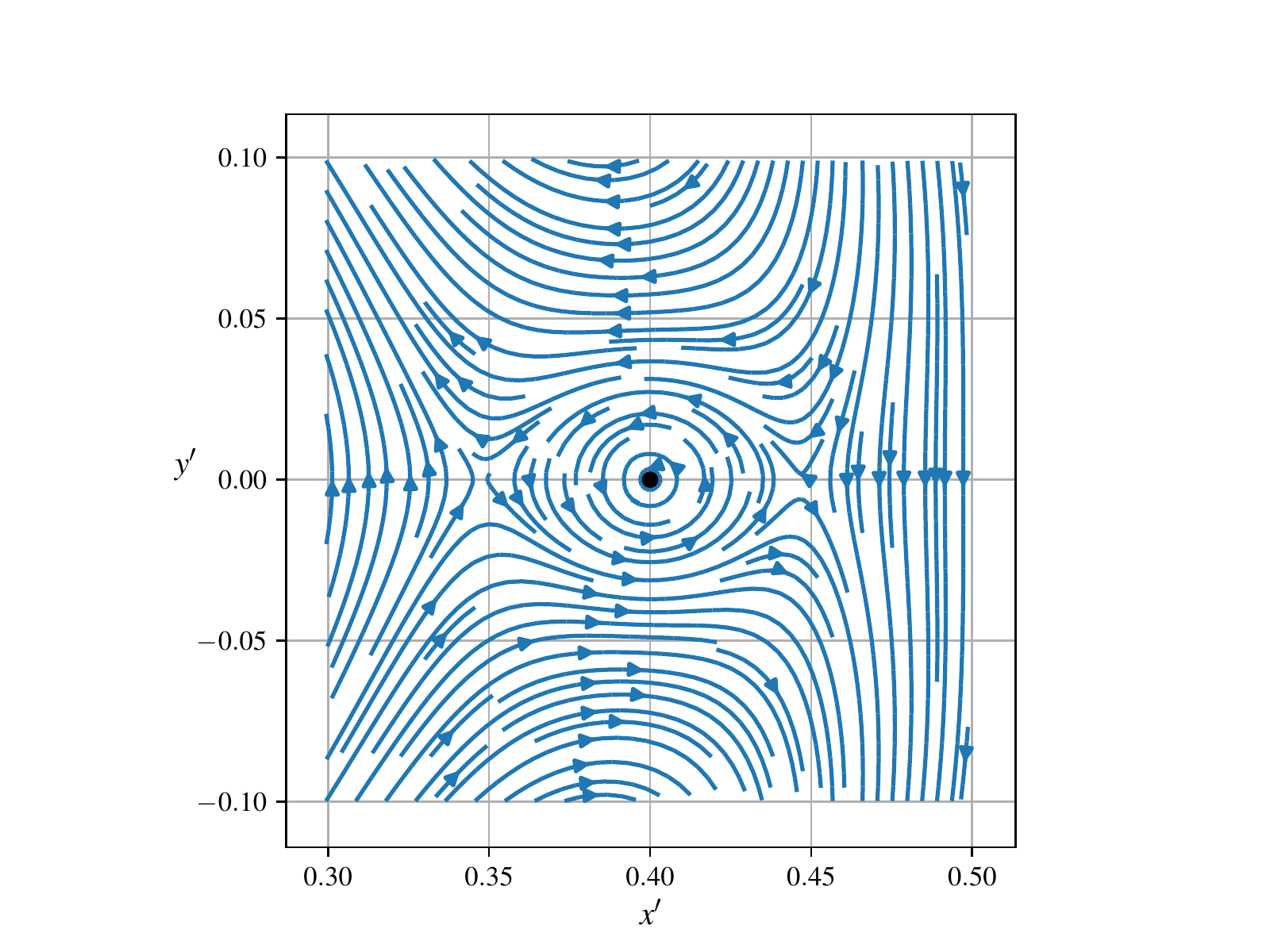}
\includegraphics[width=0.48\linewidth,trim={1cm 0 2cm 0},clip]{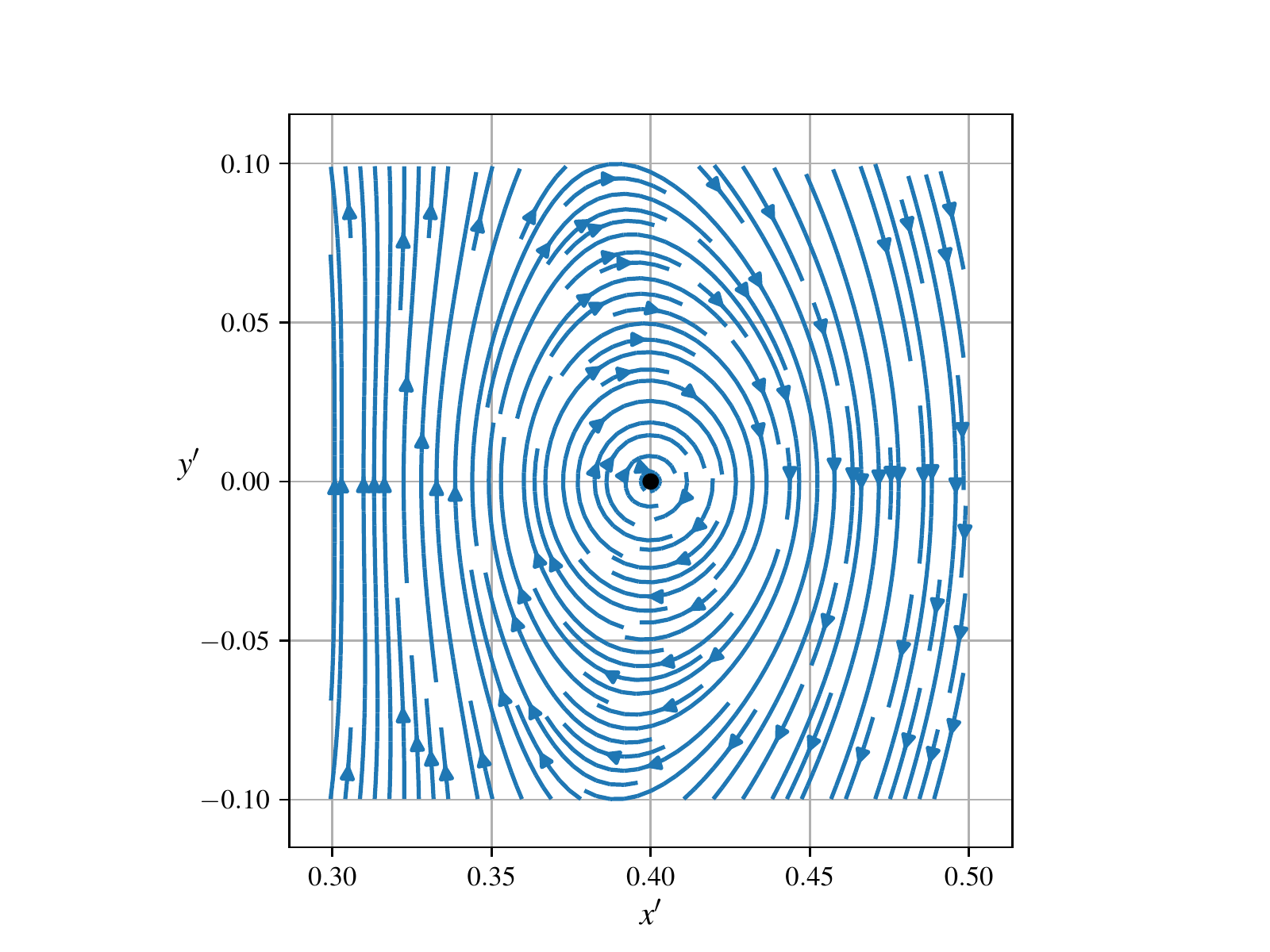}
\caption{Local stream lines of the advection field near positive (left) and negative (right)
point-like vortices in a rotating frame $(x',y')$.}
\label{fig_vic_stream_lines}
\end{figure}

\begin{figure}
\centering
\includegraphics[width=0.48\linewidth,trim={1cm 0 1cm 0},clip]{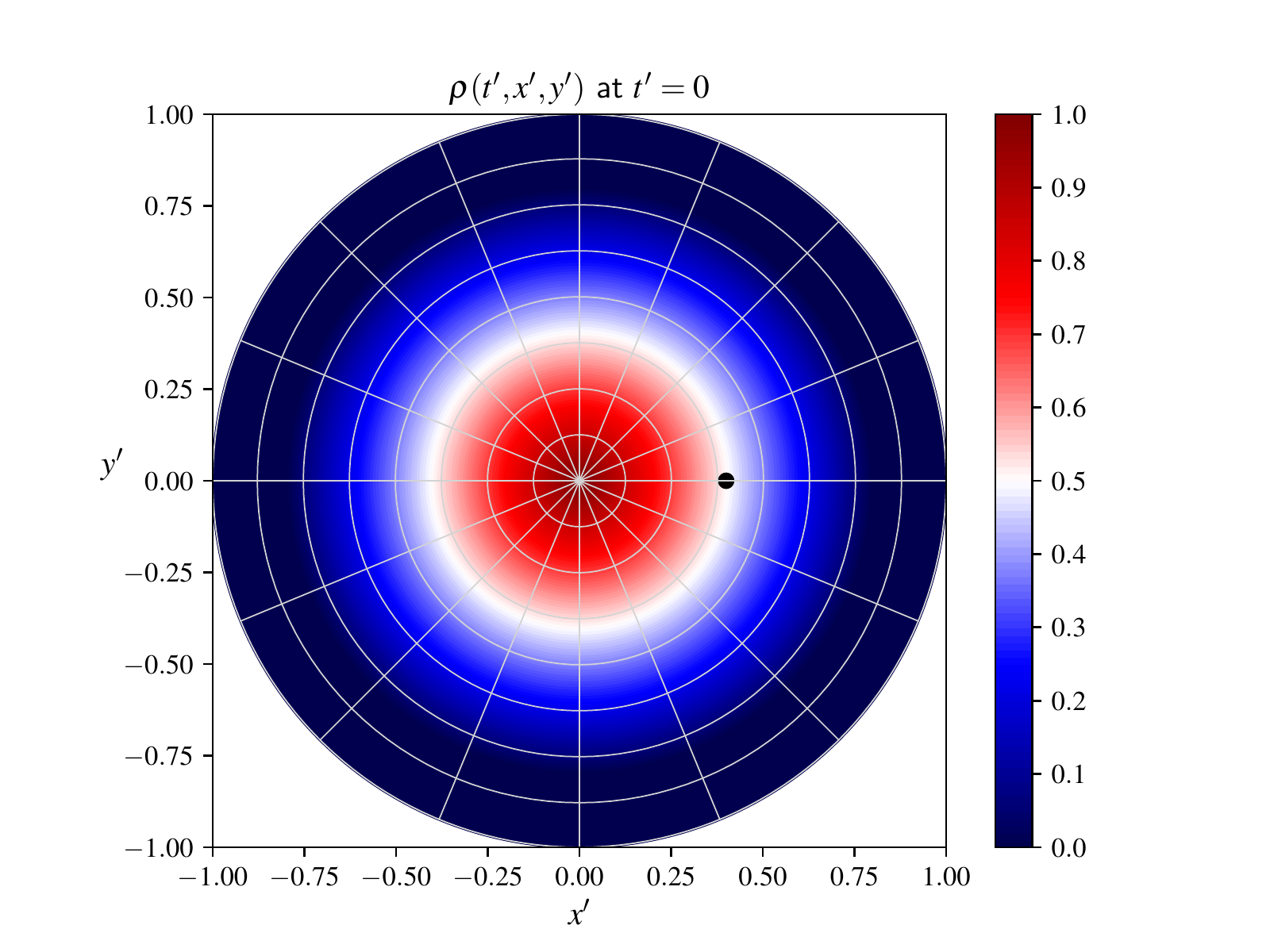}
\includegraphics[width=0.48\linewidth,trim={1cm 0 1cm 0},clip]{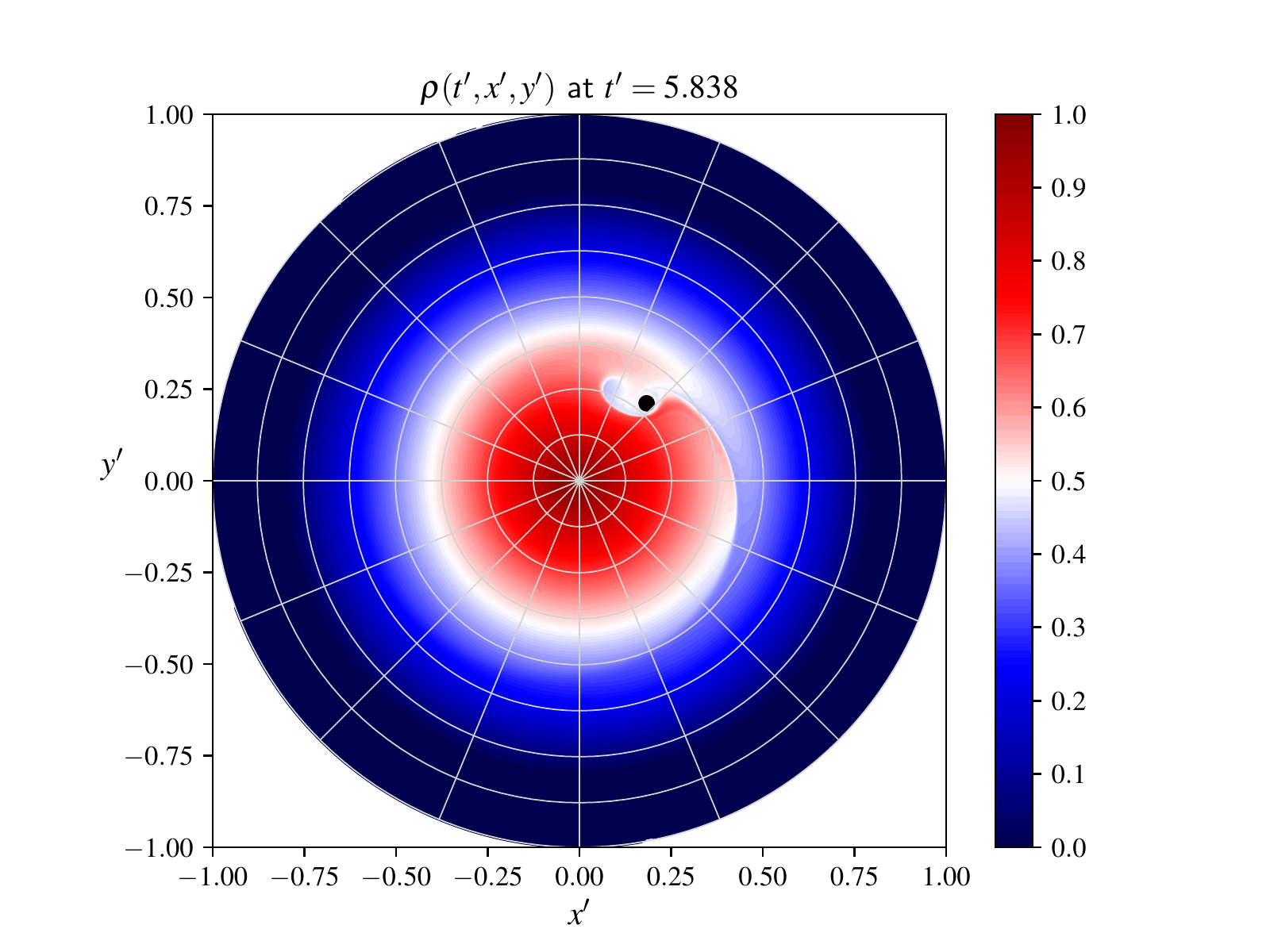}
\includegraphics[width=0.48\linewidth,trim={1cm 0 1cm 0},clip]{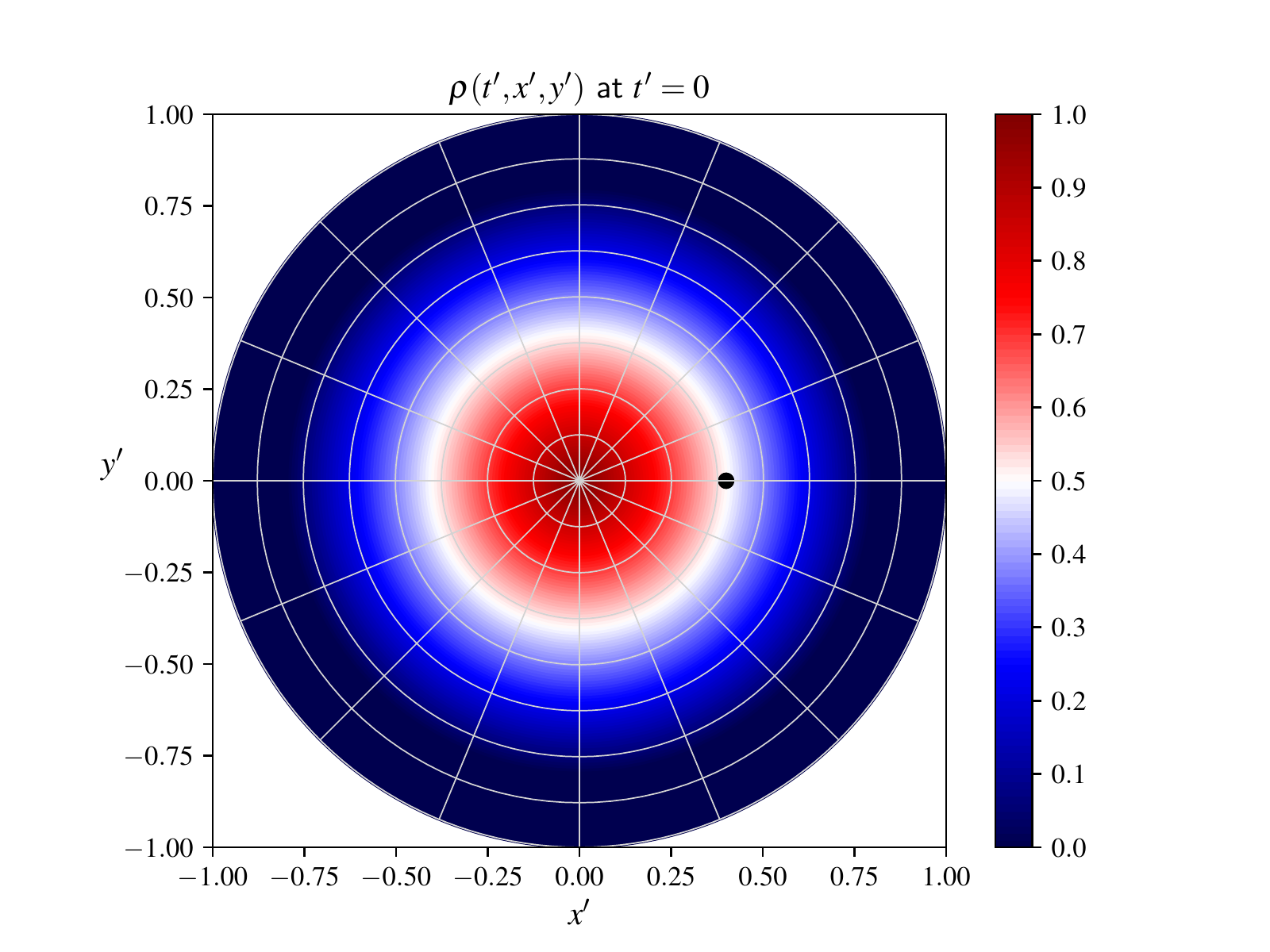}
\includegraphics[width=0.48\linewidth,trim={1cm 0 1cm 0},clip]{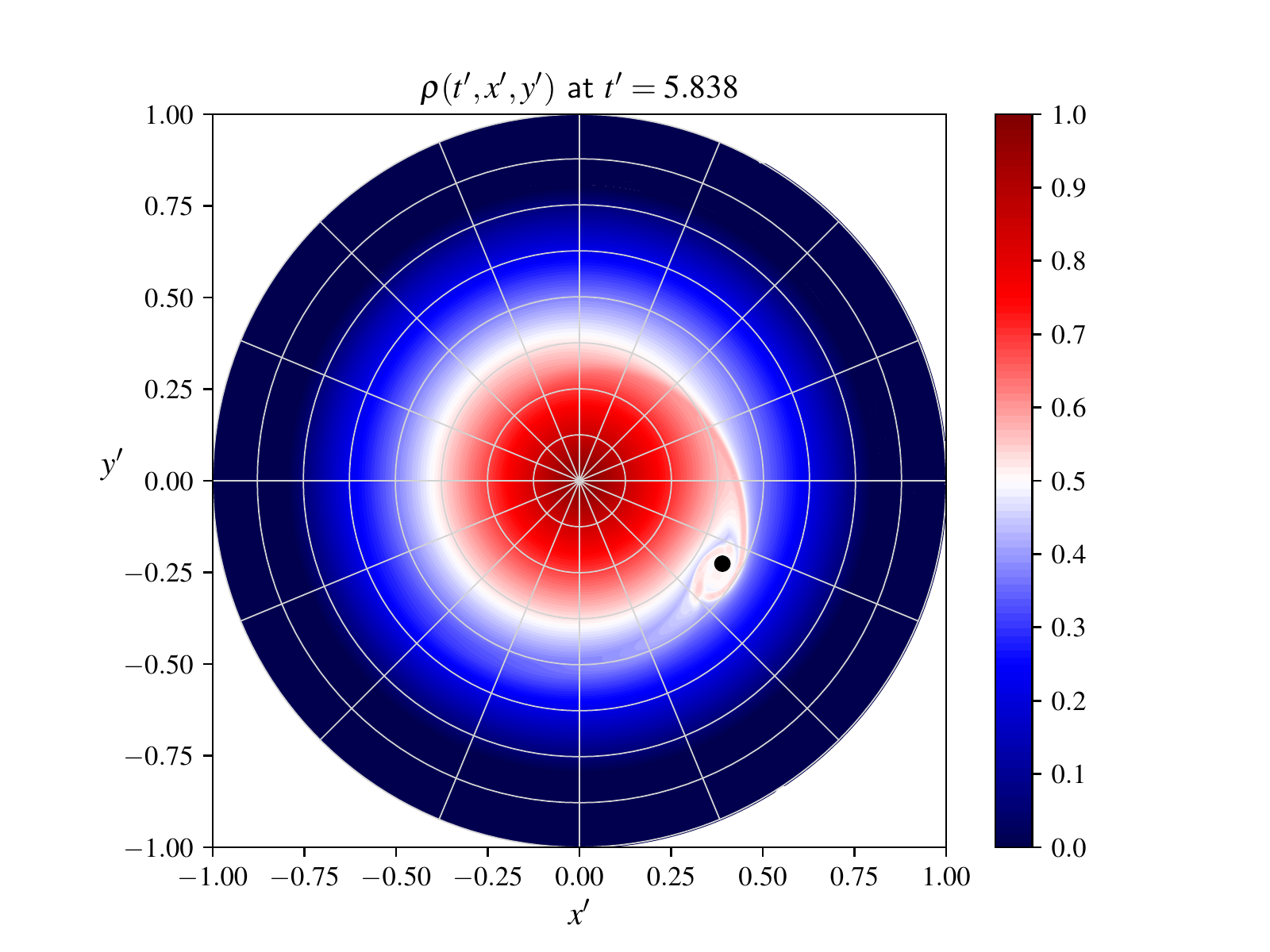}
\caption{Dynamics of a positive (top) and negative (bottom) point-like vortex on
a domain defined by a circular mapping: contour plots of the vorticity
at times $t'=0$ and $t'=5.838$ in a rotating frame $(x',y')$.}
\label{fig_vic}
\end{figure}

\begin{figure}
\centering
\includegraphics[width=0.48\linewidth,trim={0 0 0 0},clip]{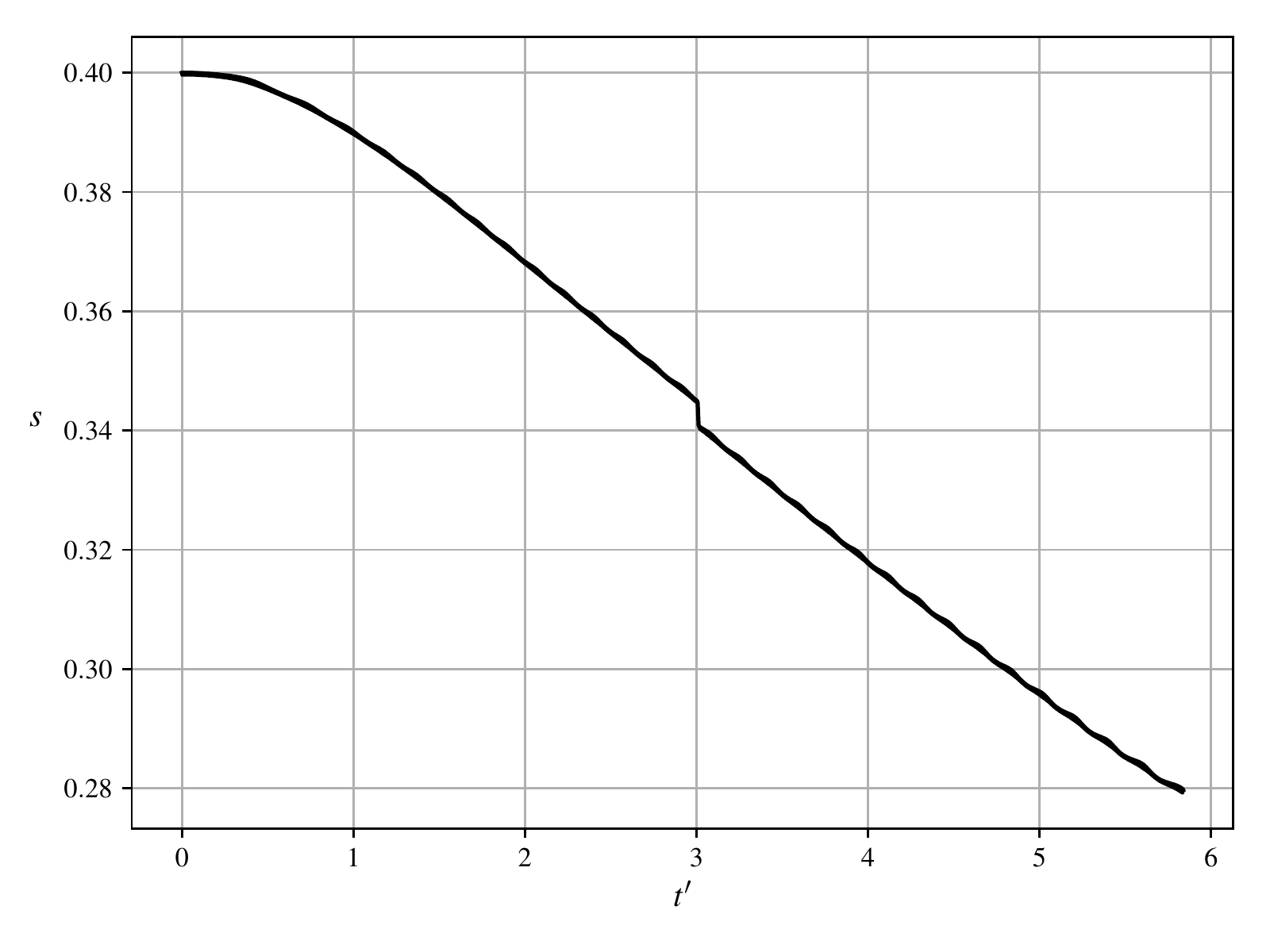}
\includegraphics[width=0.48\linewidth,trim={0 0 0 0},clip]{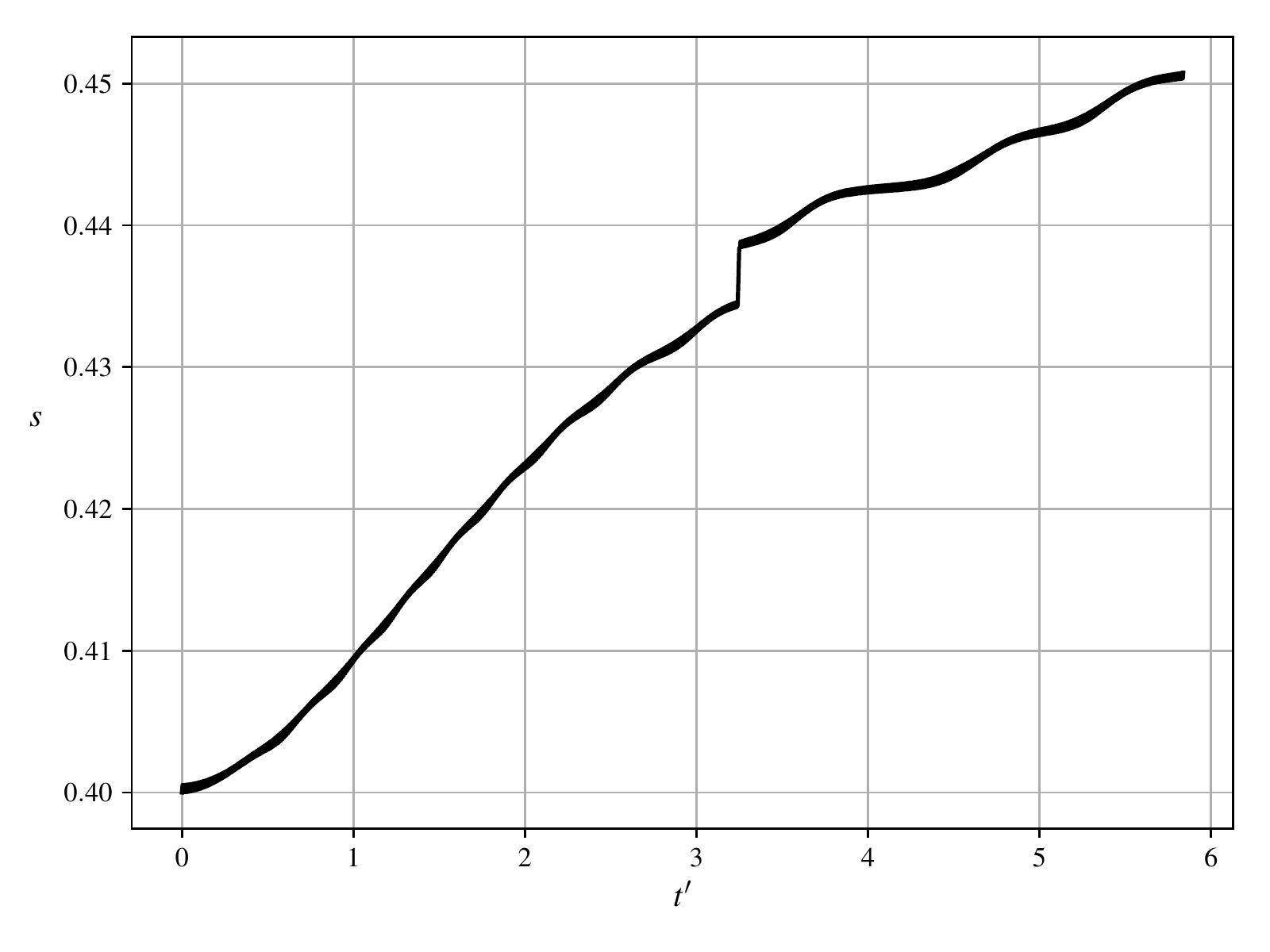}
\caption{Time evolution of the radial position of a positive (left) and negative
(right) point-like vortex.}
\label{fig_vic_radial_position}
\end{figure}

\begin{figure}
\centering
\includegraphics[width=0.48\linewidth,trim={0 0 0 0},clip]{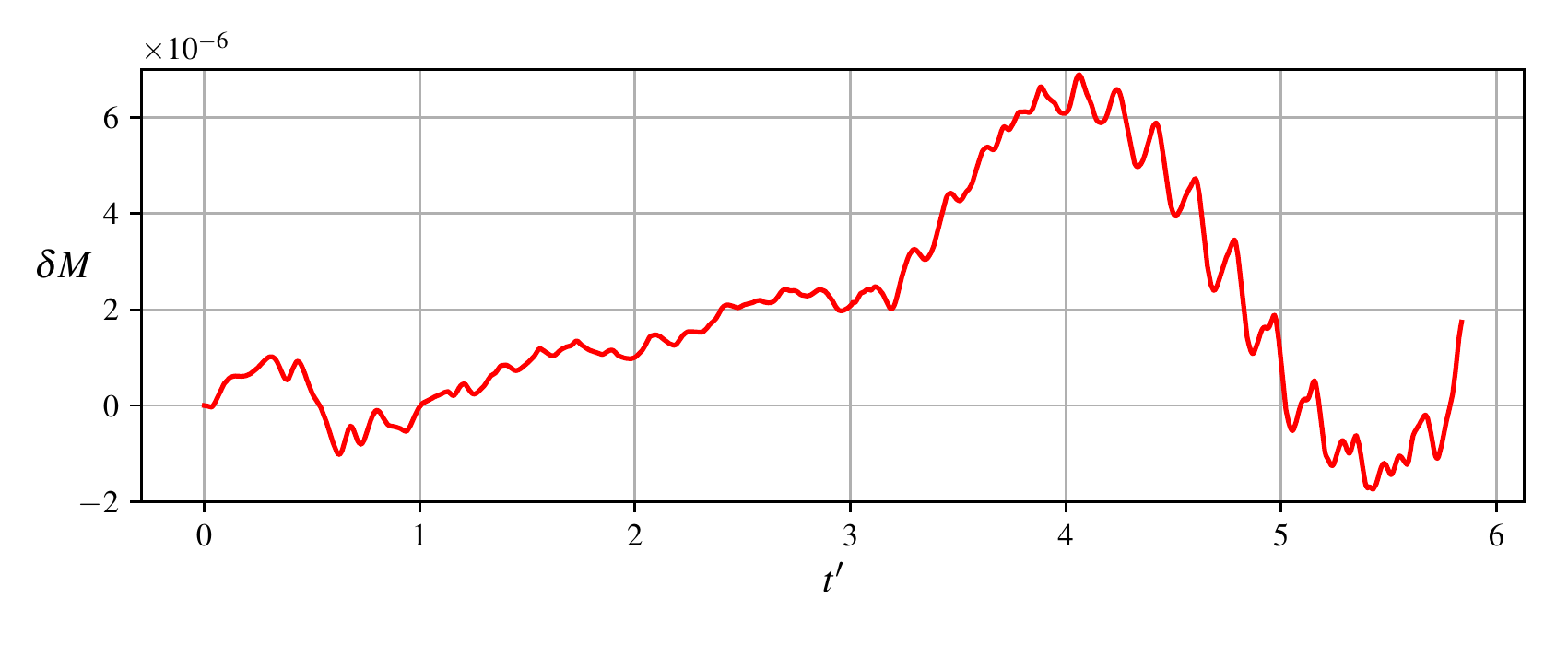}
\includegraphics[width=0.48\linewidth,trim={0 0 0 0},clip]{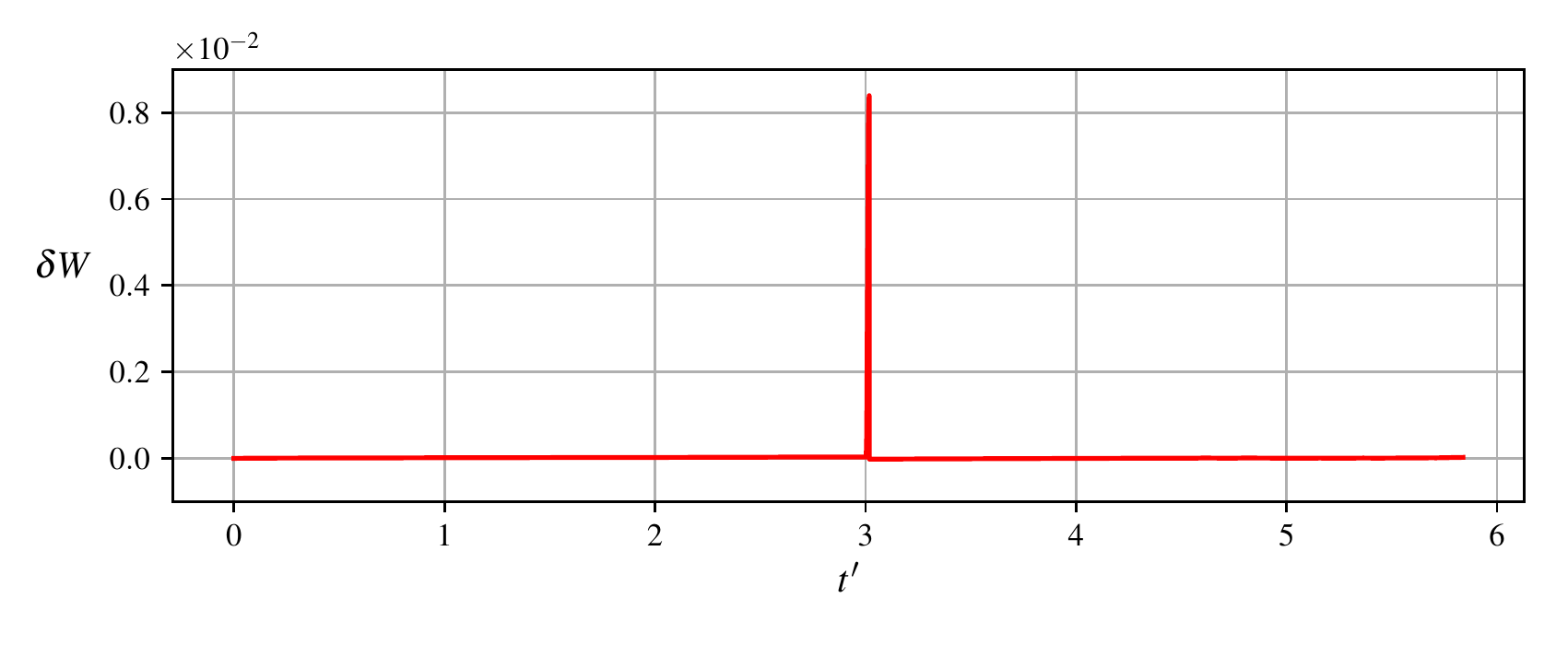}
\includegraphics[width=0.48\linewidth,trim={0 0 0 0},clip]{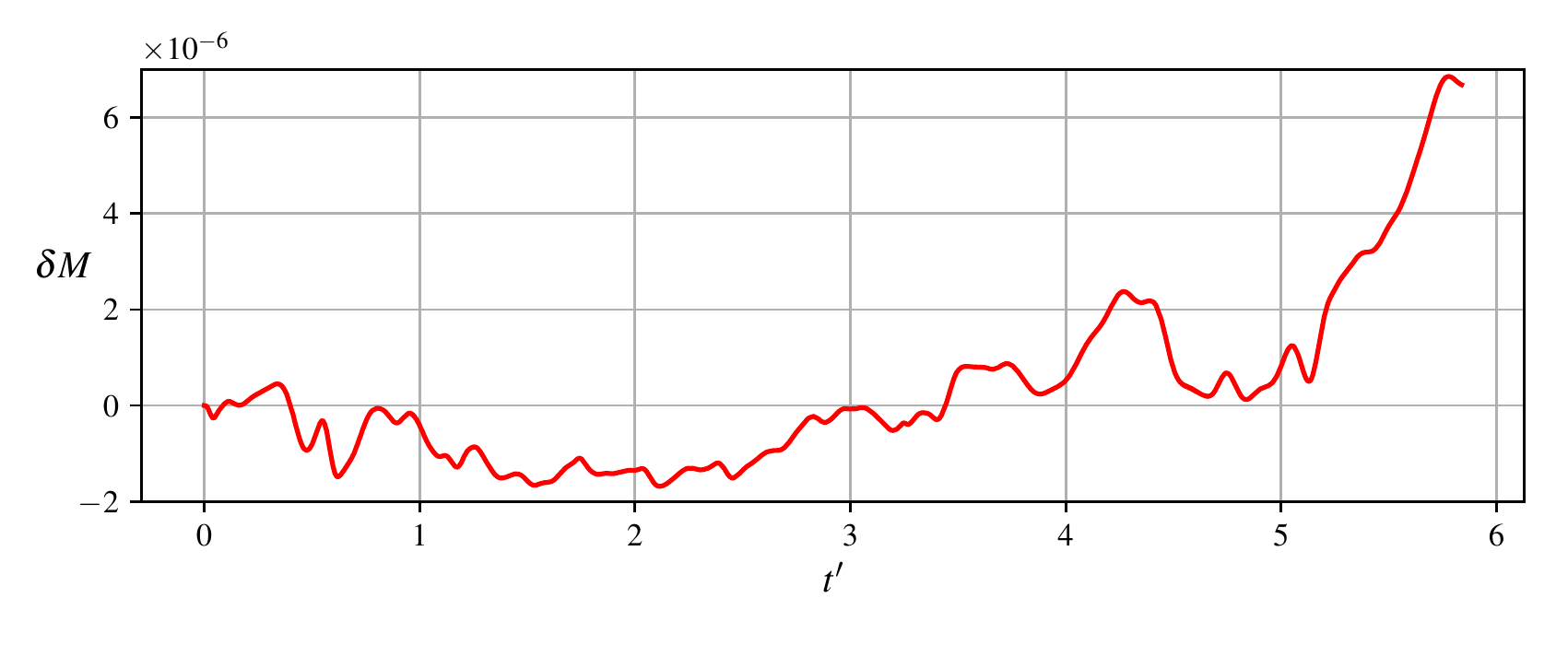}
\includegraphics[width=0.48\linewidth,trim={0 0 0 0},clip]{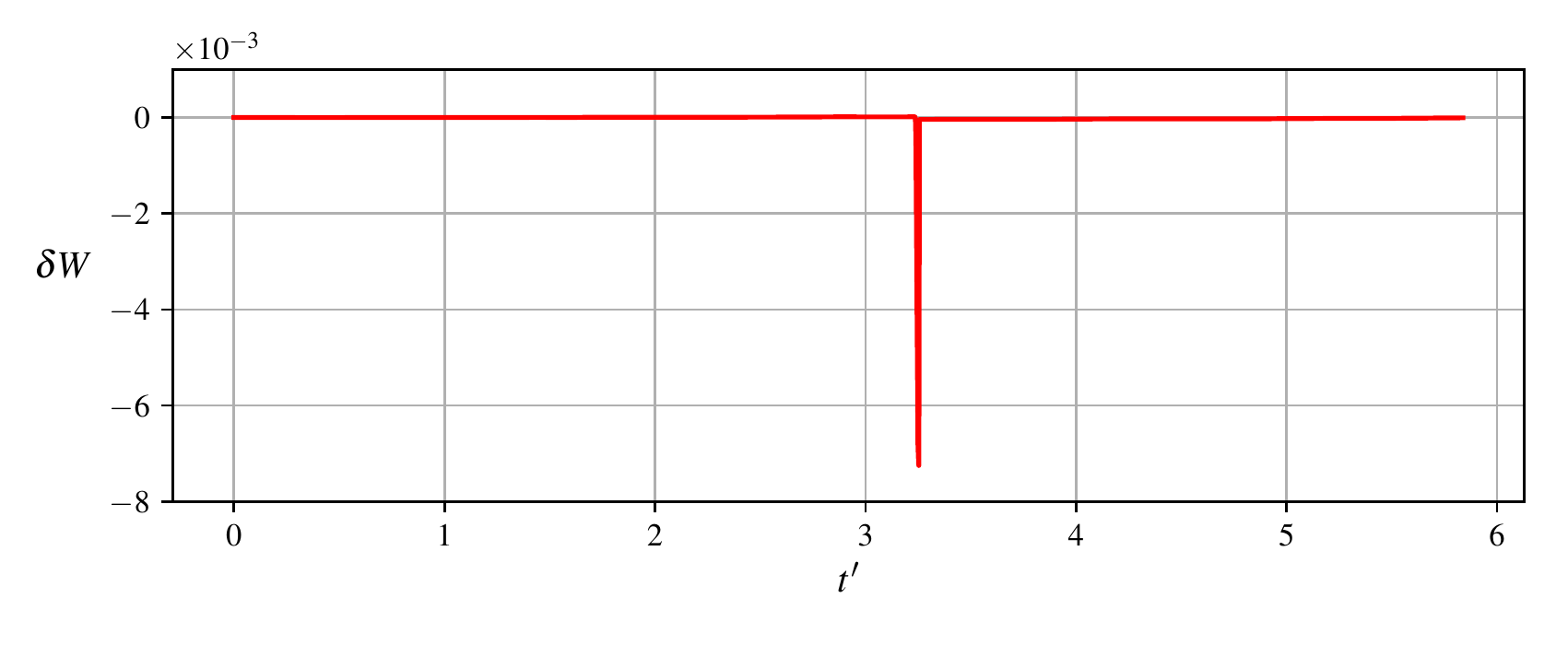}
\caption{Dynamics of a positive (top) and negative (bottom) point-like vortex on
a domain defined by a circular mapping: time evolution of the relative
errors on the total mass (left) and energy (right).}
\label{fig_vic_invariants}
\end{figure}

\begin{figure}
\centering
\hspace{-1em}
\includegraphics[width=0.48\linewidth,trim={1cm 0 1cm 0},clip]{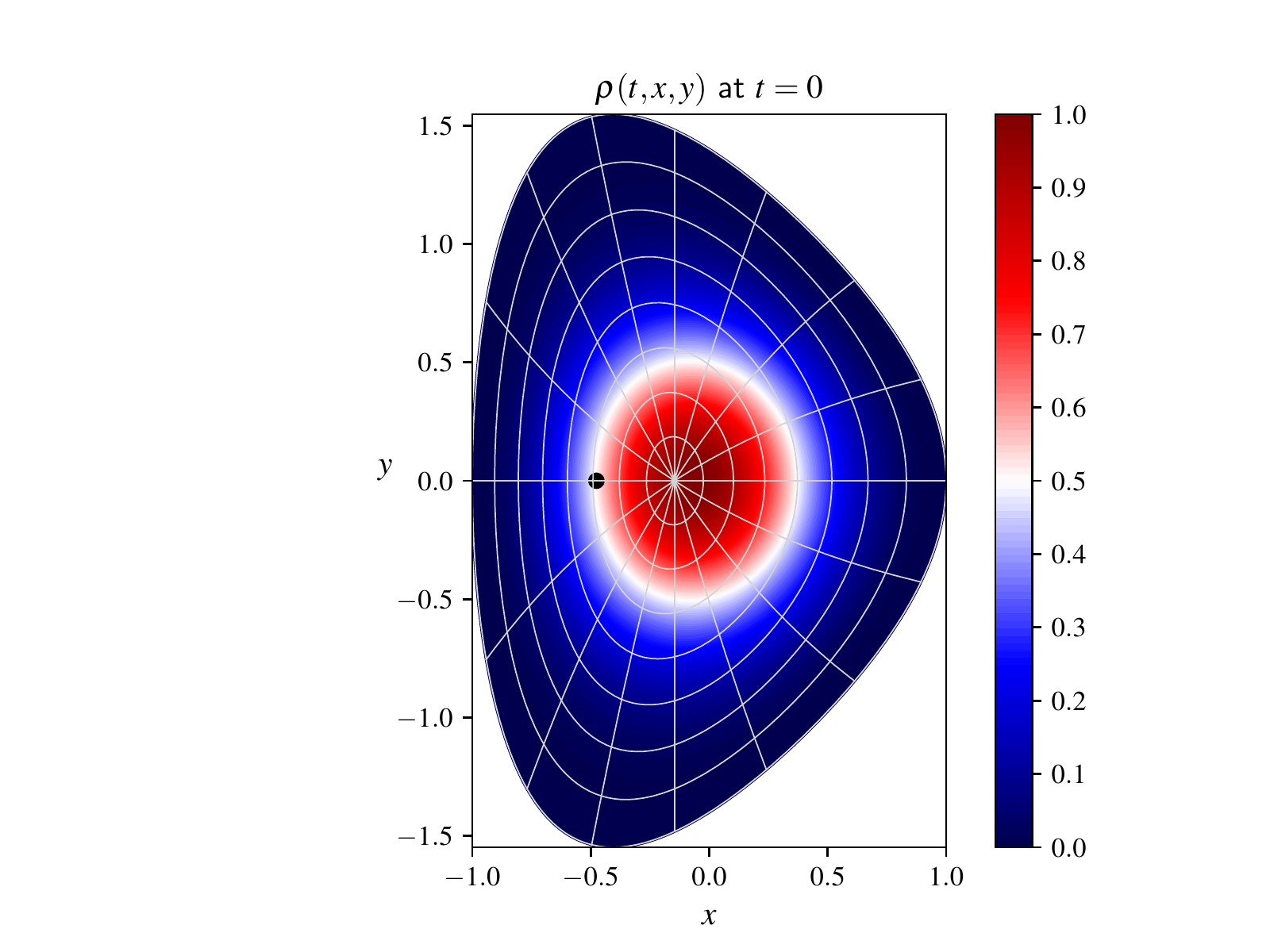}
\includegraphics[width=0.48\linewidth,trim={1cm 0 1cm 0},clip]{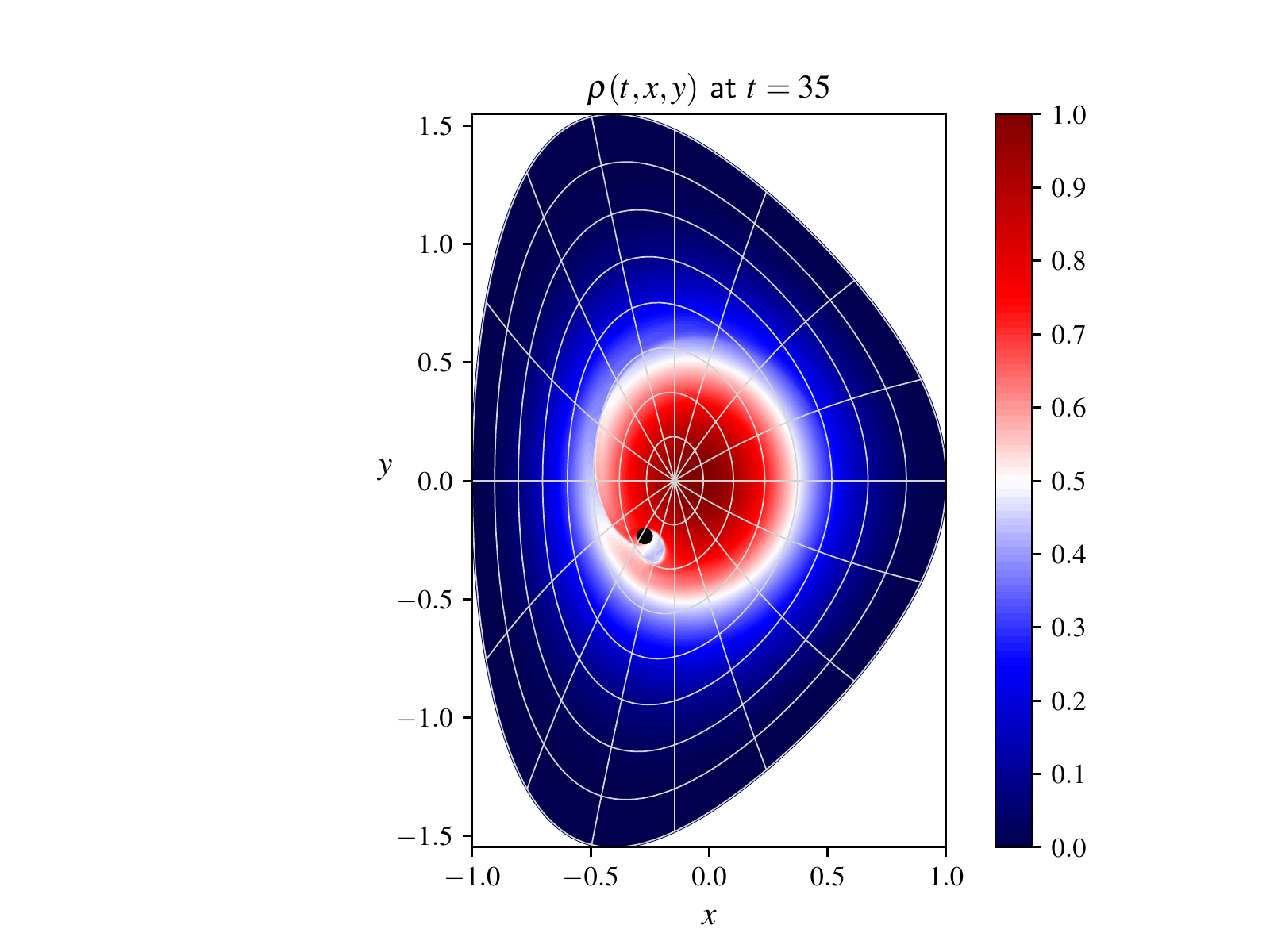}
\\ \hspace{-1em}
\includegraphics[width=0.48\linewidth,trim={1cm 0 1cm 0},clip]{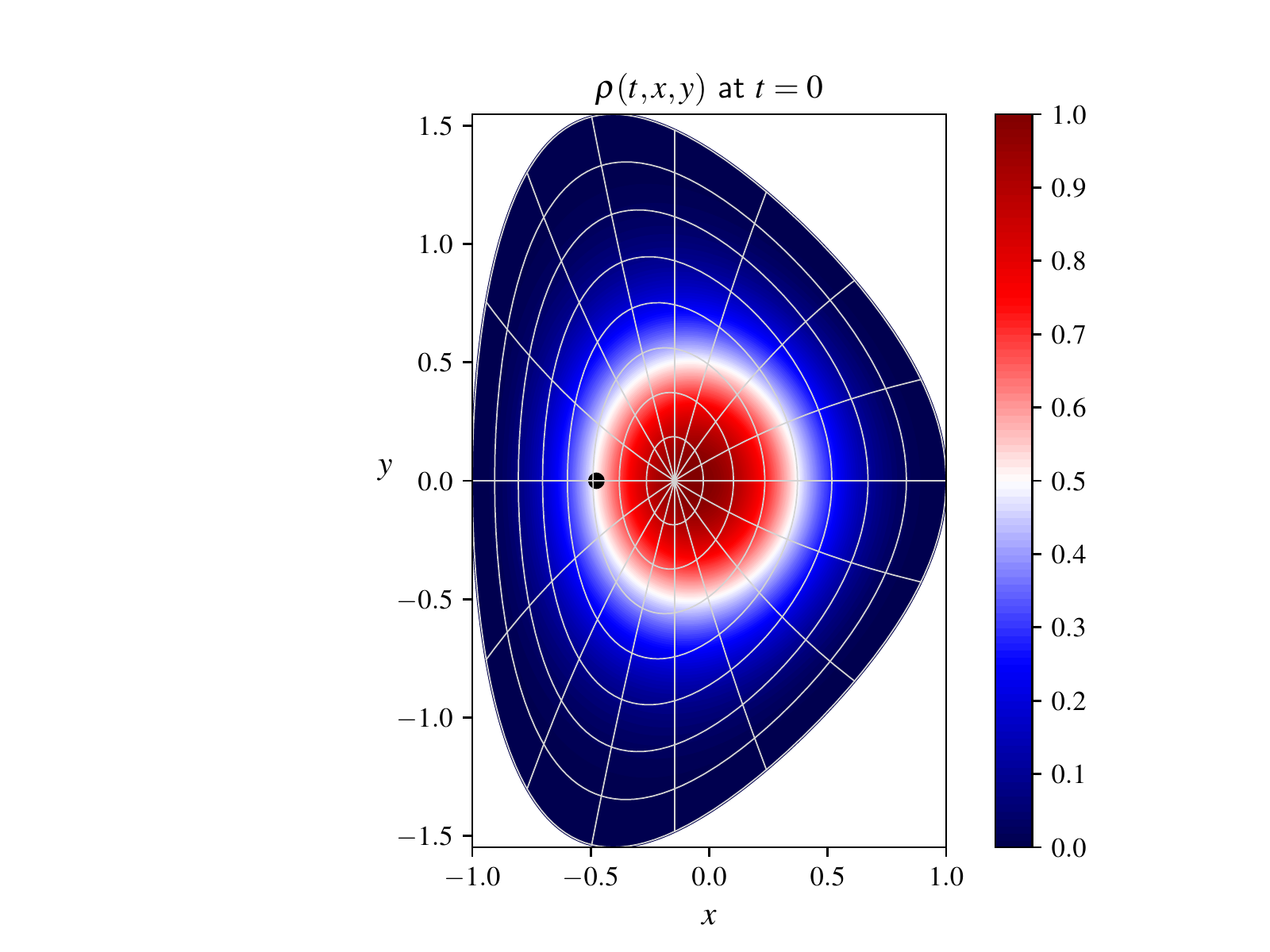}
\includegraphics[width=0.48\linewidth,trim={1cm 0 1cm 0},clip]{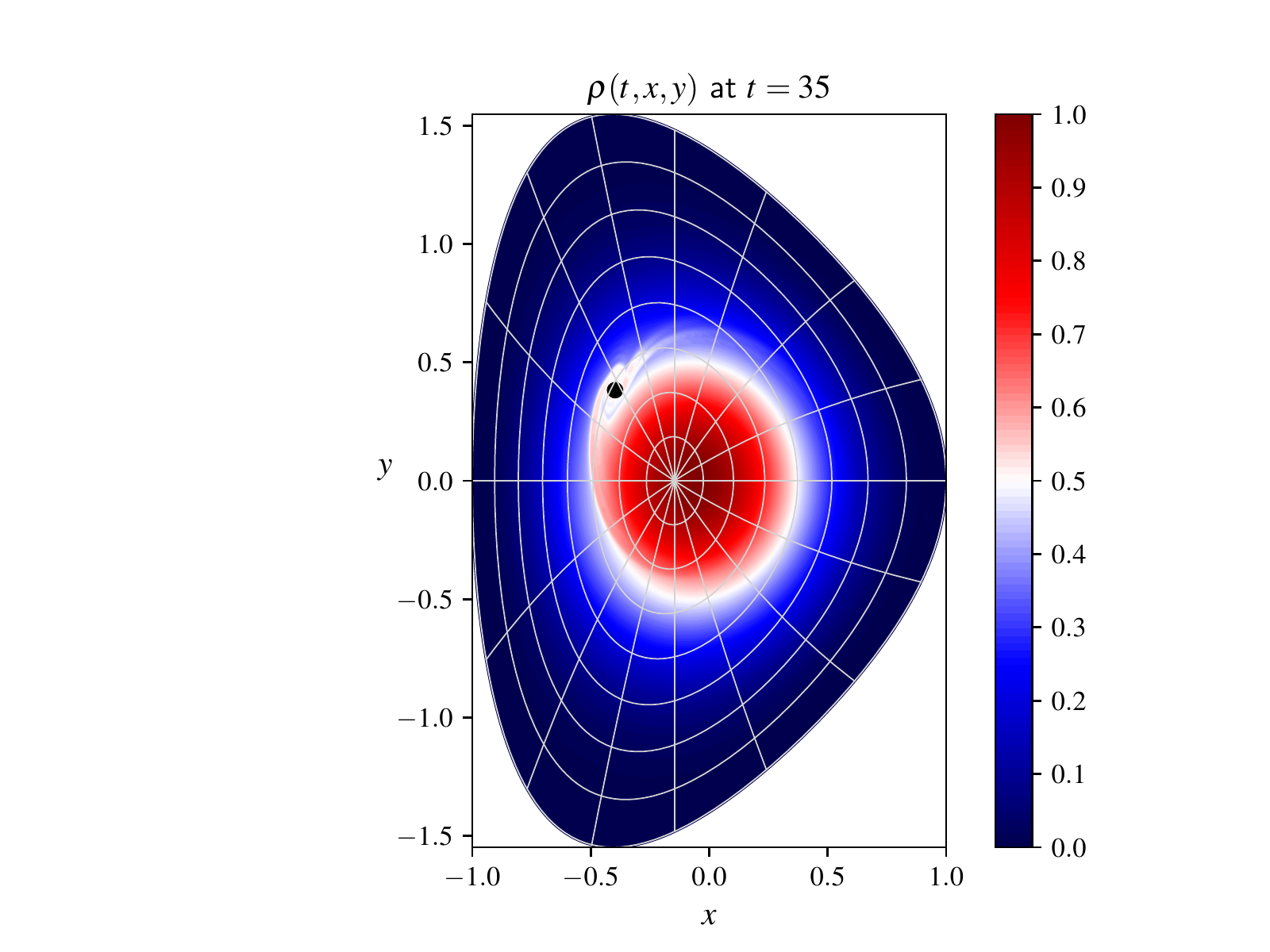}
\caption{Dynamics of a positive (top) and negative (bottom) point-like vortex on
a disk-like domain defined by mapping~\eqref{czarny} with the parameters
in~\eqref{czarny_params}: contour plots of the vorticity at times $t=0$ and $t=35$.}
\label{fig_vic_map}
\end{figure}

\begin{figure}
\centering
\includegraphics[width=0.48\linewidth,trim={0 0 0 0},clip]{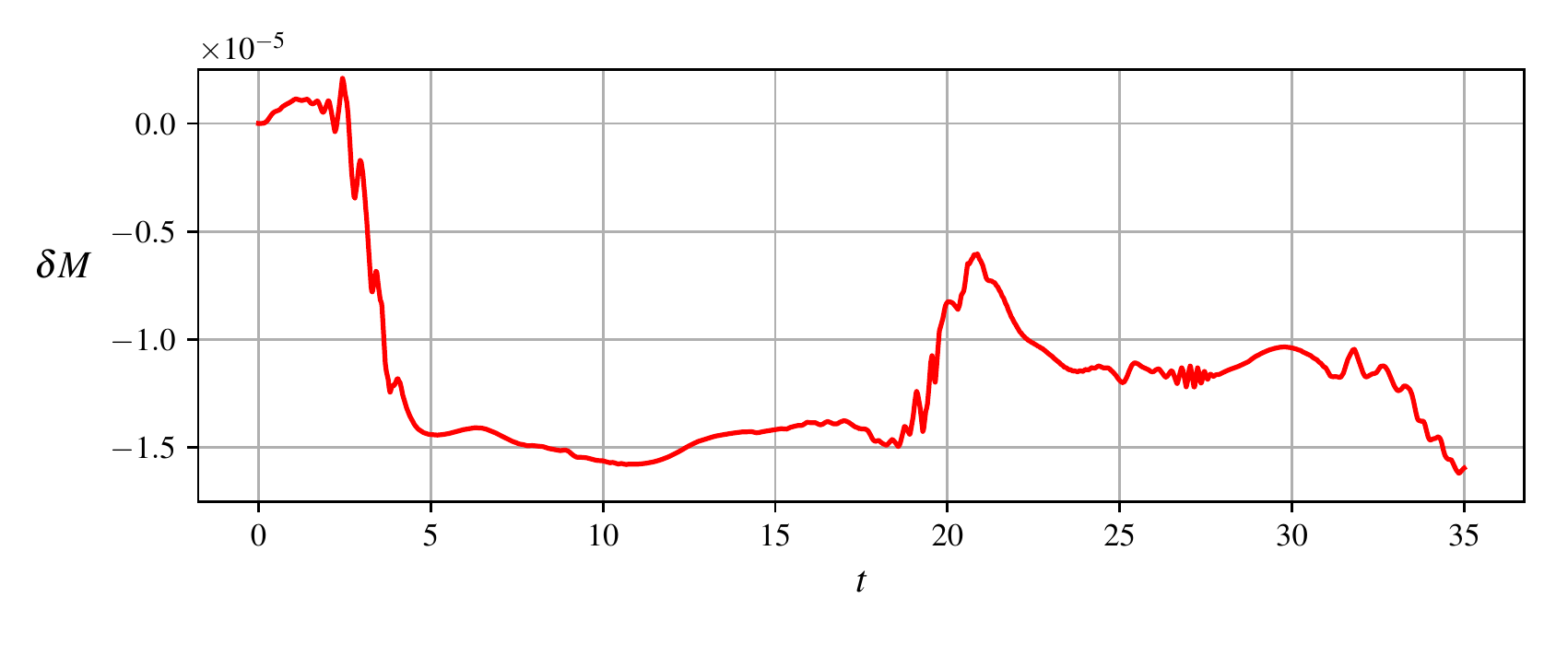}
\includegraphics[width=0.48\linewidth,trim={0 0 0 0},clip]{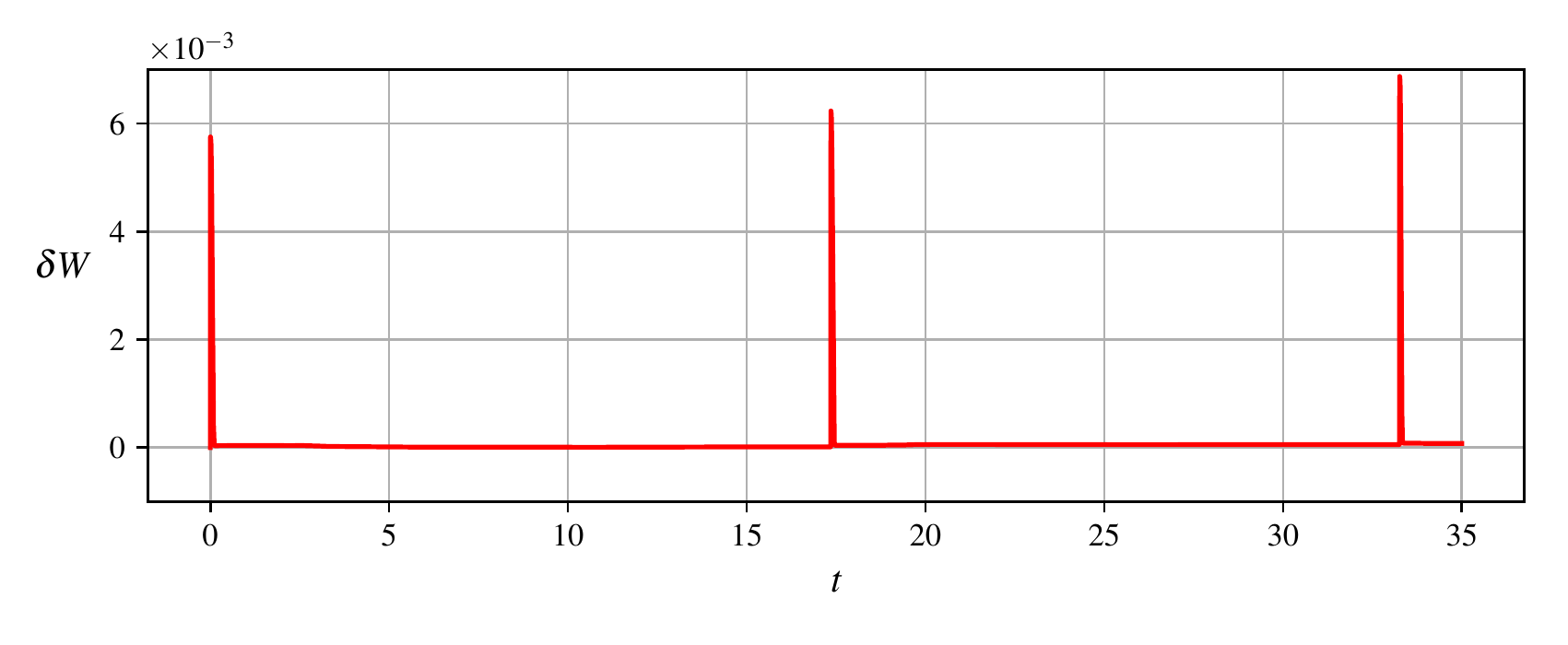}
\includegraphics[width=0.48\linewidth,trim={0 0 0 0},clip]{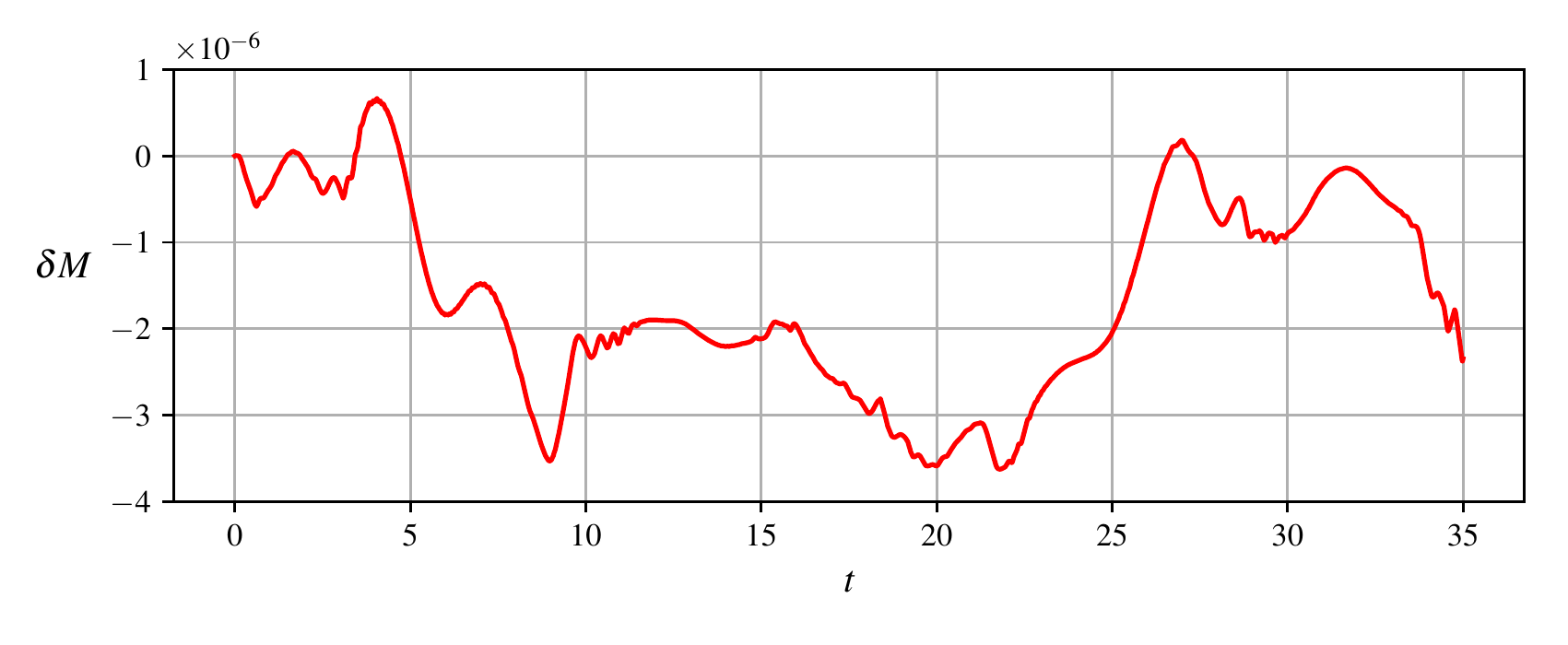}
\includegraphics[width=0.48\linewidth,trim={0 0 0 0},clip]{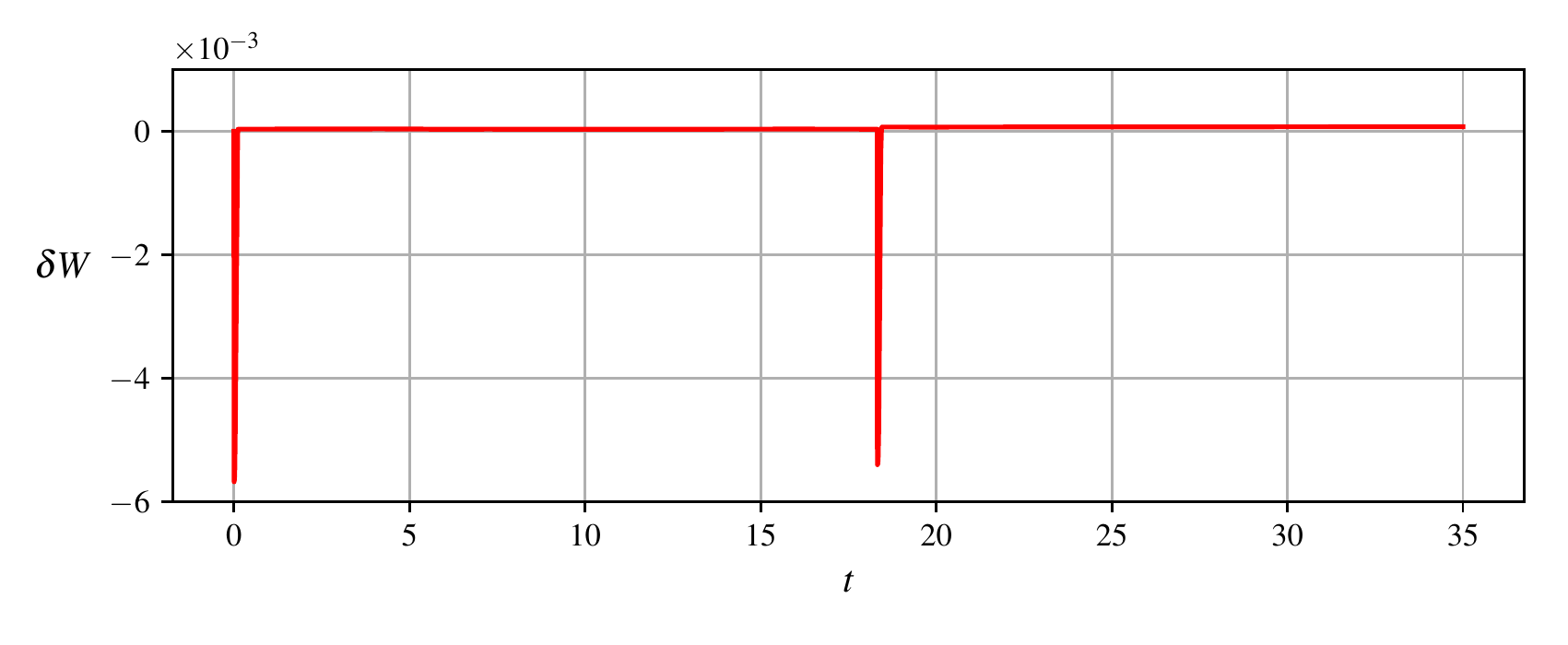}
\caption{Dynamics of a positive (top) and negative (bottom) point-like vortex on
a disk-like domain defined by mapping~\eqref{czarny} with the parameters in~\eqref{czarny_params}:
time evolution of the relative errors on the total mass (left) and energy (right).}
\label{fig_vic_map_invariants}
\end{figure}

We also investigate the dynamics of point-like vortices (or point charges) on a
non-uniform equilibrium, following the discussion in \citep{SchecterDubin2001}.
The numerical tests presented in this section show that the numerical
approaches suggested in this work can be applied straightforwardly in the context of
particle-in-cell methods. This makes our numerical strategy interesting also for
numerical codes based on such methods, as for example many codes developed for
the simulation of turbulence in magnetized fusion plasmas by means of gyrokinetic
models \citep{Ethieretal2005,Wangetal2006,Kuetal2009,Bottinoetal2010}.
The examples discussed here can be considered as limit cases of usual particle-in-cell simulations,
as we will include only one single point-like vortex (or point charge) in the system.
Since our strategy turns out to work well for this extreme scenario, we do not expect
issues when dealing with the usual case of large numbers of particles.
The point-like vortex contributes to the total charge density as described
in equation \eqref{eq_strong_poisson}. Moreover, the position of the point-like
vortex is evolved following the same advection field $(-E^y,E^x)^T$ responsible for
the transport of $\rho$. Integration in time is performed with the second-order explicit
scheme described in section \ref{sec_explicit}. For a domain defined by a
circular mapping, we consider an equilibrium vorticity of the form
\begin{equation*}
\wh\rho_0(s):=
\begin{dcases}
1-1.25\,s & s\leq 0.8 \,, \\
0 & s > 0.8 \,,
\end{dcases}
\end{equation*}
identical to the one considered in \citep[section IV]{SchecterDubin2001}.
Figure \ref{fig_vic_stream_lines} shows the local stream lines of the advection field
near positive and negative point-like vortices at the initial time in a rotating frame
where the point-like vortices are initially at rest. This is obtained in practice by
rotating given coordinates $(x,y)$ at time $t$ as
\begin{equation*}
\begin{aligned}
& x' := x\cos(-\omega t) - y\sin(-\omega t) \,, \\
& y' := x\sin(-\omega t) + y\cos(-\omega t) \,,
\end{aligned}
\end{equation*}
and by transforming the advection field $(-E^y,E^x)^T$ to the rotated
field $(E^y+\omega\,y,E^x-\omega\,x)^T$,
where $\omega=0.3332$ represents the angular velocity of the background at $t=0$
and $s=0.4$.
Figure \ref{fig_vic} shows results for a point-like vortex
of intensity $q=\pm 0.0025$ at the initial position $s=0.4$ and $\theta=0$, again
viewed in a rotating frame. Time is here normalized as $t'=0.1668\,t$ (as in
\citep{SchecterDubin2001}, where $t'$ is denoted as $T$). The results shown in Figure~\ref{fig_vic}
are in agreement with the ones shown in Figures 7a and 10a of \citep{SchecterDubin2001}.
As explained in \citep{SchecterDubin2001}, positive point-like vortices drift transverse
to the shear flow, up the background vorticity gradient, while negative point-like
vortices drift down the gradient. Figure \ref{fig_vic_radial_position} shows the
time evolution of the radial position of the vortices, in agreement with
the results shown in Figures 7b and 10b of \citep{SchecterDubin2001}.
The simulation is run with $n_1\times n_2=256\times512$ and time step $\Delta t = 0.005$.
The time step is chosen small enough to resolve the oscillations due to the self-force
experienced by the point-like vortices. Moreover, the computational mesh needs
to be finer than the previous test cases in order to capture correctly the complex
nonlinear dynamics of the interaction between the point-like vortices and the background
vorticity. For a rough comparison, the vortex-in-cell simulations discussed in \citep{SchecterDubin2001}
require as well a large computational rectangular grid of size $1025\times1025$.
Special techniques may be used to reduce
self-force effects on non-uniform meshes (or even unstructured meshes) \citep{Bettencourt2014},
but they are not considered in this work. For the conservation of mass and energy we get
\begin{equation*}
\max_{t'\in[0,5.838]}\delta\Mcal(t')\approx 6.9\times10^{-6} \,, \quad
\max_{t'\in[0,5.838]}\delta\Wcal(t')\approx 8.4\times10^{-3}
\end{equation*}
for the positive point-like vortex and
\begin{equation*}
\max_{t'\in[0,5.838]}\delta\Mcal(t')\approx 6.9\times10^{-6} \,, \quad
\max_{t'\in[0,5.838]}\delta\Wcal(t')\approx 7.3\times10^{-3}
\end{equation*}
for the negative point-like vortex. The time evolution of the relative errors on
these conserved quantities is shown in Figure \ref{fig_vic_invariants}.
Similar simulations on mapping \eqref{czarny}, initialized with an equilibrium
vorticity obtained with the numerical procedure described
in section \ref{num_equil} with $f(\phi)=\phi^2$ and $\rho_\text{max}=1$, show the same
qualitative behavior: a positive point-like vortex drifts towards the center
of the domain, while a negative point-like vortex drifts towards the boundary
(Figure \ref{fig_vic_map}). The final time $t=35$ corresponds to the normalized
time $t'=5.838$ considered before.
For the conservation of mass and energy we get
\begin{equation*}
\max_{t\in[0,35]}\delta\Mcal(t)\approx 1.6\times10^{-5} \,, \quad
\max_{t\in[0,35]}\delta\Wcal(t)\approx 6.9\times10^{-3}
\end{equation*}
for the positive point-like vortex and
\begin{equation*}
\max_{t\in[0,35]}\delta\Mcal(t)\approx 3.6\times10^{-6} \,, \quad
\max_{t\in[0,35]}\delta\Wcal(t)\approx 5.7\times10^{-3}
\end{equation*}
for the negative point-like vortex. The time evolution of the relative errors on
these conserved quantities is shown in Figure \ref{fig_vic_map_invariants}.

\section{Conclusions and outlook}
We presented a comprehensive numerical strategy for the solution of
systems of coupled hyperbolic and elliptic partial differential equations on disk-like
domains with a singularity at a unique pole, where one edge of the rectangular logical
domain collapses to one point of the physical domain. We introduced a
novel set of coordinates, named pseudo-Cartesian coordinates,
for the integration of the characteristics of the hyperbolic equation of the system.
Such coordinates are well-defined everywhere in the computational domain, including
the pole, and provide a straightforward and relatively simple solution for dealing
with singularities while solving advection problems in complex geometries.
They reduce to standard Cartesian coordinates in the case of a circular mapping.
Moreover, we developed a finite element elliptic solver based on globally
$\Ccal^1$ smooth splines \citep{Toshniwaletal2017}. In this work we considered only $\Ccal^1$
smoothness, but higher-order smoothness, consistent with the spline degree, may be
considered as well. We tested our solvers on several test cases in the simplest
case of a circular domain and in more complex geometries. The numerical methods presented
here show high-order convergence in the space discretization parameters, uniformly across
the computational domain, including the pole. Moreover, the techniques discussed
can be easily applied in the context of particle-in-cell methods and are not
necessarily restricted to semi-Lagrangian schemes, which were here discussed in more
detail. The range of physical problems that can be approached following the ideas
presented in this work includes the study of turbulence in magnetized fusion
plasmas by means of Vlasov-Poisson fully kinetic models as well as drift-kinetic
and gyrokinetic models, and turbulence models for incompressible inviscid Euler fluids
in the context of fluid dynamics.

\section*{Acknowledgments}
We would like to thank Eric Sonnendr\"ucker for introducing us to the idea of using
$\Ccal^1$ smooth spline basis functions and for constantly supporting this work,
Ahmed Ratnani and Jalal Lakhlili for helping us with the implementation of the finite
element elliptic solver and the choice of the data structure to be used for that purpose,
Omar Maj and Camilla Bressan for helping us with the problem of finding numerical
equilibria on complex mappings. We would like to thank also the anonymous reviewers
involved in the peer-review process for carefully reading our manuscript and for
giving valuable and helpful comments and suggestions in order to improve the quality
and clarity of this article. This work has been carried out within the framework
of the EUROfusion Consortium and has received funding from the Euratom research and
training program 2014-2018 and 2019-2020 under grant agreement No 633053. The
views and opinions expressed herein do not necessarily reflect those of the European
Commission. This work has been carried out within the EUROfusion Enabling Research
project MAGYK. Simulation results in section \ref{sec_pointlike} have been obtained
on resources provided by the EUROfusion High Performance Computer (Marconi-Fusion)
through the project \textit{selavlas}.

\bibliographystyle{unsrtnat}
\bibliography{refs}

\end{document}